\documentclass[11pt,preprint]{aastex}
\usepackage{lscape}

\def \etal{{et~al.\null}}

\def\lea{\mathrel{<\kern-1.0em\lower0.9ex\hbox{$\sim$}}}
\def\gea{\mathrel{>\kern-1.0em\lower0.9ex\hbox{$\sim$}}}
\newcommand{\lta}{{\>\rlap{\raise2pt\hbox{$<$}}\lower3pt\hbox{$\sim$}\>}}
\newcommand{\gta}{{\>\rlap{\raise2pt\hbox{$>$}}\lower3pt\hbox{$\sim$}\>}}

\begin{document}


\shorttitle{LEGUS and H$_{\alpha}$ Observations of NGC 4449}



\title{LEGUS and H$_{\alpha}$-LEGUS Observations of Star Clusters in NGC 4449: Improved Ages  and
  the Fraction of Light in Clusters as a Function of Age}

\author{\sc 
Bradley C. Whitmore\altaffilmark{1},
Rupali Chandar\altaffilmark{2},
Janice Lee\altaffilmark{3},
Leonardo Ubeda\altaffilmark{1} ,
Angela Adamo\altaffilmark{4},
Alessandra Aloisi\altaffilmark{1},
Daniela Calzetti\altaffilmark{5},
Michele Cignoni\altaffilmark{6}, 
David Cook\altaffilmark{3},
Daniel~A.~Dale\altaffilmark{7},
B. G. Elmegreen\altaffilmark{8},
Dimitrios Gouliermis\altaffilmark{9,10},
Eva K. Grebel\altaffilmark{11},
Kathryn Grasha\altaffilmark{12},
Kelsey E. Johnson \altaffilmark{13},
Hwihyun Kim\altaffilmark{14},
Elena Sacchi\altaffilmark{1},
Linda J. Smith\altaffilmark{1},
Monica Tosi\altaffilmark{15},
Aida Wofford\altaffilmark{16}}

\affil{email: whitmore@stsci.edu}


\altaffiltext{1}{Space Telescope Science Institute, Baltimore, MD, USA, 21218}
\altaffiltext{2}{Department of Physics \& Astronomy, The University of Toledo, Toledo, OH, USA, 43606}
\altaffiltext{3}{Infrared Processing and Analysis Center, California Institute of Technology, Pasadena, CA}
\altaffiltext{4}{Department of Astronomy, The Oskar Klein Centre, Stockholm University, Stockholm, Sweden}

\altaffiltext{5}{Department of Astronomy, University of Massachusetts-Amherst, Amherst, MA 01003, USA}

\altaffiltext{6}{INFN, Largo B. Pontecorvo 3, 56127, Pisa, Italy}

\altaffiltext{7}{Department of Physics \& Astronomy, University of Wyoming, Laramie, WY USA}

\altaffiltext{8}{IBM Research Division, T. J. Watson Research Center, Yorktown Heights, NY 10598, USA}

\altaffiltext{9}{Institut f̈ur Theoretische Astrophysik, Zentrum  f̈ur Astronomie der Universit ̈at Heidelberg, D-69120 Heidelberg, Germany}

\altaffiltext{10}{Max Planck Institute for Astronomy, D-69117 Heidelberg, Germany}

\altaffiltext{11}{Astronomisches Rechen-Institut, Zentrum f\"ur Astronomie der Universit\"at Heidelberg, M\"onchhofstr, 12-14, 69120 Heidelberg, Germany}


\altaffiltext{12}{Research School of Astronomy and Astrophysics, Australian National University, Canberra, Australia
ARC Centre of Excellence for All Sky Astrophysics in 3 Dimensions, Australia}

\altaffiltext{13}
{Department of Astronomy, University of Virginia, Charlottesville, VA, 22904-4325}

\altaffiltext{14}{Gemini Observatory, Casilla 603, La Serena, Chile}

\altaffiltext{15}{INAF � Osservatorio di Astrofisica e Scienza dello Spazio, Bologna, Italy}

\altaffiltext{16}{Instituto de Astronomía, Universidad Nacional Autónoma de Mexico, Unidad Academica en Ensenada, Km 103 Carr. Tijuana-Ensenada, Ensenada 22860, Mexico}

\begin{abstract}

We present a new catalog and results for the cluster system of the starburst galaxy NGC~4449 based on multi-band imaging observations 
taken as part of the LEGUS and H$_{\alpha}$-LEGUS surveys.
We improve the spectral energy fitting method used to estimate cluster ages and find that the results, particularly for older clusters, are in better agreement with those from spectroscopy.
The inclusion of H$_\alpha$ measurements, the role of stochasticity for low mass clusters, the assumptions about reddening, and the choices of SSP model and metallicity all have
important impacts on the age-dating of clusters. 
A comparison with ages derived from stellar color-magnitude diagrams for partially resolved clusters shows reasonable agreement, but large scatter in some cases. 
The fraction of light found in clusters  relative to the total light (i.e., $T_L$) in the $U$, $B$, and $V$ filters in 25 different $\approx$ kpc-size regions throughout NGC~4449 
correlates with both the specific Region Luminosity, $ R_L$, and the dominant age of the underlying stellar population in each region.
The observed cluster age distribution is found to  decline over time as $dN/d\tau \propto \tau^{\gamma}$, with $\gamma=-0.85\pm0.15$, independent of cluster mass, and is consistent with strong, early  cluster disruption. The mass functions of the clusters can be described by a power law with $dN/dM \propto M^{\beta}$ and $\beta=-1.86\pm0.2$, independent of cluster age.
The mass and age distributions are quite resilient to differences in age-dating methods.
There is tentative evidence for a factor of $2-3$ enhancement in both the star and cluster formation rate $\approx$ 100 - 300  Myr ago, indicating that cluster formation tracks star formation generally. The enhancement is probably associated with an earlier interaction event. 
\end{abstract}

\section{Introduction}
Are most stars born in clusters or in the field? Does the fraction of
stars found in clusters remain constant, change over time,  or vary with
the environment within a galaxy ?  These questions are the primary
focus of this paper.

The discovery of large numbers of massive ($\gea10^5~M_{\odot}$), young ($\lea \mbox{few}\times100$~Myr) 
``super star clusters''
in merging and starbursting galaxies led to the idea that in these galaxies, at
least, a large percentage of star formation occurred in clusters
(e.g., Meurer et al. 1995, Whitmore \& Schweizer 1995). 
Subsequently, Larsen \& Richtler (1999), using multi-band ground based observations, discovered that massive young clusters are also forming in normal spiral galaxies, albeit in smaller numbers, as appropriate for their lower star formation rate (SFR).  

In a followup paper including starbursts, spirals, and a handful of mergers, Larsen \& Richtler (2000) 
determined that the fraction of $U$ band light coming from clusters relative to the total galaxy, $T_L(U)$,
ranged from $<1$\% to $\sim15$\%, and that $T_L(U)$ increased with the ratio of far infrared to B-band flux and the optical surface brightness of the host galaxy.
Converting the IR luminosities to SFR, they found that $T_L(U)$ also correlates with both SFR and SFR per unit area ($\Sigma_{SFR}$).

More recently, following Bastian (2008), several studies (e.g., Goddard et al. 2010; Adamo et al. 2015; Johnson et al. 2016) have attempted to convert the
measurements of the fraction of {\em light} in clusters,  T$_L$, 
to the fraction of {\em stellar mass} in clusters relative to the total mass of the galaxy, ($\Gamma$), a more physically-motivated quantity but one that requires more assumptions and extrapolations (for example, extrapolating the mass function below the observational limit to include the mass from all clusters).

Observations of both starburst and spiral galaxies suggest that many or most of their young clusters disrupt soon after their formation, depositing their remaining stars into the field (e.g., Whitmore 2004, Fall et al. 2005, Whitmore et al. 2007, although see Johnson et al. 2017 for a different view). 
If this is the case, then both the fraction of light and of mass found in clusters should decline with age.  However, clusters also fade rapidly with time, which complicates the interpretation of $T_L(U)$ when mixed-age cluster populations are present, since a given cluster luminosity limit  includes clusters of very different ages.
In this situation, it is possible that a higher fraction of very young, luminous clusters are included in galaxies with higher rates of star formation (and $\Sigma_{SFR}$) relative to those with lower rates. This would artificially increase $T_L(U)$ measured for galaxies with high SFR and $\Sigma_{SFR}$.

To get around this issue in this work we take a new approach, and  measure the fraction of light emitted from clusters T$_L(\lambda)$ in the 
starburst NGC~4449, but in roughly kpc-size sub-regions designed to isolate areas that appear to be dominated by stellar populations with a narrow range in age.
A similar strategy was used in Kim et al. (2012) to study 50 regions in M83.
This approach has several potential advantages over previous works that used entire galaxies (which have more mixed-age cluster populations), since the ability to isolate regions dominated by clusters of different ages simplifies the interpretation of $T_L(U)$, although it also results in low number statistics in some regions. 
The method provides an alternative way of studying cluster formation and disruption, and is largely complementary to the approach of studying entire galaxies taken in most previous studies.

We have selected NGC~4449 for this study,  a nearby (distance = 3.82 Mpc; Annibali et al. 2008), well studied ``starburst'' galaxy with a rich
population of young, intermediate, and old clusters. It is part of the
LEGUS (Legacy Extragalactic UV Survey), which has imaged 50 nearby star-forming galaxies in five broad-band filters using the Hubble Space Telescope (Calzetti et
al. 2015). 
It has a M$_B$ magnitude of -18 and is considered a dwarf galaxy by some authors. This galaxy also has narrow-band imaging that covers the H$_{\alpha}$ line (including the adjoining [NII] lines), and is part of the H$_\alpha$-LEGUS survey (Chandar et al. 2019).  NGC 4449 is a particularly good galaxy for this study since it is possible to isolate regions that appear to be dominated by stars and clusters of a single age.
The 25 regions identified in Figure~1 will be used for this purpose.

One of the primary goals of our study is to determine if values of
T$_L(\lambda)$ depend on the ages of the stars and clusters that appear to dominate the  integrated light in  a
given region.  Accurate ages are therefore required, hence we begin by
comparing the measurement of cluster ages using a variety of
commonly used age-dating methods (i.e.,
integrated colors, spectroscopy, stellar color-magnitude diagrams, emission line ratios in HII regions). NGC 4449 is sufficiently close that we can study both the clusters and underlying stellar population directly. 
We note that in their recent review, Krumholz et al. (2019) suggest that the details of how cluster catalogs are treated can lead to different conclusions about cluster disruption; we test this suggestion in NGC~4449 by comparing the results from a variety of different age dating methods.
Finally, we also examine a number of general properties of the clusters, such as the age distributions
and mass functions. We compare the star formation history (SFH) derived from the stellar component with the cluster age distribution to help disentangle the cluster formation and disruption rates.  In addition, we compare enhancements in the age distributions of the clusters and stars to see if they are similar, which would imply a close link between the formation of stars and clusters. We also examine various properties as a function of position in the galaxy to determine if there are environmental dependencies. 

The remainder of this paper is organized as follows: 
\S 2 describes the observations and selection of clusters,
\S 3 discusses features in the cluster color-color diagram, including reddening and the effect of stochasticity,
\S 4 presents our age dating method, which includes both broad- and narrow- band photometry, and compares our age results with those from the LEGUS survey, 
\S 5 compares our ages with those determined from spectroscopy, CMDs, and HII regions,
\S 6 examines the fraction of light in clusters and how it correlates with region  and age, 
\S 7 discusses general cluster properties such as the mass functions and age distributions, 
and \S 8 summarizes the results.

\section{Observations and Reductions}

NGC 4449 has been observed with three generations of cameras onboard the {\em Hubble Space Telescope} ($HST$).  Figure~2 shows the
coverage with the Wide Field Planetary Camera 2 (WFPC2 - see Gelatt et al. 2001),
Advanced Camera for Surveys (ACS - see Annibali et al. 2008, Rangelov et al. 2011), and Wide Field Camera 3 (WFC3 - see Calzetti et al. 2015).
In this work, we focus on the ACS and WFC3 observations.
The new WFC3 observations have a scale of 0.04" per pixel.
We adopt a distance of 3.82 Mpc to NGC~4449, corresponding to a
distance modulus of 27.91 mag, as determined by Annibali
\etal\ (2008) using the tip of the red giant branch method.  Hence
1$''$ is equivalent to 18.7~pc, 
and
1 WFC3 pixel is equivalent to 0.75~pc.

Note that many of the observations have been restricted to the central
star forming portion of the galaxy (e.g, the WFC3 LEGUS observations, PI = Calzetti, proposal ID = 13364); only the ACS (F438W, F555W, F658N, F814W 
filters, PI = Aloisi, proposal ID = 10585) imaged the outer parts of the galaxy (see Figure~2).  The availability of only three broadband filters in the outer
regions affects the age dating of the clusters and stars at some
level,
a topic that will be discussed in \S 4.2 .  
Observations of
the central region (including observations from LEGUS - Calzetti et
al. 2015) provide the widest wavelength coverage, including both the
ACS filters listed above and the F275W and F336W filters from WFC3. The galaxy actually extends to much larger radii than shown in Figure~1, with evidence
of former interactions (from two different dwarf companions) in the range  100 - 500 Myr  ago (Hunter et al. 1998,
1999, Theis \& Kohle 2001,  Karachentsev et al. 2007, Martinez-Delgado et
al. 2012, and Rich et al. 2012) based on both optical and HI radio
observations. In particular, Hunter et al. 1998 find counter-rotating gas systems and
high velocity dispersions in the outer part of the optical galaxy.

\subsection{Cluster Selection and Photometry}

The initial selection of star cluster candidates in NGC~4449 followed
the basic steps described in Adamo et al. (2017) for LEGUS galaxies.  Briefly, point-like
sources were identified using SExtractor, and sources brighter than
$M_V=-6$ (after including an average aperture correction) that have a concentration index (difference in magnitudes
within 1 and 3 pixel radii) greater than 1.3 were selected as cluster
candidates. 
For reference, isolated stars have a concentration index value around 1.2 (e.g., see Adamo et al. 2017). 
One of the authors (BCW) then visually classified each candidate
cluster using the following categories, as defined in LEGUS: $1=$symmetric extended source, $2=$
asymmetric extended source, $3=$clustered grouping of
close point sources (i.e., compact association), $4=$likely artifact (e.g., individual star, close
pair of stars, background galaxies).
We define a source to be category 3 in NGC~4449 if it has at least 4 stars within a 5 pixel radius.
Out of the original 1361 candidates, 473 were classified as category 1, 2, or 3, while the remaining 888 (i.e., 65 \%) objects were considered artifacts.

In addition to classifying each source visually, a grid search of the images by one of us (BCW) identified cluster candidates that were added from the 
original LEGUS list. In general, these objects were clearly visible but were either slightly below
the $M_V=-6$ limit or were missing from the original SExtractor detection because they were slightly more diffuse than other clusters.
Each of these sources has a peak pixel count of at least $0.1$~ct~sec$^{-1}$. This flux level  was selected since it can be seen against the background level of the galaxy almost into the central region. 
Some of these added objects were in the original source catalog but were removed because they were fainter than the $M_V=-6$ cutoff.  The added sources tend to be more diffuse, and therefore have larger-than-average aperture corrections.  An additional 121 cluster candidates were identified and added to the sample, resulting in a total of 594 category 1+2+3 cluster candidates in the final catalog. This will be called the  H$_{\alpha}$-LEGUS catalog, and is somewhat different from the LEGUS catalog (used in Cook et al. 2019 for example), as described below. Figure~3 shows three examples of objects that were added (the white circles) in Region 23, along with several original category 1 and 2 objects 
for comparison. 
The sample, including the added clusters, was also vetted by Dave Cook as part of the Cook et al. (2019) study. He retained 94 \% of the added cluster candidates.

The High Level Science Product (HLSP) available from the LEGUS website 
contains the cluster categories defined in the Cook et al. (2019) study rather than from the current H$_{\alpha}$-LEGUS study. 
Unlike LEGUS, the H$\alpha$-LEGUS catalog includes narrow-band photometry, does not correct for foreground extinction (the age-dating software fits for the foreground $+$ local extinction), and applies aperture corrections that do not depend on the filter, i.e. no color-dependence is introduced.
We note that the foreground reddening is very low (i.e., E(B-V) = 0.019 according to Schlegel, Finkbeiner, \& Davis 1998)
To obtain the H$_{\alpha}$-LEGUS catalog described in the current paper the website at  https:$//$tinyurl.com$/$halpha-legus must be used.

The addition of these clusters increases the level of completeness in our sample.  The luminosity function of the original sample (i.e., before including the added clusters) for category $1+2$ candidates begins to artificially flatten near $M_V=-6.8$, due to issues with completeness.  A fit to the bright portion of the luminosity function with an extrapolation to fainter magnitudes indicates that the 50 \% completeness level occurs near $M_V=-6.4$, and that the sample is only complete at about the 20 \% level at the $M_V=-6$ cutoff. With the addition of the 121 clusters, the flattening now occurs at $M_V  \approx -6.4$ and the $M_V=-6.0$ cutoff is closer to a 
50 \% completeness level.  
The H$_\alpha$-LEGUS sample presented here has a more gradual cutoff, and includes some
very faint clusters with $M_V \approx-4$ in the outer parts of the galaxy. 
The  addition of  these clusters allows us to more completely examine the ages of clusters in the outer regions with faint backgrounds. 
More stringent criteria (i.e., $M_V=-6.4$; $\approx$ 80 \% completeness) are imposed for various subsamples when constructing the mass and age distributions, as will be discussed in \S 7.

Little effort was made to add category$=3$ compact associations (i.e., only 4 of the added 121 cluster candidates),
since this becomes quite a difficult and subjective exercise in crowded regions. In
general, the category 3 populations should be considered less certain for
this reason than category 1 and 2 sources.  While their inclusion 
provides a way to study the properties of the lower density stellar
groupings, their completeness and absolute numbers are not as well
defined.

The final number of objects in categories 1, 2, and 3 are 120 (20 \%), 261 (44 \%), and 213 (36\%) respectively.
This is very similar to the relative percentages found for other LEGUS galaxies, as reported in Grasha et al. (2017) and 
Kim et al. (2019), with a slightly larger fraction of category 2 clusters.

The visual classification was performed by BCW using the normal method of LEGUS classification described in Adamo et al. (2017) (i.e., using a DS9-based tool, the IMEXAMINE task, and the contrast control as the
primary tools), but with just one rather than three classifiers. Color images produced using the ACS F438W,
F555W, F814W, and H$_\alpha$ image from the Hubble Legacy Archive (HLA -
see Whitmore et al. 2016) were also examined during the grid search for each cluster.
This allowed us to include a visual determination 
of the morphology of
associated H$_\alpha$ emission present around each cluster.
This procedure was inspired by the results from the Whitmore et
al. (2011) study of M83 which found a strong correlation between
H$_\alpha$ morphology and cluster age. We use the following classification system for H$_\alpha$ morphology:
objects with H$_\alpha$-class $= 1$ have line emission 
on top of and largely coincident with the candidate cluster,
H$_\alpha$-class $= 2$ show a ring-like structure around the
candidate cluster, H$_\alpha$-class $= 3$ have some diffuse H$_\alpha$
in the general area that 
may or may not be associated with the
object, and H$_\alpha$-class $= 4$ sources show no H$_\alpha$ emission around them at
all. As will be seen in \S 4, the H$_\alpha$ morphology provides very useful
constraints during the age dating procedure.

Photometry was performed using apertures with radii of 5 pixels and sky values in annuli  with radii between 7 and  8 pixels. While different size apertures and assumptions about aperture corrections would affect our results at some level, our experience (e.g., Chandar et al. 2010, Whitmore et al. 2014) has been that this represents a relatively minor uncertainty. 
We do not apply any correction for foreground extinction to the magnitudes, unlike the LEGUS catalog where  small corrections (i.e., 0.03 in F814W to 0.11 in F275W from NED) were made; instead, we fit for the total extinction (foreground plus internal) for each cluster, as described in \S 4.1. These small adjustments would introduce very minor differences in the results; much smaller than the larger effects discussed in \S 4.2 and \S 4.3 (e.g., use of H$_\alpha$ measurements, different SED models, assumptions about reddening).
We apply aperture corrections to the measured magnitudes in two ways: (1) an average aperture correction determined from bright, fairly isolated clusters, and (2) a CI-dependent aperture correction:  $\mbox{Apcorr} = -4.452 + 6.4638\times\mbox{CI} - 2.3469\times \mbox{CI}^2 - 0.04518\times \mbox{CI}^3$, which can be applied over the CI range $1.3-2.23$. 
In both cases the determinations are made from the $V$ band measurement and applied to all filters, to avoid introducing uncertainties in the colors of the clusters.
Cook et al. (2019) demonstrate that the method used to determine aperture corrections has very little impact on the resulting age and mass distributions (see also Chandar et al. 2010).

Figure~4 shows a region (including, but extending outside of region 11 in Figure~1) that illustrates the object selection and classification system. The top panel shows a F555W image while the bottom shows a F438W, F555W, F814W color image from the HLA (Whitmore et al. 2016). Red circles are category 1 (symmetric),
green circles are category 2 (asymmetric), blue circles are category 3 (compact associations). The eight slightly smaller orange circles are  the clusters in this region from a study by Annibali et al. (2011), which will be discussed in \S~5.1. The final log Age values  
are shown in yellow. The diffuse green light in the bottom panel is indicative of emission line flux (i.e., H$_\beta$ at 4861 Angstroms and [OIII] at 5007 and 4959 Angstroms) 
that leaks into the F555W filter. Note that most of the clusters with this green emission have very young ages (i.e., log Age $\approx$ 6.5 - 6.7 Myr).

\section{ Color-Color Diagrams, Reddening, and Stochasticity}

As will be described in \S 4, our cluster age-dating procedure uses a SED fitting procedure to provide estimates of age, reddening and mass. However, a $U-B$ vs. $V-I$ color-color diagram also provides a useful guide to the ages of clusters, and insights  into the role of reddening and stochasticity in the age-dating procedure.

\subsection{Color-Color Diagrams}

In  Figure~5 we present the $U-B$ vs. $V-I$ color-color diagram for the full
cluster catalog (top left), and for each of the 3 cluster categories individually: category 1 or symmetric clusters (top right), category 2 or asymmetric clusters (bottom left), and category 3 or compact associations (bottom right). The 121 clusters that were added to the sample (as described in \S2.1) are shown as open circles. In general we find that their distribution roughly matches the distribution of the original cluster candidates.
A reddening vector with amplitude of $A_v = 1$ (Fitzpatrick 1999) mag is included in each panel. The solid curve in each panel shows the predicted progression from the $1/4$ solar metallicity Bruzual-Charlot (2003) model (as appropriate for young clusters in NGC~4449; Annibali et al. 2011) in color-color space for a cluster as it ages from 1~Myr in the upper left to 10 Gyr in the lower right.  

The locations of key ages from the Bruzual-Charlot models, which are used to produce the H$\alpha$-LEGUS cluster properties, are shown in the 
upper-right panel of Figure~6 
(triangles). We note that the triangles for 1 and 2 Myr have been slightly displaced from each other for clarity; in the Bruzual-Charlot models they actually have identical colors. This is why there are no clusters with age estimates of 1 Myr in the H$_\alpha$-LEGUS catalog.  We also show the predictions from the Zackrisson (2011) Yggdrasil models for the same metallicity (dashed lines) in the upper left panel of Figure~5 and in Figure~6.
The  Yggdrasil models are used to estimate the LEGUS ages.
\footnote{Note that the LEGUS catalog used here adopted a somewhat
different version of the Yggdrasil models than currently available. This results in only minor differences, with most clusters
having identical ages, and fewer than 7\%  having estimated ages
that differ by more than a factor of 2,and does not affect the age or
mass distributions presented here.}

The most notable difference between the Yggdrasil predictions and the Bruzual-Charlot ones used here, is that emission from ionized gas is included in the former (but not the latter), leading to bluer
predicted $V-I$ colors at the youngest ages.
The Yggdrasil models appear to better match the few clusters with very strong line emission, but the colors of the majority of the very young, blue clusters in NGC~4449 appear to better follow the predicted colors of the Bruzual \& Charlot models (i.e., they have values $V-I$ $\approx$ -0.2) .

Category 1 objects are found in two distinct knots in the color-color diagram, old
globular clusters farthest to the bottom right with $V-I$ in the range
1.0 to 1.2, and a second group just above them and to the left. This second group has  $V-I$ values in the range $\approx0.4$ to 0.6, indicative of ages in the  few hundred Myr range. 
We will discuss this second population in more detail in \S 7.3.  There is also a sprinkling of very young clusters with $U-B$ $\approx -1.5$,
indicative of very young ($\approx$few Myr) ages.

Category 2 clusters also fall in two knots in color-color space. The first
is similar to the few hundred Myr old knot found in the category 1 objects, but
extends to slightly younger ages. The second enhancement is similar to the young
distribution in the category 1 diagram, but with roughly a factor of three more objects.

Category 3 objects (``compact associations'') consist of essentially all
young objects, but with a longer extension to the red due 
primarily to the random presence of red supergiants (i.e., stochasticity), which is more important
in these typically lower mass objects - e.g., see Fouesneau et al. 2012. This stochasticity will be discussed in more detail in \S 3.3.

Six snapshot images show typical objects in different parts of Figure~5 ranging from reddish old globular clusters and whitish intermediate-age clusters in category 1 to emission dominated  (greenish) compact associations in the upper left of category 3.  Note that this very young object is better fit with the Yggdrasil models, as expected since these broad-band colors include nebular line$+$continuum emission, while the Bruzual-Charlot models used here do not. However,  this does not appear to affect the age estimates very much since  the seven bluest points in $V-I$ have a median log Age value = 6.0 (i.e., 1 Myr) using LEGUS ages and 6.5 (i.e., 3 Myr) using  H$_\alpha$-LEGUS ages. This is because the inclusion of the narrow-band F658N filter in the H$_\alpha$-LEGUS fitting procedure compensates for the lack of nebular emission in the predicted broad-band colors from the Bruzual-Charlot models, as will be discussed in \S 4.

\subsection{Constraints on Reddening Towards Clusters  in NGC 4449}

In this section we use the color-color diagram to set constraints on the maximum amount of reddening allowed by the SSP age-dating algorithm that will be discussed in more detail in \S 4.  
Constraints on the expected range of extinction values towards optically visible clusters can help to improve the age dating results for a given galaxy. Most SED fitting routines allow any  value of Av, hence a cluster with the colors of an old globular cluster can be appropriately fitted with an age of 10 Gyr and Av $\approx$0 mag, or erroneously fitted with an age $\approx$10 Myr and 
Av $\approx$1.0 mag, because of the degeneracy between age and reddening in broad-band filters. Spectroscopic observations can often be used to remove this degeneracy, as will be shown in \S 5.1 .

In Figure~6  we estimate the highest likely values of reddening in NGC 4449 using the clusters embedded in H$\alpha$ (those with H$\alpha$-class = 1),  by
estimating the amount of reddening towards the clusters that fall redward of the models. We then use this value to set constraints on the maximum reddening allowed by the SSP fitting routine.  Note that a
number of strong H$\alpha$-emitting clusters fall {\em blueward} (to the left) of the Bruzual-Charlot model; we find that these clusters are assigned young ($\approx$few~Myr) ages no matter what assumption we make for $E(B-V)$, and therefore do not consider them when setting constraints on the maximum allowed reddening. As might be expected, most (69 \%) of these strong H$\alpha$ regions are found in the blue boxes in Figure~1, with nine in or near Region 11, and four in Region 15.
Twenty-two \% are found in the yellow boxes, with three each in regions 13 and 24.
Dwarf and lower mass irregular galaxies often have lower extinction (and hence lower reddening) than more massive galaxies
(e.g., Zaritsky et al. 2002 find little or no extinction in the Magellanic Clouds except around the youngest stars).
Hence we might expect NGC 4449 to have relatively low values of reddening as well.

Previously, Whitmore et al. (2011) found evidence for moderate extinction towards very young, embedded clusters in the more massive spiral galaxy M83, based on the locations of strong, H$\alpha$-emitting clusters in the color-color diagram. These values are included as open circles in Figure~6 for comparison with NGC 4449. 

This figure reveals a key difference between the colors of very young clusters in NGC~4449 and M83: in M83 very young, embedded clusters follow the reddening vector nearly all of the way down to the end of the model tracks, but in NGC~4449, the distribution of colors appears to be fairly horizontal rather than following the reddening vector diagonally down and to the right.  
As discussed further in Section~4, this left-ward $\sim$horizontal scatter in $V-I$ probably results from the contamination of gaseous emission lines around young stars in the F555W filter, 
which are included in the Yggdrasil models but not in the Bruzual-Charlot models. 
Only two NGC 4449 data points in Figure~6 are slightly low, and these are consistent with ages of 7 Myr or less. These ages are compatible with expectations for regions with H$\alpha$ emission,  hence there is no need for reddening to explain their location in the color - color diagram, unlike the case for M83. 
We conclude that the clusters in our NGC~4449 catalog appear to have very low total reddening (foreground plus internal), with $E(B-V) \lea 0.2-0.3$~mag.
We adopt an upper limit  $\approx3\times$ larger 
(i.e., E(B-V) = 0.75)
when age-dating our clusters, as described in the \S 4.1. 

Figure 7 shows the reddening values from the LEGUS age-dating solution (upper panel), the reddening values using the H$_\alpha$-LEGUS algorithm 
and limit of E(B-V) $<$ 0.75 (middle panel), and a hybrid using the 0.75 mag limit for the youngest clusters and a value of 0.0 mag for clusters with age estimates greater than 10 Myr (note: this is done in two iterations; the first where reddening is allowed to vary to determine the age and the second where E(B-V) is set to 0.0 mag for the older clusters).  This latter strategy is what is actually used in the final H$_\alpha$-LEGUS catalog, as will be discussed 
in more detail in \S 4.1.  We note that most of the clusters with E(B-V) values greater than 0.4 in the final H$\alpha$-LEGUS fits are those discussed earlier, with strong Halpha emission pushing their colors blueward of the model.

A consistency check is possible by comparing our E(B-V) values with those for HII regions in NGC 4449 based on Balmer decrement observations (Annibali et al. 2017). They find values ranging from 0.10 to 0.24 for six HII regions. This is consistent with our estimate of $E(B-V) \lea 0.2-0.3$~mag from Figure 6 for the objects with strong H$_\alpha$, and also with the mean value of E(B-V) = 0.16 for the 104 clusters with H$_\alpha$-LEGUS ages less than 10 Myr  in Figure 7. 
Annibali et al. (2017) also make estimates of E(B-V) for older planetary nebulae in NGC 4449. While these estimates have larger uncertainties, 4 of the 5 values are consistent with E(B-V) = 0.0. 

Note that the LEGUS  solution (top panel) has a large number of clusters with E(B-V) $\approx$ 1. Essentially all of these objects are actually old globular clusters with overestimated values of E(B-V), based on comparisons with either spectroscopic (Annibali et al. 2018) or integrated photometry (Annibali et al. 2011) observations. This topic will be revisited in \S 5.1.

While our results suggest that it is appropriate to restrict the range of reddening and extinction that is considered in our fitting algorithm for NGC~4449, we note that more massive and metal rich galaxies such M83 or the Antennae require a higher reddening limit (e.g., Whitmore et al. 2011, Whitmore et al. 2010).
 
 \subsection{Effect of Stochasticity} 
 
 If reddening is a relatively minor effect in NGC 4449 then why are there so many points well to the right of the models in Figure 5? In Figure 8 we isolate 79 objects with these colors and in Figure 9 we show part of Region 8, where 10 of these objects reside (i.e.,the yellow circles). {\it In all 10  yellow circles we find that the reddish V-I colors are caused by the presence of red stars in the aperture}. A visual examination of all 79 objects shows that 76 of them have bright red stars in the aperture!
 We note that Johnson et al. (2012 - Figure 12) found a similar distribution of objects in M31.

 This effect is often called stochasticity; the random presence of at least one red supergiant in a low-mass cluster or association. For low mass clusters  the chance of containing a single red supergiant is often less than 50 \%, hence there are no red stars in the aperture
 for some associations. Five of these all blue-star compact associations  are shown by blue circles in Figure 9,  (note that the 5 pixel apertures used to measure the photometry are roughly half the size of the circles shown in Figure 9). 
 The locations in the color-color diagram for these regions with only blue stars are shown by the squares in the top panel of Figure 8. As expected, all five are well to the left of the objects where the red
 stars are found.
 This stochasticity introduces a large random component in the age dating of low-mass clusters, as discussed in several papers (e.g., Maiz Apellaniz 2009, Fouesneau \& Lancon 2010, Fouesneau et al. 2012, Krumholz et al. 2015).

More specifically,  stochastic effects can result in underestimated ages from most SED fitting procedures, since the algorithm assigns a large reddening vector to bring it into better correspondence with the models. This is shown in the middle panel of Figure 8, where LEGUS assigns Log Age = 7 for nearly all of the objects in this part of the diagram. The H$\alpha$-LEGUS age estimates are older, with mean values around 7.5 (i.e., 30 Myr). This is more realistic since there is essentially no H$\alpha$ in the region, indicative of ages greater than 10 Myr.  The bottom panel of Figure 8 shows that age differences estimated by H$_\alpha$-LEGUS are interpreted as large reddening values in the LEGUS estimates. 
  A more detailed discussion of stochasticity, and a potential method of reducing its effects by the "stacking" of objects,  is included in Hannon et al. (2019).

  Hence, the lack of any clear evidence for strong extinction in NGC 4449 (i.e., Figure 6) and the fact that stochasticity can result in the underestimate of cluster ages (i.e., Figure 8) leads us to adopt a "zero-reddening" for cluster ages greater than the 10 Myr solution for the H$_\alpha$-LEGUS catalog, as shown in the bottom panel of Figure 7. 
  We note that  this is also compatible with several recent findings showing that the dust is generally cleared around young clusters in only a few Myrs (e.g., Whitmore et al. 2011,
Hollyhead et al. 2015, Grasha et al. 2017, Matthews et al. 2018).
  In addition, the adoption of the zero reddening solution for older clusters  is  similar to the procedure used by Annibali et al. (2011), who assumed no internal extinction for all clusters in NGC 4449.
  
  In principle, it might be possible to limit the effects of stochasticity by only including relatively high mass clusters. However, as shown in Figure 10, this does not work particularly well, since there are similar fractions of sources with colors in the "stochastic zone" (i.e., $U-B < -0.6$ and $V-I > 0.7$) for  the fairly massive clusters (i.e., greater than 10,000 solar mass) as there are for the lower mass clusters (less than 3,000 solar mass). We also note that only 2 of the 122 added clusters are in the stochastic zone, primarily because very few category = 3 (compact associations) were added.  One way to limit the effects of stochasticity is to only include category 1 and 2 sources (see Figure 5), as we and several other LEGUS studies have done in various parts of the analysis.

\section{Age Results From SSP Fitting}

In this section, we estimate ages for the H$\alpha$-LEGUS catalog of clusters in NGC~4449, and compare the results with those from the LEGUS project. We incorporate the results from \S 3, and 
explore the separate impacts that different filter combinations, assumptions about reddening, SSP models, and  assumptions about metallicity  have on the results.

\subsection{The H$\alpha$-LEGUS Method for Estimating Cluster Ages, Extinctions, and Masses} 

We find the best fit combination of age and extinction for each cluster by comparing the magnitudes measured in five broad-band filters (UV, U, B, V, and I), and one narrow-band filter (H$\alpha$), with predictions from the Bruzual \& Charlot (2003) stellar evolution models.  The narrow-band filter contains nebular line plus stellar continuum emission. The Bruzual \& Charlot models used here do not include nebular emission, but do predict the number of Lyman continuum photons.  We use this to predict the H$\alpha$ line luminosity as a function of age, using equation~(9) in Leitherer \& Heckman (1995), and combine it with the predicted stellar continuum to get a total (line$+$continuum) predicted magnitude for this filter.
 
The measured and predicted magnitudes are compared by performing a least $\chi^2$ fit where each filter is weighted by  $W_{\lambda} = [\sigma_{\lambda}^2 +
(0.05)^2]^{-1}$, where $\sigma_{\lambda}$ is the photometric uncertainty, and assuming a fixed metallicity of $Z=0.004$ ($\sim1/4$~solar), a Chabrier (2003) initial stellar mass function (IMF), and a Galactic extinction law (Fitzpatrick 1999).   The mass of each cluster is determined by multiplying the predicted $M/L_V$ at the best fit age, with the extinction corrected $V$-band luminosity of the cluster, using an assumed distance modulus of $\Delta(m-M)=27.91$.  Our final cluster catalog is called 'H$\alpha$-LEGUS' in what follows.

The method for including the narrow-band H$_\alpha$ measurements is updated here, over the one described in Chandar et al.  (2010), based on the
additional information provided by the H$_\alpha$ morphological classification discussed in \S~2.1 which allows us to characterize the presence or absence of associated line emission beyond  the 5 pixel radius used directly for the $H_\alpha$ measurement.
For H$\alpha$-morph class 1 and 2 the actual magnitude measured for the narrow-band filter is used (including both line and continuum emission). 
For H$_\alpha$-morph classes 3 and 4, which have little or no associated H$_\alpha$ line emission, respectively, the 
F658N filter is effectively treated as a measure of the $R$-band continuum.

We make two additional updates to our age-dating method based on the discussion of reddening in \S 3.2 and \S 3.3, and the graphics in Figure 6.
The first is to set an upper limit of E(B-V) $\leq$ 0.75 mag,
reflecting the low extinction in this galaxy even for the youngest clusters. The second
is to adopt a zero reddening ($E(B-V) = 0$) age-solution for clusters with estimated ages older than 10~Myr (based on an initial iteration where the reddening is allowed to vary). 

With these revisions included, the age estimates for older clusters from our $H_\alpha$-LEGUS catalog are in much better agreement with the spectroscopic determinations (see \S 5.1 and Table 1). In addition, we find a larger, more reasonable number of clusters with ages $\gea10^9$~yr (77 instead of just 5).

To this point, we note that integrated colors are significantly worse at providing age estimates of ancient globular clusters than integrated spectroscopy, at least in part because of the age-metallicity degeneracy.  We estimate the age we would determine for the bluest known, most metal-poor  Galactic globular clusters, which are confirmed to have ages $\approx13$~Gyr from their main sequence turnoffs (VandenBerg et al. 2013), by comparing their colors of $U-B\approx0.0$ and $V-I\approx0.8$ to the $Z=0.004$  Bruzual \& Charlot model.  We find that these colors would give a  predicted age of $\approx9.1$ in log Age; we use this value as a lower limit for candidate globular clusters in NGC~4449. We note that this is consistent with our results from Table 1, where we find that all of the confirmed old globular clusters from Annibali et al. (2018) have H$_\alpha$-LEGUS ages greater than 9.2 in log Age.

Our final age estimates have random uncertainties of $\approx0.3$ in log~Age, or a factor of 2.  There are also systematic uncertainties near log~Age $\approx7.0$, when the model colors loop back on themselves, leading to 'gaps' in the age-mass diagram (e.g., see the upper right panel of Figure 6). See discussions in Chandar et al. 2010 for further discussion of error estimates. 

\subsection{Comparison of Age Results from H$\alpha$-LEGUS vs. LEGUS}

In this section, we take a detailed look at the age results from H$\alpha$-LEGUS, and compare them with the ages determined as part of the LEGUS survey. 
It is important to remember that there are a number of differences in the methods used to estimate the ages in the two cluster catalogs:
\begin{itemize}

\item the catalogs use different fitting codes: the procedure used for H$\alpha$-LEGUS is described above (\S 4.1), and that for LEGUS is described in Adamo et al. (2017),

\item  the catalogs use different filter combinations, with and without the narrow-band H$\alpha$ measurement,

\item the treatment of reddening is different,
as discussed in \S 3, 

\item the catalogs use different methods for making aperture corrections: H$\alpha$-LEGUS  applies an aperture correction that does not vary from filter-to-filter and hence does not affect the colors,
whereas LEGUS applies independent aperture corrections to each filter (see Adamo et al. 2017),

\item the catalogs use different SSP models: Bruzual \& Charlot (H$\alpha$-LEGUS) vs. Yggradasil (LEGUS).

\end{itemize}

\noindent In this section we compare the age results from the catalogs generated by H$_\alpha$-LEGUS and LEGUS\footnote{The ages are from version 1 of the LEGUS catalog release, and are included in our publicly available H$\alpha$-LEGUS cluster catalog.}. 
 In Section~4.3 we examine the impact of different assumptions {\it one at a time} by using
  the same fitting code (i.e., the H$_\alpha$-LEGUS code described in \S 4.1). The same photometry is used to assess the impact that different combinations of filters, reddening, SSP models, and assumed metallicities have on the results.

In 
Figure~11 we compare our H$\alpha$-LEGUS results 
with those from the 
LEGUS HLSP (High Level Science Product) catalog available from the LEGUS public website.  The filled circles show results when photometry in all filters is available (i.e., 
UV, UBVIH$\alpha$ for H$\alpha$-LEGUS and UV, UBVI for LEGUS), and the open circles show results when no UV or U band photometry is available (i.e., BVIH$\alpha$ for H$\alpha$-LEGUS and BVI for LEGUS). 

It is important to note that the standard procedure for 
LEGUS is to only include age estimates when four or more broadband
filters are available.  However, we have relaxed this constraint for our study of NGC 4449 since there are a number of clusters in the outer parts of the galaxy with only BVI (and H$\alpha$) observations.
While age-dating that does not include the UV or U band filters can result in larger uncertainties in general, if zero internal extinction is appropriate (e.g., for nearly all the  
clusters in the outer portions of NGC~4449 where only BVI observations are available), good age estimates are possible, as will be shown below.

It is illustrative to examine clusters that fall in different parts of this diagram. 
Four representative cluster snapshots are shown for this purpose. 

The top image shows an example of a cluster near the top of the most prominent vertical chimney,
with an age of log~Age $ \approx$ 6.7 from LEGUS, and log~Age $ \approx$ 9.4 from H$_\alpha$-LEGUS. 
This cluster, and essentially all others in the top of this chimney, are old globular clusters based on their appearance, colors, and spectra (the spectra are discussed further in \S 5.1).  Hence the older H$_\alpha$-LEGUS ages are more accurate.
If we follow the chimney down farther, to log Age(H$\alpha$-LEGUS)  $>$ 7.9, we find a larger fraction of  Category = 2 clusters coming in. There are essentially no category 3 objects in the chimney (i.e., 33 of the 34 are category 1 or 2).  
Hence this chimney is caused by effects related to the inclusion of H$_\alpha$ and differences in the treatment of reddening, as will be discussed in \S 4.3, not by stochasticity which is mainly relevant for category 3 objects. 

While many of the clusters in this chimney do not have UV or U band photometry, a number do; therefore it is not only the lack of information in these bluer filters that drives the discrepancy between the H$\alpha$-LEGUS and LEGUS ages.

The second snapshot down shows an example of a cluster further down in the most prominent chimney, with an estimate log~Age $ \approx$ 6.7 from LEGUS, and log~Age$ \approx$ 8.8 from H$_\alpha$-LEGUS.
The older age appears to be more appropriate for this and most of the other objects in this part of the chimney since  
the cluster is diffuse and whitish instead of blue, and there is no evidence of H$_\alpha$ emission.

The third snaphshot down shows one of the many ($\approx$ 100) clusters which are assigned young best-fit ages from both methods. These are generally very blue, often with evidence of H$_\alpha$ emission in the vicinity as in this particular snapshot (i.e., the diffuse, green emission).  The main difference in the results for the few clusters that show strong line emission (those that follow the Yggdrasil model extension along the top left in the color-color diagrams shown in Figure~5 and in Figure 6) is that H$_\alpha$-LEGUS returns best fit ages of log~Age $\sim6.4$, while LEGUS returns best fit ages of log~Age $= 6.0$.

The bottom snapshot
shows an example where both LEGUS and H$_\alpha$-LEGUS
find intermediate ages, with log~Age $ \approx$ 8.0 from LEGUS, and log~Age $ \approx$ 9.0 from H$_\alpha$-LEGUS.
It is unclear whether the H$_\alpha$-LEGUS or LEGUS ages are more appropriate based on the appearances of these clusters. 

We now look at the overall comparison in Figure 11 in more detail.
The first obvious difference is the much larger number of clusters in H$\alpha$-LEGUS with ages $>10^9$~yr, as noted above. 
The number of clusters represented by each of the four snapshots are 17 (top snapshot - LEGUS $<=$ 7.0 and H$\alpha$-LEGUS $>=$ 9.0 in log age), 39 (second snapshot down - LEGUS $<=$ 7.0 and H$\alpha$-LEGUS between 8.0 - 9.0 in log age), 98 (third snapshot down - LEGUS $<=$ 7.0 and H$\alpha$-LEGUS  $<=$ 7 in log age), 20 (bottom snapshot - LEGUS between 7.5 and 8.0 and H$\alpha$-LEGUS $>=$ 8.3 in log Age).

Hence there are 76 (i.e., 17 + 39 + 20 from above) clusters (i.e., 13 \% of the 592) in these three  "chimneys".
The larger number (i.e., 98) of clusters represented by the third snapshot down demonstrates that the overall agreement is actually fairly good; the outliers are spread out more and hence look more dramatic in the figure.  

Another way to quantify the differences between ages derived in LEGUS and H$_\alpha$-LEGUS is to normalize  by the mean offset between the two systems and then look for discrepancies greater than a factor of 3 (i.e., 0.5 in log age) for clusters with H$_\alpha$-LEGUS ages that are less than log Age = 9 (i.e., where SED ages are less reliable - see Chandar et al. 2019). 27 \% of the clusters fall in this outlier category using this method of comparison.
{\it Hence, again, while there are some important differences, overall the agreement between the ages from H$_\alpha$-LEGUS and LEGUS is actually fairly good. }

In Section~7 we find that the mass and age distributions based on the H$\alpha$-LEGUS and LEGUS catalogs give similar results, despite the differences discussed in this section. {\it This demonstrates that the mass and age distributions are fairly resilient to the detailed differences in age dating.}

\subsection{Age Results from Different Filters, Reddening Assumptions,  Models, and Metallicities}
 
 In this section we assess the impact that different combinations of filters, different assumptions for reddening, different SSP models, and different assumed metallicities have on the results; we consider each parameter in turn.

\subsubsection{Impact of Using Different Filter Combinations}

So far, we have focused on the effects that different assumptions about reddening can have on age-dating clusters in NGC 4449. However, an equally important effect (actually more important in dusty galaxies where one cannot assume minimal reddening) is the use of H$\alpha$ in the SED fitting of young cluster populations, to help  break the age/extinction degeneracy.
We focus on that question in this section.

We note that it is 
just as important to know if there is {\em no} line emission as it is to measure the line emission when it is present. 
For example, this is one way to distinguish between young and old clusters in the prominent chimney in Figure 11. We also  examine the relative importance of including the UV and U filters in this section. 

Here, we examine  only the impact on the results from different combinations of filters by re-running the age-dating, and using the same input photometry, fitting code, SSP model ($Z=0.004$ from Bruzual \& Charlot), and allowing $E_{B-V}$ as a free parameter in the fit, so that only the combination of filters is different.

In Figure~12 we compare the results between the following four filter combinations:

\begin{itemize}
    \item UV, UBVI, H$\alpha$ (all 6 filters)
    \item UV, UBVI (5 filters, drop H$\alpha$)
    \item UBVI, H$\alpha$ (5 filters, drop UV)
    \item UV, BVI, H$\alpha$ (5 filters, drop U)
\end{itemize}

\noindent Given four sets of results, six comparisons can be made.
The main result, which appears in three of the six panels, is that 
dropping the H$_\alpha$ filter has the strongest impact on the age results.
{\it The other panels show that dropping a broad-band filter such as the UV or U band but keeping H$\alpha$ does not significantly impact the age results compared with all 6 filters.} 

We also note the similarity between Figure 11 and the three panels in Figure 12 that include the "drop H$_\alpha$" filter combination. This demonstrates that one of the primary causes of the difference in   H$_\alpha$-LEGUS and LEGUS age estimates is the inclusion of the H$_\alpha$ filter. The other primary difference is the treatment of reddening. After inspecting clusters visually, we confirm that 
the presence and absence of line emission is important for accurate age dating. 

Based on these figures and the differences between the H$\alpha$-LEGUS and LEGUS age results discussed in detail in Section~4.2, it appears that adding a measure of the line emission is much more powerful for age-dating star clusters in star-forming galaxies than adding another broad-band filter at short wavelengths.  Both the UV and the U band appear to work equally well for age-dating clusters.

\subsubsection{Impact of Different Reddening Criteria}

After the combination of filters, the next strongest effect in our age-dating procedure comes from our new assumption that reddening only affects cluster colors for the first $\sim10$~Myr in the case of NGC 4449.
As described in \S~4.1, our final cluster ages come from the best fit combination of age and reddening in the regime log~Age $ < 10$~Myr, and from the best-fit zero reddening solution for older ages.  We discussed the justification for this in \S~3.2, and in \S~5.1 we will  show that this assumption leads to significantly better estimates for the oldest clusters, based on comparisons with spectroscopically determined ages.

In Figure~13, we compare our final H$\alpha$-LEGUS ages with those found when we allow values of $E(B-V) < 0.75$ at all ages in the left panel, and $E(B-V)=0$ at all ages in the right panel.  When the reddening is allowed to be a free parameter at all ages, we see that some clusters are assigned younger ages because they are best fit with a combination that includes some reddening (those below the
1-to-1 
line).  We find that $\sim14$\% of the clusters are significantly affected (at a level of 0.5 in log Age or more) by this effect. 
The estimated ages of clusters with ages below $10$~Myr are identical in this case, as expected. 
In the right panel, we see that the assumption of $E(B-V)=0$ has a very small impact on the estimated ages of clusters younger than 10 Myr.

Overall, we find that both the addition of H$\alpha$ photometry and applying a maximum $E(B-V)$ during age-dating are important, and act primarily to prevent older clusters from being misclassified as younger ones.  The age estimates of a similar number of clusters are affected in each case. There are however, some differences.  Including H$\alpha$ in the age-dating procedure prevents older clusters with little reddening from erroneously being assigned a very young age ($<10$~Myr) plus high reddening.  By itself, however, including H$\alpha$ does not prevent older globular-like clusters from being assigned ages of $\approx100$~Myr.  A more accurate estimate of the ages of these older clusters depends on restricting the maximum allowed value of $E(B-V)$, regardless of whether or not H$\alpha$ is included in the fit.  

\subsubsection{Impact of Assumed SSP Model}

We now explore how using different SSP models affects the results, using the same photometry, code (Chandar et al), metallicity ($Z=0.004$ ), and set of filters (UV, UBVI). We retrieved the Yggradrasil models in 2019 from their website.  These assume a covering fraction of 0.5, and are somewhat different from those used as part of the LEGUS project, which used a different interpolation scheme.

The results are shown in Figure~14, where we compare ages from the Bruzual \& Charlot models (x-axis) and those from the Yggdrasil models (y-axis).
 When 
 using the Yggdrasil models there are more striations because of the lower age sampling.  We also notice that unlike the LEGUS results but similar to ours when using the Bruzual \& Charlot models, when the Yggdrasil models are used in our fitting code almost no clusters are assigned ages as young as log~Age =6.0. This suggests that the absolute age values assigned to the youngest clusters may vary between models and fitting methods.

Overall, we find that the results are fairly similar (with over 80\% of the sources having estimated ages within a factor of 3), but with some notable differences. 
There are two areas of the diagram where the ages deviate significantly: one where the Yggdrasil models give ages older by more than 0.5 in log~$\tau$ or a factor of 3 (47 clusters), 
and one where the Bruzual \& Charlot models give ages older by a similar amount (12 clusters).

In the cases where the Yggdrasil models give older ages, a visual inspection indicates that most clusters have blue colors, suggesting that they are quite young, consistent with ages of several Myr from the Bruzual-Charlot models, but inconsistent with the older log few$\times10$~Myr ages from the Yggdrasil models. This is likely related to the fact that the predicted colors from the Yggdrasil models at ages $<10$ Myr dip
below the measured colors of very young clusters in NGC~4449.

\subsubsection{Impact of  the Assumed Metallicity}

While we assume $1/4$ solar metallicity for cluster age dating in NGC~4449, it has been suggested that half-solar may be a better match to the abundance of the current gas (Annibali et al. 2017). 
In Figure~15, we compare the results when the 
H$\alpha$-LEGUS age dating procedure (with all 6 filters) is run with half-solar metallicity instead.

The results show that there is a tendency towards slightly younger absolute ages (by $\approx0.1$ in log~Age) when our default metallicity $Z=0.004$ is assumed, compared with the higher metallicity $Z=0.008$.   Note that the relative age estimates are similar for both metallicities.
We find that only 25 out of 594 ($\sim4$\%) clusters have ages that differ by more than a factor of 2 or log~Age = 0.3. 
We conclude that if subsolar metallicities are used, the exact value that is assumed for NGC~4449 has a relatively small impact on the age results. 

In a similar way, it might be more realistic to assume an even lower metallicity for the old globular clusters (e.g., Annibali et al. 2018 estimate 1/10 solar). We would expect this to have a similar effect as our experiment comparing 1/2 and 1/4 metallicity shown in Figure 15, but in the opposite direction.

\subsubsection{Summary of Age Comparisons Taken one at a Time}

To summarize this section, by using the same fitting code (H$_\alpha$-LEGUS), and letting only one item vary at a time, we find that the {\it addition of the H$\alpha$ filter appears to be more important than the addition of UV photometry} for breaking the age/extinction degeneracy when age-dating a population of clusters in actively star-forming galaxies. Other effects, in order of importance for the case of NGC 4449, are assumptions about reddening, the choice of SSP model, and the adopted metallicity.

These differences in age-dating methods can lead to measureable systematic differences, as demonstrated in this section. Howerver,  we will find in \S 7  that overall they have relatively small impacts on the mass and age distributions.

\section{Comparison with Independent Age Dating Methods}

While it is useful to make comparisons between similar methods of age-dating clusters, such as between H$_\alpha$-LEGUS and LEGUS, it is equally important to make comparisons with completely independent methods,
as we do in this section. 

\subsection{Comparing with Age Estimates from Integrated Spectroscopy}

Age estimates from absorption lines measured from integrated, low resolution spectra in the range 3,200 to 10,000 Angstroms have been made for 11 clusters in NGC~4449 by Annibali et al. (2018), seven of which are in common with our sample. In Table~1 we compare the spectroscopic age estimates with those determined from LEGUS, 
H$_\alpha$-LEGUS, and
Annibali (2011 - integrated colors) results.
{\it In all cases we find the
Annibali et al. (2018) spectroscopic ages to be older than the
photometric ages, and especially for the first three LEGUS  values.}  Two of the three discrepant clusters have no U-B values in Table~1, indicating that they have only ACS BVI measurements (i.e., they are in the outskirts of the galaxy as seen in Figure~2). 
 All of the H$_\alpha$-LEGUS values in Table 1 are also below the Annibali et al. (2018) spectroscopic ages, but none by more than an order of magnitude.
We also compare with integrated light age estimates using BVI from Annibali et al. (2011) in Table 1, finding better agreement with the spectroscopic ages, but still slightly lower values for the photometrically determined ages.

For LEGUS, the difference between the ages derived from integrated photometry and spectroscopy appears to be mostly due to the fact that the SSP fitting routine prefers the combination of a young age plus high reddening over an old age with low reddening.  For H$_{\alpha}$-LEGUS, the lack of detected H$_{\alpha}$  pushes the algorithm to an older age solution, although they are still lower than the spectroscopic age estimates. 
We also note that the well-known age-metallicity degeneracy affects the age estimates,  since clusters with ages $\gea$ Gyr generally have lower metallicities than the one assumed for younger clusters, resulting in ages lower sometimes by $\sim0.6-0.7$~dex than found via spectroscopy (the effect is significantly smaller at younger ages).

Comparisons between ages derived from LEGUS (including cases with only three filters, BVI, which is non-standard for LEGUS and must be done with caution; i.e. only for clusters where there is no evidence of reddening), 
H$_\alpha$-LEGUS, and Annibali et al. (2011 - using BVI integrated colors) are shown in Figure~16.

\begin{tiny}
\begin{landscape}
\begin{deluxetable}{lccccccccccc}

\tablecaption{Comparison of log Age Values \label{tab:alpha}}
\tablewidth{0pt}
\tablehead{
\colhead{ID/alias}  & \colhead{LEGUS} & \colhead{E(B-V)} & \colhead{H$\alpha$-LEGUS}  
& \colhead{Annibali (2011)}  & \colhead{Annibali (2018)}  
& \colhead{RA}  & \colhead{DEC}  & \colhead{V-I}  & \colhead{U-B} \\

& & &     & photometry & spectra & & & \\

&  (log Myr) & (mag) & (log Myr) &  (log Myr) & (log Myr) 
& & & (mag) & (mag)}

\startdata

582/CL3 & {\it 6.85} & 0.86  &  9.40 &  9.85  & 9.95   &   187.06849 &  44.12486 &      1.253  & 0.035 \\

  592/CL77 & {\it 6.78}  &  0.91 &  9.65  &   9.86  &10.08  &    187.05641&  44.14404 & 1.135 & --- \\
 
  32/CL79 & {\it 6.70} & 0.92  & 9.26 & 9.91 & 10.04 & 186.99944 &  44.07870 & 1.070 & --- \\
 
13/CL76 &  9.48      & 0.02  &   9.40  &        10.08   &10.04  &      187.01615  &    44.07090   &       1.165 & --- \\
 
28/CL67  &  8.70 & 0.00   & 8.86   &         8.49   &   ---    &               187.03891 &    44.07748 &    0.712  & -0.025 \\
 
153/CL20 &  9.60   & 0.03  &   9.70   &          9.16  & 10.04 &        187.07843  &   44.08878 &   1.219  &  -0.011\\
 
417/CL8 &  9.30   & 0.00  &  9.41   &        9.71   &   ---    &         187.07828  &    44.10641  &  1.015 & -0.010\\

\hline
\enddata

\tablecomments{ 1. Values with discrepancies greater than 1.4 from the Annibali (2011, 2018) values are shown in italics.}

\end{deluxetable}
\end{landscape}
\end{tiny}

The comparison between LEGUS and the Annibali et al. (2011) models look very similar to the Figure~11 comparison between H$_\alpha$-LEGUS and LEGUS, presumably because  Annibali et al. (2017)  assume that there is no reddening internal to NGC~4449 itself, similar to the assumption we make in the 
H$_\alpha$-LEGUS method for log Age $>$ 7.0 clusters. This results in older age estimates for many of the  clusters in both cases.

Indeed, the  right panel comparison between Annibali (2011) and the 
H$_\alpha$-LEGUS
 method is quite good, with a slope near unity, small scatter (RMS = 0.42), and a small offset (+0.13 mag). This follows the good agreement we found between H$_\alpha$-LEGUS and Annibali (2011) in the much smaller sample shown in Table 1.

A comparison of all the clusters in the Annibali (2011) photometric study with the LEGUS sample shows that  16 of the 25 outliers (i.e., in the  vertical chimney in the left panel of Figure 16)  turn out to be clusters with only BVI measurements. However, the other 9 clusters in the chimney do have measurements in all 5 filters, hence this problem is not caused exclusively by the lack of the U filter observations. All filter combinations (i.e., both the open and filled circles) are in agreement in the right panel comparison between Annibali (2011) and H$_\alpha$-LEGUS.

\subsection{Comparing with Age Estimates from Color Magnitude Diagrams}

Recent papers by Sacchi et al. (2018) and Cignoni et al. (2018) provide another potential comparison with our cluster age estimates. These authors use the stellar component of NGC 4449 to determine star formation histories in several large regions in the galaxy. Their results are also used in \S 7 to help separate the effects of cluster formation and disruption. In the current section we 
use the PSF-fitting photometry of resolved stars from the stellar catalog provided by  
 Sabbi et al (2018) to
estimate ages for 10 each of the category = 2 (asymmetric) and category = 3 (compact associations) objects in our catalog using resolved stars. As can be seen from the color-color diagram in Figure 5, most of the category = 1 clusters are older, and the individual stars are too faint to be detected. Hence this procedure was not attempted for category = 1.

Given the extreme crowding conditions and the small size of these samples, we applied an 
isochrone fitting technique to
 the CMDs, instead of a full statistical derivation of the cluster SFH. Category 2 and 3 clusters which appeared to have extended halos of resolved stars
were selected for this exercise.
Stars within a radius of 20 pixels (= 15 pc)  of the objects were evaluated, using the cluster's
appearance to help determine where most of the stars were likely to be associated with the cluster (i.e. the density was higher than the surroundings).  There
were typically about a dozen stars that appeared to be associated with a cluster. 
In some cases, especially in the outer annuli, it is likely that some of the stars are in the background rather than in the clusters. However, to the extent that stars in the surrounding region have the same age (i.e., they are both part of a larger association) this will generally give the same result.
One of the
primary concerns for this approach is the presence of blends since many of these regions
are very crowded. For this reason it is only possible to provide upper  or
lower estimates in some cases.

Figure~17 shows an example of how the age dating is done using the CMD
for compact association C3-3144-6144 (alias: cluster 401 in Table 2).  Note the
enhancement of 9 blue stars on the left side of the upper panel
(i.e., within a radius of 15 pixels). These are only compatible with 
the 5 or 10 Myr isochrones.  The 6 stars to the right cover a variety
of potential ages and are likely to be foreground or background stars. The
bottom panel, consisting of the annulus just outside of 15 pixels,
allows us to distinguish between the 5 and 10 Myr isochrones (assuming
minimal reddening so the points do not move around much on the CMD)
with two stars along the 10 Myr isochrone. Hence
this compact association is assigned an age of 10 Myr in Table 2.

The resulting comparisons with our H$_\alpha$-LEGUS  ages are shown in
Figure~18.  Keep in mind that some
of the CMD estimates are upper or lower limits. 
While the scatter is relatively large for the
comparison in some cases, it does appear that the CMD ages are compatible with
our integrated light age estimates in general, and the approximate mean values (the X's in Figure 18) are in fairly good agreement.  Note that the mean values are calculated without taking into account the fact that many of the points are upper and lower limits. The mean position, (i.e. the 'X') would almost certainly be closer to the 1-to-1 line in the left panel if estimates without upper and lower limits could be made since there are eight lower limits and only two upper limits.

A careful examination of the snapshots in Figure~18 provides important insights into the age
dating for both methods, and the classification of category 2 (asymmetric clusters) as compared to category 3 (compact associations). The two images on the left of each panel have diffuse weak emission-line flux (the green color), and hence are given slightly younger ages using the H$_\alpha$-LEGUS method. The two snapshots close to the one-to-one line have no emission and no dominant red stars.  The agreement between the two methods is very good in theses cases. The right snapshot  in the right panel shows a case where two very bright red supergiants have mislead the integrated light measurement into considering it an older cluster. However, these can be well fit as evolved stars with young ages in
the isochrone fitting algorithms  (i.e., this is cluster 401 shown in Figure 17 and discussed above). This is a good example of the effects of
stochasticity for clusters/associations with masses less than a few $\times$
$10^3$ $M_{solar}$ (e.g., see Fouesneau et al. 2012), as discussed in \S 3.2. 

The upper right snapshot in the Category 2 (left) panel shows a case where the CMD age estimate is probably uncertain due to the presence of faint foreground/background stars and the true age is  much older than the CMD estimate of log Age = 8.2. In fact, both Annibali et al. (2011) - integrated light) and Annibali et al. (2018 - spectra) consider this object to be an old globular cluster with log age $>$ 10 Gyr, as does our H$_\alpha$-LEGUS determination. Note in Table 2 that this object is only assigned a lower limit (i.e., $>$ 170 Myr) by the CMD method.

A similar comparison between CMD and integrated light age estimates was performed by Larsen et al (2011) for relatively nearby,
partially resolved clusters in NGC 1313, M83, and three other galaxies which are at similar distances to NGC 4449.
As here, there was reasonably good agreement between the two methods of estimating
ages, although crowding was identified  as a primary difficulty.

While estimating CMD ages for clusters and compact associations is inherently difficult at the 
distance of a few Mpc, it is reassuring that the mean ages are in reasonably good agreement with the mean ages from our integrated light determinations, as shown in Figure 18. Although the scatter is large, we note that 16 of the 20 points are within 1 dex of the 1-to-1 line.

\begin{tiny}
\begin{landscape}
\begin{deluxetable}{lcccccccccccc}

\tablecaption{Comparison between  CMD and H$_\alpha$-LEGUS Ages  \label{tab:alpha}}
\tablewidth{0pt}
\tablehead{
\colhead{ID}  & \colhead{log Age (CMD) } & \colhead{log Age (H$_\alpha$-LEGUS)  }  
& \colhead{x-pos} & \colhead{y-pos}  
& \colhead{RA}  & \colhead{DEC}  \\
&  (Myr) & (Myr) & (pix) & (pix)}

\startdata

Cat = 2 \\
55 &  $<$ 7.48   &  6.70 &  5135.5  &      4059.0     &            187.05546  &   44.08153         \\
60  &  $>$ 8.08  & 8.76 & 7520.6     &  4128.9       &    187.01891  & 44.08223      \\
193 &  $>$ 8.11   &  8.41 & 4198.0     &  4953.4        &     187.06983 & 44.09136      \\
334 & $>$  8.11  & 8.81 & 3206.6  &   5650.1   &           187.08502  &   44.09904         \\
380 & $>$  7.70  & 8.01 & 3984.7   &  6000.9    &        187.07310    &  44.10290          \\
414 & $>$  8.00 & 9.16 & 3099.1  &    6285.0    &          187.08668  &    44.10602     \\
436 & $>$  8.00 & 10.06 &4196.0   &  6494.7 &           187.06987   &  44.10834       \\
442 & $<$  7.00  &   10.00 &5079.0     &   6524.0      &        187.05633   &  44.10866             \\
 557 & $>$  8.23 & 10.30 & 7892.4 &    7382.3 &            187.01320  &     44.11810   \\
587 &   $>$ 8.18  &  8.36 &  5931.0    &   8204.4 &            187.043270 & 44.12716   \\
 &  mean = 7.89 +/- 0.39 (rms)  & mean =  8.57 +/- 1.14 (rms)  \\
\\
Cat = 3 \\
87 & $<$ 7.00 &   6.70 &  5321.0 &    4338.0 &  187.05262 &  44.08460   \\
101 & $<$ 7.00  &  7.18  &   6219.0  &    4436.4    &        187.03886  &    44.08569     \\
171 & 7.30  &  7.30  & 5830.0  &   4838.2    &        187.04482   &  44.09011       \\
190 & $>$ 7.30  & 8.11 & 5974.9   &   4946.0   &         187.04260   &   44.09130       \\
191 & $>$ 7.60 & 7.16 &  5772.5  &   4948.1    &         187.04570   &  44.09132       \\
312 & 7.40   & 7.42  &  5128.0  &   5503.4     &      187.05558   &  44.09743       \\
401 & 7.00    & 8.96 &  3144.0   &     6144.0    &       187.08599   &  44.10447       \\
413 & $<$ 7.00 & 7.18  &    5798.0  &     6279.1 &      187.04531  &   44.10597     \\
449 &  7.40  & 7.38 & 3920.3    &    6558.0    &        187.07409  &   44.10903       \\
538 & $<$ 7.30  & 6.70 &  4409.4  &       7230.0   &         187.06660  &   44.11643       \\
 &  mean =  7.23 +/- 0.22 (rms)  & mean = 7.41 +/- 0.67 (rms)   \\

\hline
\enddata

\end{deluxetable}
\end{landscape}
\end{tiny}

\subsection{Comparing with Age Estimates from HII Regions}

Sokal et. al., (2015) have used a combination of optical and infrared observations (i.e., Spitzer IRAC 3.6$\mu$m, 4.5$\mu$m, 5.8$\mu$m, and 8.0$\mu$m observations and Herschel Space Telescope observations) of the giant HII region S26 (a strong  thermal radio continuum source and the brightest object in our Region 7 - a snaphot of this object is shown in the bottom right panel, on the left side, of Figure 5) and estimate an age of 3.1 $\pm$ 0.3 Myr for this object. The region has strong Wolf-Rayet features which are consistent with this age estimate. 

The LEGUS age estimate for S26 is 3.0 Myr while the H$_\alpha$-LEGUS age estimate is 5.0 Myr. There are a total of eight objects in Region 7, all with fairly similar age estimates. The LEGUS age estimates range from 1.0 to 5.0 Myr, with one outlier at 15 Myr. The  H$_\alpha$-LEGUS ages are all between 3.0 and 5.2 Myr.  Reines et al. (2008) have estimated ages for 11 HII regions in NGC 4449, all in the range 2 to 6 Myr. We conclude that the age estimates for HII regions from both LEGUS and H$_\alpha$-LEGUS are in quite good agreement with those from HII regions in NGC 4449.

We also note that Sokal et. al. (2015) estimate a reddening  value of E(B-V = 0.13 mag for S26, in good agreement with the values discussed in \S 3.2 for regions with strong H$_\alpha$. This measurement is also compatible with  earlier optical studies of S26 and other HII regions in NGC 4449 by Reines et al. (2008) and (2010).

\section {The Specific Luminosity, ($T_L$), in 25 Regions}

Having improved our age estimates as described above, 
in this section we measure the fraction of light, $T_L$, coming from clusters relative to the total light 
within 25 regions within NGC~4449.  This quantity was first measured in the $U$ band for young cluster systems in 21 nearby star-forming galaxies by Larsen \& Richtler (2000), and is defined $T_L = 100\times~L_{clusters}/L_{galaxy}$. It is sometimes called the specific luminosity. Here, we measure the fraction of light in the three available broad-band filters (F435W, F555W, F814W) from the ACS observations, which cover the largest FOV in the galaxy (see Figure~2).

Figure~1 shows the 25 regions in NGC 4449 that are used to measure $T_L$.
These are color-coded based on their appearance, with red for
regions which appear to be dominated by older clusters, yellow for intermediate-age, and blue for regions dominated by young clusters and compact associations. The youngest regions are easily identified by their green color in Figure~1, which is due to the presence of nebular line emission (i.e., H$_{\beta}$, [OIII] 5007, [OIII] 4959) from HII regions in the F555W filter.
Most of the regions appear to be dominated by a stellar population
with a particular age (for example the outer regions have very little star formation and only old clusters), with the exception of region 17 (the nuclear region) and region 24, which clearly have a mix of both young and old
clusters.
We assign an 'age' to each region from the average value of  Log~Age of all clusters in the box; this value is given in Table 3 for each region in Figure~1.

For the cluster component we sum the luminosity of all the detected clusters within each region, where an average aperture correction has been applied to the photometry of the cluster.  We find that there is no significant change to our results if a size-based aperture correction is used to determine cluster luminosities instead (see also Cook et al. 2019). 
For the stellar component we use two different methods; the first is to add the luminosity from the individual stars from the stellar catalog provided by  
 Sabbi et al (2018) and available at:

\begin{verbatim}
https://archive.stsci.edu/prepds/legus/photometric_catalogs/ngc4449.html
\end{verbatim}

\noindent and divide by the area of the region. Hence, this provides an estimate of the specific Region Luminosity determined from individual stars,
and will be denoted $R_L$(star). 

The second method is to estimate the  total luminosity of the region from the broadband image, not just from detected stars in the stellar catalog.
This is again divided by the area of the region to
provide an estimate of the specific Region Luminosity from the total luminosity, and will be called $R_L$(total), as discussed in the next section. Table 3 includes our determinations of R$_L$ and T$_L$ using both the stars method and the total method.

Figure~19 shows our estimates of the fraction of light in clusters in the three different filters, $T_L$(F435W) (top panels), $T_L$(F555W)(middle panels), and $T_L$(F814W) (bottom panels) versus specific Region Luminosity, in this case determined by adding up the flux from individual, detected stars in the region [i.e., R$_L$(stars)].
The left set of panels are restricted to category 1 and 2 clusters only, while the right panels include categories 1, 2, and 3.
As discussed in \S 2.1, category 3 sources tend to be young, are the most difficult to select, and our source list for this type of object is likely incomplete.
The different symbols show regions with different ages, as found in Table 3.

All panels in Figure~19 show increasing trends for $T_L$ values with specific Region Luminosity;  the correlation is strongest for the F438W ($B$) filter and when category 3 objects are included.
Correlations are found with significance ranging from 2.7
to 3.0 sigma for the Cat $= 1+2$ fits, and 4.3 to 5.9 sigma for the Cat $=
1+2+3$ fits. 

Our results are similar to the original results from Larsen \& Richtler (2000) for spiral galaxies, and to those from Billett et al. (2002) for dwarf galaxies, since we also find higher values of $T_L(\lambda)$ for regions with higher luminosities.
However, we can take our results one step further.
As shown in Figure~20, by breaking the sample into regions dominated by clusters of different ages, we see that there is a correlation between $T_L$ and log Age. 

These results are also similar to results  found for $\Gamma$, the fraction of stellar mass in clusters, at different ages, but considering entire galaxies (Chandar et al 2017).
For 8 galaxies which span a wide range of SFR and $\Sigma_{SFR}$, they found that $\Gamma$ for the youngest $\lea 10$~Myr clusters has a typical value of $\approx24\pm9$\%, which drops to $\Gamma\approx2\pm1$\% by a few hundred Myr.
In fact, the
values of $\Gamma$ and T$_L$ are quite similar, ranging from $\approx20-30$\% for the youngest regions
to just a few percent for the oldest regions.
Hence, the strong apparent correlation between T$_L$ and specific Region Luminosity, R$_L$, is likely due to the fact that regions dominated by older clusters tend to be fainter than regions dominated by young clusters, since the decrease in
$\Gamma$, or T$_L$, is a natural result of the destruction of the clusters with time.   

The correlations with age in Figure~20 are similar or slightly stronger than in Figure~19. 
Correlations are found with significance ranging from 2.5 to 2.9 sigma for the Cat $= 1+2$ fits and 4.3 to 6.8 sigma
for the Cat $= 1+2+3$ fits. 
Hence the correlations of T$_L$ with specific Region Luminosity, $R_L$, and with Log Age are similar. 

A  correlation matrix analysis (using parameters $T_L$,  $R_L$(stars), B and V magnitudes, reddening, B-V, and age for the LEGUS star clusters of category 1 and 2) leads to a similar conclusion. The average Pearson correlation coefficients between $T_L$ and $R_L$,  and between $T_L$ and Log Age, are roughly the same, having values of about 0.6 in the former case, and 0.5 in the latter case.

We use a second method to estimate the specific Region Luminosity  as a check on these results, since it is  possible that that we have significantly  underestimated the total luminosity by only including {\em detected} stars, and excluding stars in very crowded regions.
Within each region, we estimate the total counts as follows.  We determine the mean pixel value, subtract off the background level, then multiply by the total number of pixels, and then divide by the area of the region.  We make the assumption that the number of background pixels dominates over those that have individual sources, and therefore adopt the median pixel value as the background level.
In the inner region the "background" probably includes many old red stars from the bulge component. Hence our total counts estimate should be thought of as representing the dominant younger stellar population.

Figure 21  shows that the two methods lead to relatively similar results.
Both show clear trends
between T$_L$ and log Age, as shown by the  two linear fits.  However, since the scatter is relatively large, the correlation over smaller time spans (e.g., less than log Age = 8) is uncertain. 

A possible complication is that the mass-to-light ratio for stars change as they age, which may contribute at some level to the correlation shown in Figure~20. However, this will happen for both the cluster population and the field star population (much of which comes from disrupted clusters),
hence this effect should largely cancel out. The fact that similar correlations are seen in all filter bands, including F814W, also suggests that this is not a major issue. 

Different cluster completeness levels between the old and young regions might also cause some of the correlation. As we will see in the next section, clusters with ages less than 10 Myr can be detected over three decades in log Mass, from 10$^3$ to 10$^6$ solar masses, while clusters with ages around 1 Gyr are only complete over two decades in log Mass, from 10$^4$ to 10$^6$ solar masses. For a power law with index 2, (appropriate in the case of both mass and luminosity for clusters) each  decade includes the same fraction of the total. Hence the young clusters would be a factor 1.5 (i.e., 3 decade compared  to 2 decades) more complete
than the older clusters. Since the average 
value of T$_L$ is $\approx$ 0.06 for clusters with ages 1 Gyr in Figure 21, a completeness correction would increase the 
value to about 0.09, still well below the value of T$_L\approx$ 0.20 for clusters with ages around 10 Myr.

Another possible complication is the fact that the clusters and stars that form in a given region will eventually move out of it. While this is likely to contribute minimally for very young populations where the clusters will not have time to move out of the box, at some age it will become more important. We can estimate this age by assuming an average random velocity of $\sim3$ km$/$sec (Massey et al. 1995). This would allow a typical cluster to move a distance equal to the radius of an average region ($\approx$ 300 pc for the intermediate-age regions) in about 100 Myr.
For much older populations (e.g., several Gyr)  essentially all of the stars and clusters will have moved out of the box they were born in, but other stars and clusters formed at approximately the same distance from the galaxy center will have moved in.
Therefore, regions dominated by old clusters and stars should not be significantly affected\footnote{Dynamical friction is quite weak at the larger radii where we have defined 'older' regions, so should not have much impact on the locations of old clusters in the outer regions dominated by older stars.}.
Hence, while the effect may be present for some of the intermediate-age regions in our sample (e.g., 21, 23  with log Age $\approx$ 8.5), it should not dominate in general. As argued above, these motions will have little impact on regions dominated by young or by old stars, which are the primary driver of the observed trend for $T_L$ with age.

\section{General Cluster Properties in NGC 4449}

\subsection{Cluster Mass Functions}

In Figure~22 we present the mass-age diagrams of the clusters in NGC~4449.
The upper panels show our results for categories 1 and 2 (left) and categories 1, 2, and 3 (right), using the H$_\alpha$-LEGUS age estimates.  
The bottom panels show the results when using the LEGUS age estimates.
The similarity in the diagrams
show that even with fairly important differences in the age dating procedures, as discussed in \S 4, the resulting changes in the mass and age distributions are likely to be relatively small. This result is confirmed when comparing the slopes in the mass and age functions, as discussed below. 

The points that are circled in the upper left panel are from added clusters, as discussed in \S 2.1. We note that only four of the added clusters are massive enough to be included within the limits used to construct the mass functions and age distributions, which are shown by the dotted lines in Figure~22. 

As is generally the case, the mass function of star clusters can be approximately described by a power law, $\psi(M)\propto M^{\beta}$.  In Figure~23, we show the mass functions using H$_{\alpha}$-LEGUS ages for category 1, 2, and 3 clusters in the top panels, and for category 1 and 2 in the bottom panels, divided into three different age intervals: $<10$~Myr (left), $10-100$~Myr (middle), and $100-400$~Myr (right).  
The distributions have an equal number of clusters in each bin (as recommended by Maiz Apellaniz \& Ubeda 2005), and are not sensitive to the exact number used.  

The  best fit values for $\beta$ are mostly between $\approx-1.7$ and $-2.1$. 
The mean value after the high and low values are removed is $\beta\approx-1.86$.
If LEGUS ages are used instead of H$_\alpha$-LEGUS ages, 
we find values $\beta\approx-1.9$, identical to those found by Cook et al. (2019) for the composite LEGUS dwarf sample, which includes NGC~4449.

The results for $\beta$ for the different age intervals are mostly similar within the uncertainties, indicating that there is no apparent change in the shape of the mass function over the age-mass ranges studied here.
The mass function for cat$=1+2$ clusters with ages log~Age $=7-8$ appears a bit flatter, but this is the range where the biases in the age-dating are strongest, and small number statistics are also playing a role.

We have also checked the mass function of clusters in three different radial bins: $R_{gc} <  1.06$ kpc (47$\arcsec)$, $1.06-1.82$~kpc ($47-97.5\arcsec$), and $R_{gc} > 1.82$~kpc ($>97.5\arcsec$).
Although the statistics are poor in some cases, the mass function in the different radial bins and in the three age ranges studied above, are also reasonably well described by a single power law with an index $\beta\approx-2$, i.e., there is no clear trend as a function of radius.

A number of studies have reported that the upper end of the cluster mass function drops off compared with a power law (e.g., Gieles et al. 2006; Larsen et al. 2011; Johnson et al. 2017; Messa et al. 2018), and that this upper mass cutoff may correlate with the SFR of the host galaxy (Johnson et al. 2017).  However, Mok et al. (2019) applied a maximum likelihood fitting method to the cluster population in NGC~4449 and did not find evidence for a cutoff mass, consistent with the distributions shown here in Figure~23.

\subsection{Cluster and Star Age Distributions}

The mass-age diagrams in Figure~22 give a preview of the cluster age distributions.  If the age distribution was flat (i.e., a power law slope $\approx$ 0), as for the hypothetical case where clusters formed at a constant rate and none were disrupted, there would be a factor of 10 more clusters in a given mass interval for each full dex in log Age, since the bin size is a factor of 10 larger for each dex. This would result in 
a strong horizontal gradient (at a given value of log Mass) in Figure~22. 
However, we find that the horizontal gradient in log Age is relatively uniform, which would suggest  a decline in the age distribution of roughly a factor of 10 each decade of log Age to compensate for the larger bins. This 
corresponds to a slope in the age distribution, when fit with a power law, of $\approx -1$ (see Whitmore et al. 2007 Figure~3 for a graphic illustration).  There does appear to be a slight 
enhancement of clusters around an age of a few 100 Myr, however, which will be discussed in \S 7.3. 

Plots of the age distributions of star clusters (i.e., dN/d$\tau$ diagrams)  are constructed by counting clusters in equal bins of log~$\tau$ from clusters within a given mass range.
These can be described by a power law, $\chi(\tau) \propto \tau^{\gamma}$.  In Figure~24 we present the cluster age functions for category 1, 2, and 3 (right panels), and for category 1 and 2 only (left panels),
in three different mass ranges, being careful to stay above the completeness limits, which are shown in Figure~22.

All of the distributions decline more-or-less continuously, (although there is some evidence for an enhancement around a few hundred Myr as mentioned above), with $\gamma$ values between $-0.74$ and $-0.95$, i.e., around $-1$ or slightly flatter, as expected since the horizontal gradient is relatively uniform in Figure~22.
This decline is  approximately independent of the mass of the clusters, since the fits in different mass ranges are within the uncertainties.
Given the range presented here, we find $\gamma =-0.85  \pm$ 0.15,  very similar to result found by Rangelov et al. (2011).
We have checked, and find a similar value of $\gamma$ for the NGC~4449 cluster catalog published by Cook et al. (2019).  Overall, we find that the shape of the age distribution of clusters in NGC~4449 appears to be similar regardless of the exact method used to select the clusters or the specific age intervals used in the analysis.

Making similar fits using the LEGUS rather than the H$_\alpha$-LEGUS ages results in $\gamma$ values between $-0.62$ and $-1.02$,
with a mean of $-0.82 \pm 0.17$, very similar to the value of -0.85 using H$_{\alpha}$-LEGUS ages.
Hence, even though differences in detail can be seen in the four mass - age diagrams in Figure 22 (i.e., using subsamples with different categories or age-dating methods), the slopes of the age distribution derived from the data are relatively resilient. If we limit the sample to only Category 1 clusters from the
H$_\alpha$-LEGUS catalog, the slopes are slightly shallower, with values between $-0.60$ and $-0.69$ and a mean of $-0.66$. The shallower slopes are due to the large number of Category 1 objects with ages of a $\sim$few hundred Myr age range (resulting from an enhancement in formation), as seen in Figure~5 and discussed in Section 7.3. 

An observed logarithmic cluster age distribution with an index of 
$-0.85$ indicates that clusters are destroyed at a rate of ($1 - 10^{-0.85}) \times 100 = 86 \%$ each decade of time, similar to the results for a number of spiral,
merging, and dwarf galaxies (Whitmore et al. 2007, Chandar et
al. 2010, Fall \& Chandar 2012, Bastian et al. 2012, Cook et al. 2019).  We note however, that some works have found significantly flatter age distributions as well (e.g., Silva-Villa \& Larsen 2011; Fouesneau et al. 2014).

We have examined the age function of clusters in three different radial bins: $R_{gc} <  1.06$ kpc (47$\arcsec)$, $1.06-1.82$~kpc ($47-97.5\arcsec$), and $R_{gc} > 1.82$~kpc ($>97.5\arcsec$), using two different mass ranges,
categories 1+2+3, and excluding the youngest datapoint since there are very few clusters in most of the samples for this bin. The resulting values of $\gamma$ range from 
an average of -0.89 $\pm$ 0.07 for the central region; to -0.96 $\pm$ 0.09 for the intermediate region; to -1.13 $\pm$ 0.24 in the outer region. 
Hence there are no clear trends for the slopes of the age distributions to vary with radius from the center in NGC 4449, although we note that the statistics are fairly poor when breaking the sample into these smaller subsamples.

The observed cluster age distribution is the  product of the formation and disruption histories of the clusters.  In order to determine the cluster disruption history we need independent information about the formation history, i.e., has the formation rate been relatively constant during the relevant period of time, as is generally assumed.
The stellar formation history of NGC~4449 has been determined from the LEGUS data, as described in detail in Sacchi et al. (2018), and is a possible proxy for the cluster formation rate under the assumption that star and cluster formation track one another closely. Chandar et al. (2017) find that this is a good assumption for the eight galaxies examined in that paper, for example. 

In  Figure~25 we show the composite SFH of the galaxy, as well as the history in the three different galaxy radii defined above. While there is a modest enhancement in the star formation history in the last 10 Myr, resulting in NGC 4449 being known as a starbust galaxy,  beyond 20 Myr there is no systematic trend for the SFR to increase or decrease over the past several hundred Myr. 
A fit to the total SFH beyond 20 Myr, which is the focus of the current discussion, gives a power-law index of $\gamma_{\rm form}\approx 0.49 \pm 0.56$, i.e. broadly, the formation rate is essentially flat, without any systematic increases or decreases over the last several 100 Myr.

Hence, with the assumption that the stellar and cluster {\it formation} rates track each other, this means that the observed age distribution, with $\gamma=-0.85$, primarily reflects the {\it disruption history} of the clusters.
There are, however, some variations at the factor of $\approx2-3$ level in the star formation rates over shorter time intervals. In particular we note the enhancement that occurred a few hundred Myr ago, for both the stellar and cluster populations. This will be discussed in more detail in the next section.

\subsection{An Enhancement in the Star and Cluster Formation Rates a Few Hundred Myr Ago}

Here we estimate the level of enhancement in the star and cluster formation rate a few hundred Myr ago. As discussed in \S 2, there is evidence that NGC 4449 had an interaction with one or more companions roughly 100 - 500 Myr ago (Hunter et al. 1998, 1999, Theis \& Kohle 2001, Karachentsev et al. 2007, Martinez-Delgado et
al. 2012, and Rich et al. 2012), probably resulting in the enhanced star and cluster formation rates. The enhancement in the cluster population is seen  in the color-color diagram (i.e., the category 1 clusters in Figure~5) and the mass-age diagrams (Figure~22) and therefore is not due to systematic biases in the age dating or to binning.

The SFHs determined from independent data sets by McQuinn et al. (2010) and Sacchi et al. (2018) for NGC~4449 both show a similar enhancement in the rate of star formation a few hundred Myr ago, although there may be differences in the exact timing of the enhancement.

Figure~25 shows both the star formation history (based on the analysis of individual stars from Sacchi et al., 2018 but extracted for the radial bins discussed in \S 7.1 and \S 7.2) and the cluster formation history, based on the data in Figure~24. To obtain the cluster formation history, we  divide the observed cluster age distribution by the best fit, smooth power-law (i.e., $\gamma$ = -0.85), to remove the effects of disruption. This leaves behind variations in the cluster formation history.  We repeat this procedure for age distributions with three different bin widths (0.4, 0.5, and 0.6 in log~Age), and for two different sets of bin centers, for a total of six realizations.
Each one shows an enhancement in the cluster formation history, which range from factors of $\approx 1.5$ to 4, around ages of $\approx~\mbox{few}\times 100$~Myr. The mean value for these six realizations is a factor of 2.2 $\pm$ 0.7.

 An enhancement is also seen in the star formation history at similar ages, with a mean value of 1.8 $\pm$ 0.6 when comparing the three radial bins in the time range from 100 to 300 Myr, normalized by the two adjoining ranges (i.e., 30 - 100 Myr and 300 - 1000 Myr). If only the two inner bins are used, which might be more appropriate since there is no apparent recent star or cluster formation in the outer regions of the galaxy, the enhancement for the stars is 2.1 $\pm$ 0.4, even closer to the estimate for clusters. Hence, while the statistics are relatively poor, there does appear to be some evidence for similar levels of enhancement for both the clusters and stars in the range 100 - 300 Myr old. 
This suggests a roughly
constant  values of $T_L$ (and presumably $\Gamma$) as a
 function of increasing SFR.

There are few galaxies where direct comparisons between the star and cluster formation histories have been made.  Two of the most promising galaxies for such studies are the Small and Large Magellanic Clouds.  Although not yet definitive, recent studies of the cluster populations suggest that there was an enhancement in the populations $\approx~\mbox{few}\times100$~Myr ago (e.g.,
Glatt et al. 2010, 
Bitsakis et al. 2017, 2018), during the time of the last closest approach.  The star formation histories of both galaxies also show an enhancement around this same time period (e.g., Harris \& Zaritsky 2004, 2009), although the strength of the enhancement for the stars and the clusters is not yet well determined. Future studies will be needed to determine if these enhancements had similar strengths.

We note that the precision of the cluster (and stellar) age dating does not allow us to quantify the duration of the enhanced formation period very well.
It might be an enhancement of a factor of 2 over the period
from 100 - 300, or a factor of 20 enhancement over a 20 Myr old period
around 200 Myr ago. 
The main conclusion here is that we observe an enhancement in both the star and cluster formation rates in NGC~4449 at a similar $\sim$factor of 2 - 3 level, a few hundred Myr ago. Hence, it appears that the cluster and stellar formation rates are closely related, and  T$_L$, and presumably $\Gamma$, are relatively constant during the burst.

Figure~26 shows a portion of NGC~4449 (region 16  and slightly to the west - see Figure~1) where nearly all of the clusters formed during this burst (i.e., 18 of the 22 clusters in this region have log Age in the range from $8 - 9$ Myr). 
Region 23 and just eastward, on the opposite side of the nucleus at a similar distance, shows a similar distribution with 29 of 40
clusters in the log Age = 8 - 9 age range. No other regions show such a clear enhancement over this time period. This suggests that the original plane of the galaxy interaction that caused the starburst a few Myr ago
 was probably oriented in roughly an east-west direction. A more
 detailed study of these two regions (i.e., determining the star formation histories)
 might provide a more statistically significant comparison between cluster and star populations during a burst.

\section{Conclusions}

In this paper we used data from the LEGUS (i.e., Calzetti et al. 2015) and H$\alpha$-LEGUS (Chandar et al. 2019) projects to address a number of
questions concerning the ability to accurately age-date star clusters, and to characterize the cluster population and its relationship to the stellar population in the 
starburst galaxy NGC 4449 . Our main results are included below. 

1. There is fair agreement between various age dating methods (integrated
light, spectroscopy, CMD, HII regions), but also a systematic bias toward
underestimating cluster ages when using integrated colors, especially for old
($\approx$ 10 Gyr) globular clusters. This primarily results 
from the flexibility within the algorithms
to trade off reddening and age to find the best fit model.
One way to mitigate this 
effect, at least for a  galaxy with relatively little dust such as NGC 4449, is to fix the reddening to be zero for clusters  with ages greater than 10 Myr. 
While this method may slightly underestimate
the reddening (and hence overestimate the age) in some cases, it results in ages that are much closer to those determined from absorption line strengths in integrated spectra (i.e., Annibali et al. 2018). 
Another way to mitigate much, but not all of  this effect, is to only use age estimates with four or more filters, including a U or UV filter (i.e., the default in the LEGUS study). 
This was not possible in the outer parts of NGC 4449 for our study due to the lack of $UV$ or $U$ observations. 

2. Inclusion of  H$_\alpha$ in the SSP fitting also helps mitigate the bias toward underestimating ages for integrated light age estimates, and provides the single most effective improvement when considering the effects of adding different filters (i.e., it is more effective than adding a U or UV filter). Other primary effects, in order of importance for NGC 4449, are assumptions about reddening, choice of the SSP model to use (i.e., Bruzual - Charlot or Yggdrasil), and metallicity. 

3. Effects of stochasticity (i.e., the random inclusion of a red supergiant star in the aperture) for low mass systems can affect the position in the color-color diagrams dramatically, and hence affect the age-dating.
Caution is therefore required when including low mass systems. Setting limits to the maximum reddening allowed for the age-dating algorithm, or fixing the reddening at zero for older clusters for galaxies with low reddening can help mitigate the effects of stochasticity. Other methods are to take a more probabalistic approach toward age dating (e.g., Fouesneau et al. 2012, Krumhlotz et al. 2015, Ashworth et al. 2017) or  stack the data for low-mass clusters (e.g., Hannon et al. 2019).

4. A correlation between the fraction of light in clusters ($T_L$) and the specific Region Luminosity (R$_L$) is found. This is similar to the finding by Larsen (1999) for entire galaxies.  The
underlying relation appears to be between $T_L$ and log Age. This is consistent with the destruction of star clusters as
a function of time which varies as dN/d$\tau=\tau^{-0.85}$, and hence is
similar to the results from Chandar et al. (2015, 2017) for the fraction of {\it mass} in clusters (i.e., $\Gamma$ ) vs. log
SFR relation.

5.  The mass and age functions in NGC 4449 are similar to other star forming
galaxies, both spirals and dwarfs (i.e., slopes of $\beta$ = -1.86 $\pm$ 0.2 and $\gamma$ = -0.85 $\pm$ 0.15 respectively), regardless of the details of the photometry and age dating. This
supports the quasi-universal model explaining the demographics of star clusters (e.g., Whitmore et al. 2007, Fall \& Chandar 2012).  The mass and age distributions do not appear to depend on galactic radius (i.e., environment), although the low number statistics do not provide very stringent limits. The effects of the different age-dating methods discussed in items 1 through 3 above are relatively minor, i.e., the determination of the mass and age functions are quite resilient.

6. A factor of $\approx$ 2 - 3 enhancement in both the cluster and star formation rates is observed from 100
to 300 Myr ago, probably associated with one or more interaction events (i.e., Hunter et al. 1999,  Theis \& Kohle 2001, Karachentsev et al. 2007, Martinez-Delgado et
al. 2012, and Rich et al. 2012).  
This suggests that $T_L$ 
 is roughly constant as a function of increasing SFR during a burst.

Future studies of other H$_\alpha$-LEGUS galaxies will allow us to determine whether these results are typical of star-forming galaxies in general.

\acknowledgements 
Based on observations made with the NASA/ESA Hubble Space Telescope,
obtained at
the Space Telescope Science Institute, which is operated by the
Association of Universities
for Research in Astronomy, Inc., under NASA contract NAS 526555. These
observations
are associated with program \# 13364. R.C. acknowledges support from NSF grant 1517819.
M.C. and M.T. acknowledge support from the INAF PRIN-SKA 2017 program 1.05.01.88.04.
We thank Francescca Annibali for useful discussions
We also thank the referee for many useful comments that improved the paper.

{\it Facilities:} \facility{HST}.

\begin{figure}
\begin{center}
\includegraphics[width = 6.0in, angle= 0]{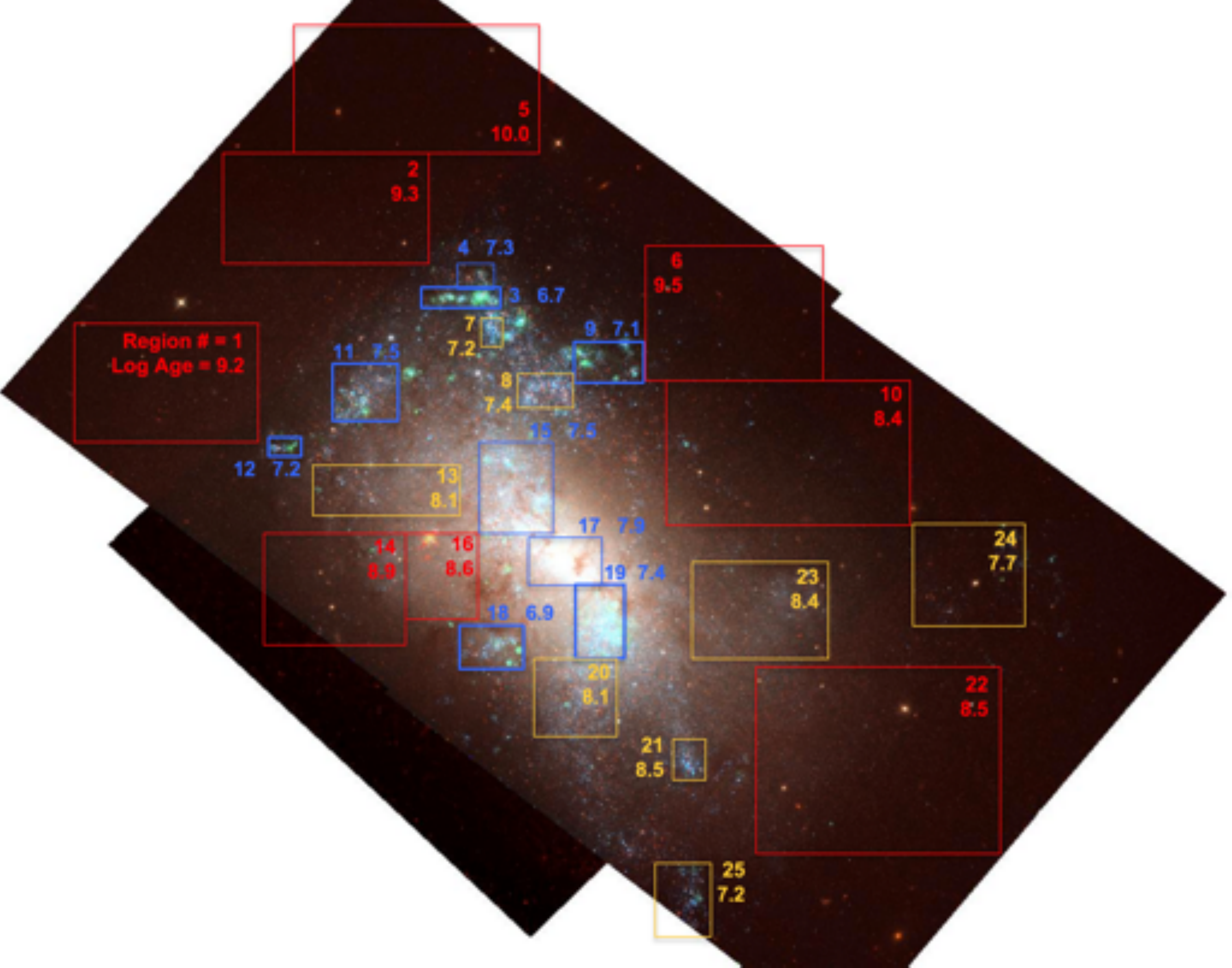}
\end{center}
\caption{Figure showing the LEGUS mosaic image of NGC 4449; the 25
  color-coded regions selected for analysis in the text; and
  the ID numbers  and log of the mean cluster ages for the clusters in these regions as derived in this paper. The color coding is red = old regions - i.e., very
  few or no blue stars or emission line regions [NOTE: emission-line HII regions are greenish in this image due to the presence of H$_\beta$, and OIII [4959, 5007] in the F555W filter]), yellow = intermediate
  age (dominated by blue stars but no emission line regions) and blue = young (dominated by
  emission lines). The sizes of most boxes were chosen to isolate regions that appear to be dominated by
stars and clusters of a single age (e.g., see region 3). Larger regions were used in the outskirts since the stars and clusters are uniformly old (i.e., no blue stars or HII regions). }
\end{figure}

\begin{figure}
\begin{center}

\includegraphics[width = 6.5in, angle= 0]{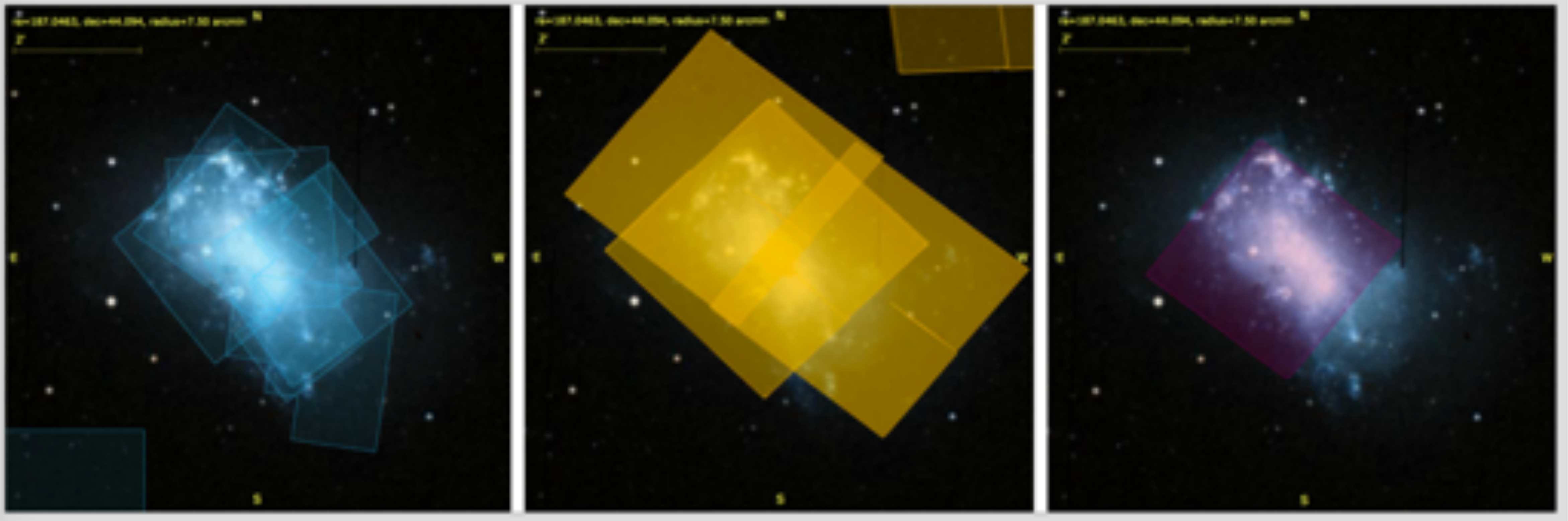}
\end{center}
\caption{Footprints for HST observations of NGC 4449 using the WFPC2 (left), ACS (center) and WFC3 (right); from the Hubble Legacy Archive (HLA; see Whitmore et al. 2016). }
\end{figure}

\begin{figure}
\begin{center}
\includegraphics[width = 6.0in, angle= 0]{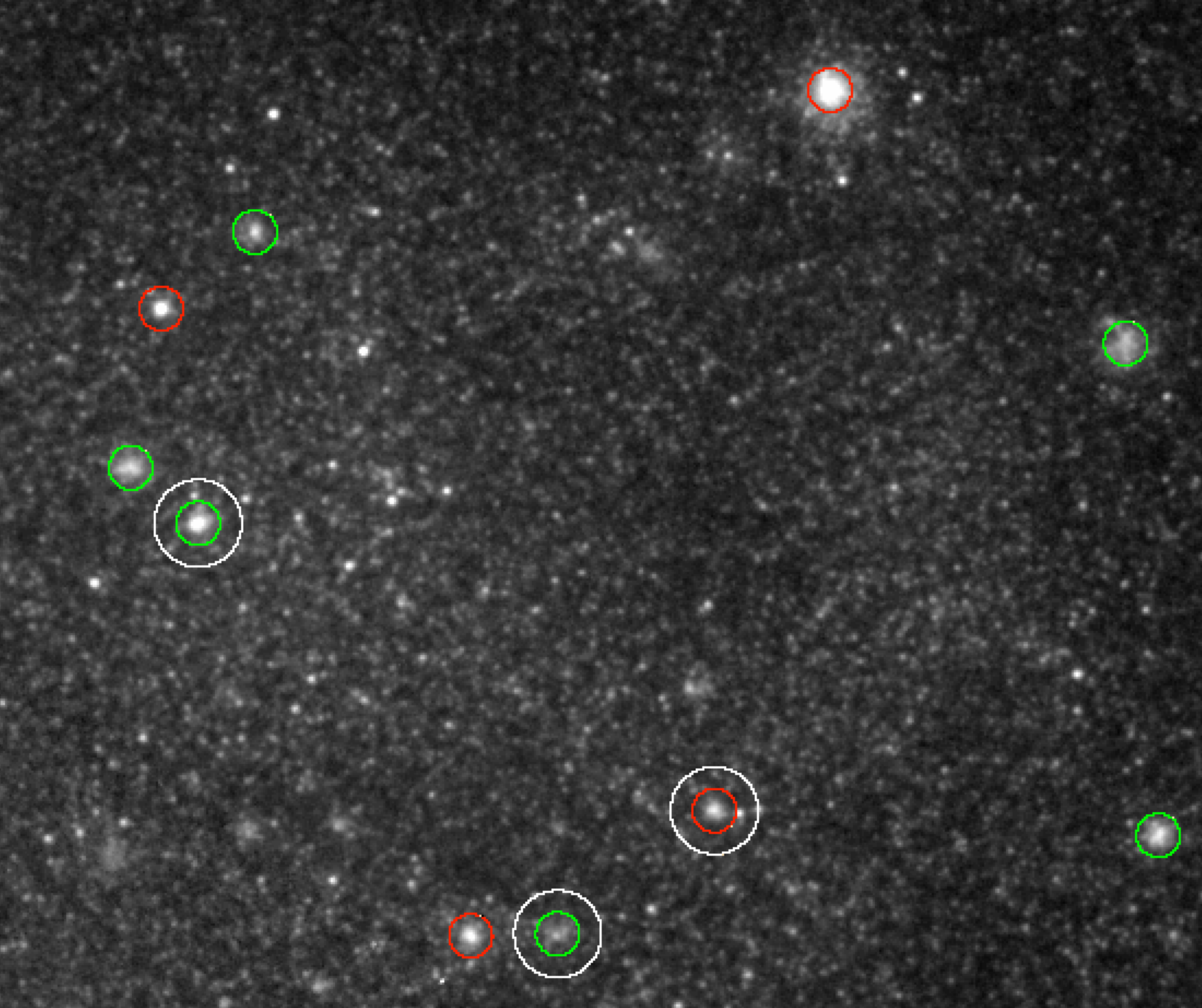}

\end{center}
\caption{White circles show examples in region 21 of three of the 121 objects that have been added to the sample. Several original category 1 (red)  and 2  (green) clusters are included for comparison. There are no category  3 objects in this region.  }

\end{figure}

\begin{figure}
\begin{center}
\includegraphics[width = 6.0in, angle= 0]{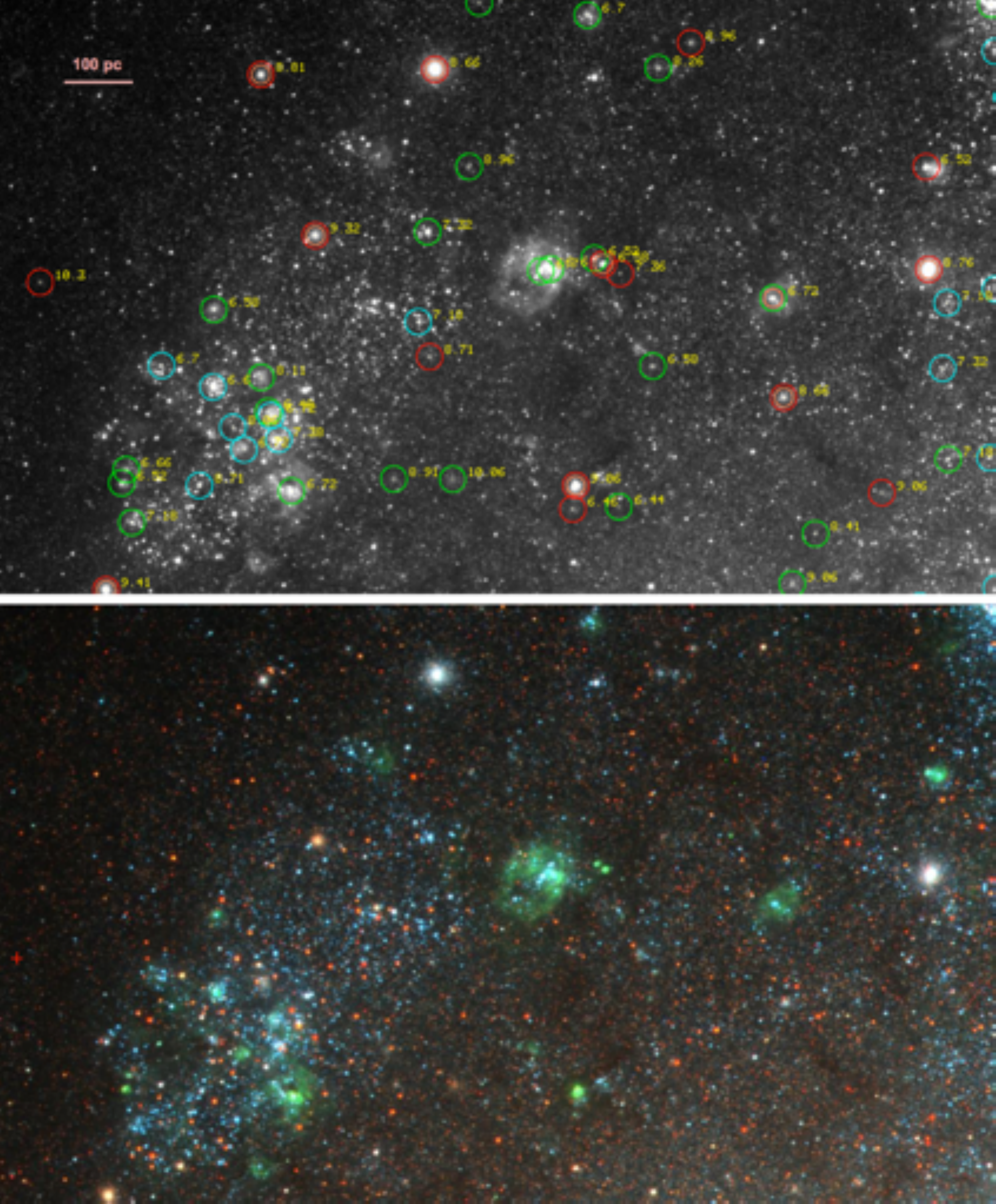}

\end{center}
\caption{Image of region 11 (and slightly beyond) showing the selection of category 1 (symmetric - red circles) and category 2 (asymmetric - green circles ) clusters, and category 3 (clustered point sources - blue circles) compact associations.  Note that the circles have radii of 15 pixel rather than the 5 pixels use for the aperture photometry. The log Age values derived from this paper  are included in yellow. The smaller orange circles show the clusters from Annibali et al. (2011) in this region.  The bottom panel shows the color image from the HLA; green colors are indicative of emission line flux. Note that bright clusters from Annibali with log Age $\approx$  8.7 (part of the enhancement that will be discussed in \S 7.3) are white, while older clusters with log Ages $\approx$ 9.4 (i.e., old globular clusters - see discussion in \S 4.1) are redder. The bar in the upper left shows the scale. }

\end{figure}

\begin{figure}
\begin{center}

\includegraphics[width = 7.0in, angle= 0]{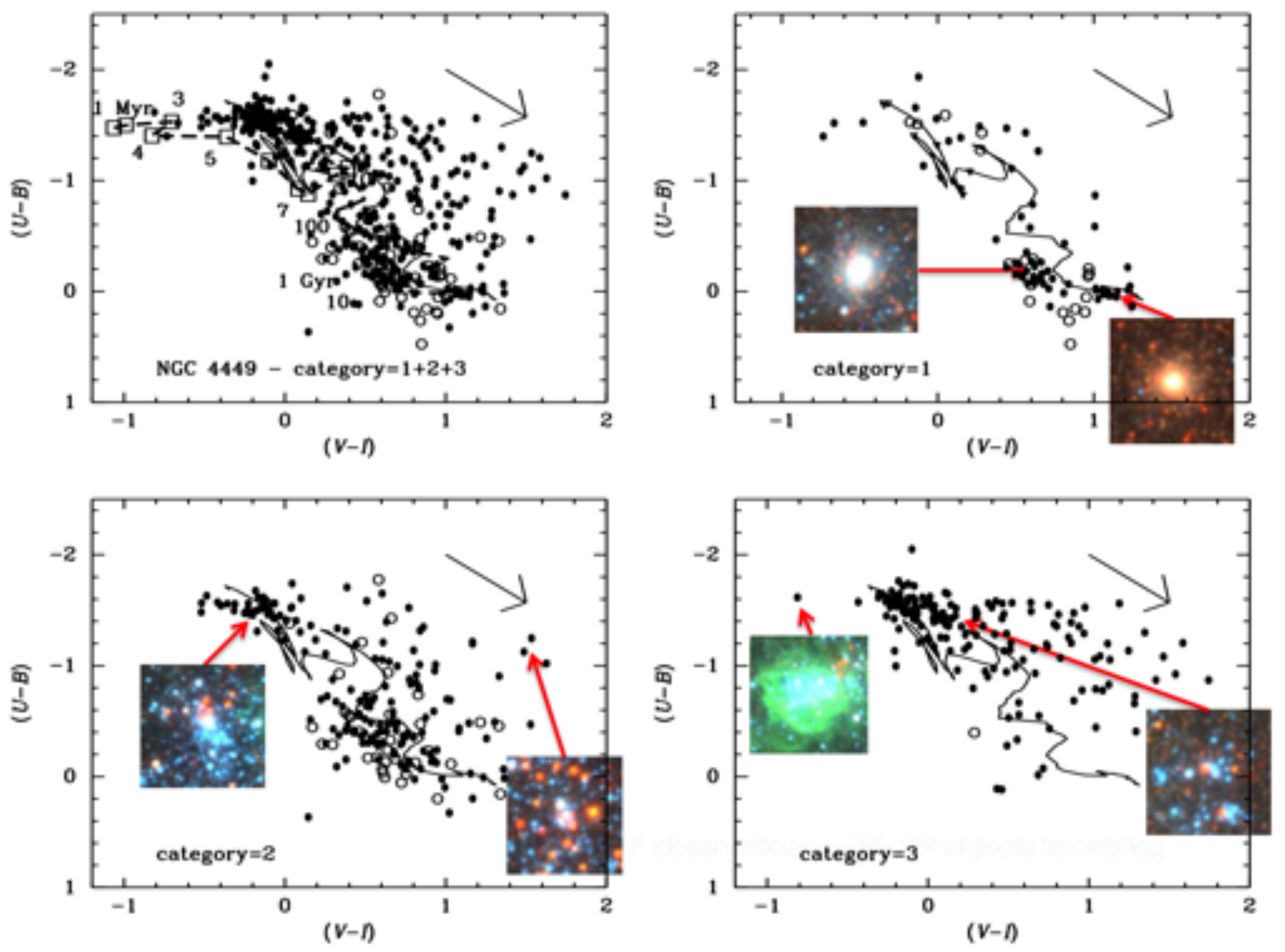}

\end{center}
\caption{V-I vs. U-B color-color diagrams for all (upper left), category = 1 (upper right - symmetric clusters), category = 2 (lower left - asymmetric clusters) and category 3 (lower right - compact associations) in NGC 4449.  The solid lines show Bruzual-Charlot (2003) cluster models  while the dashed lines in the upper left 
shows Yggdrasil (Zackrisson et al. 2011) models, both with 1/4 solar metallicity.  
The numbers are the ages 
for 
the Yggdrasil models (with ages 1 - 10 Myr shown as squares). 
Ages for the Bruzual-Charlot models are included in Figure~6.
The arrow shows a A$_V$ = 1.0 reddening vector. Open circles represent 'added' clusters (see \S
2.1) while filled circles represent clusters from the original LEGUS list.  Snapshots are shown for various clusters.
See \S 3.1 for discussion.}
\end{figure}

\begin{figure}
\begin{center}

\includegraphics[width = 7.0in, angle= 0]{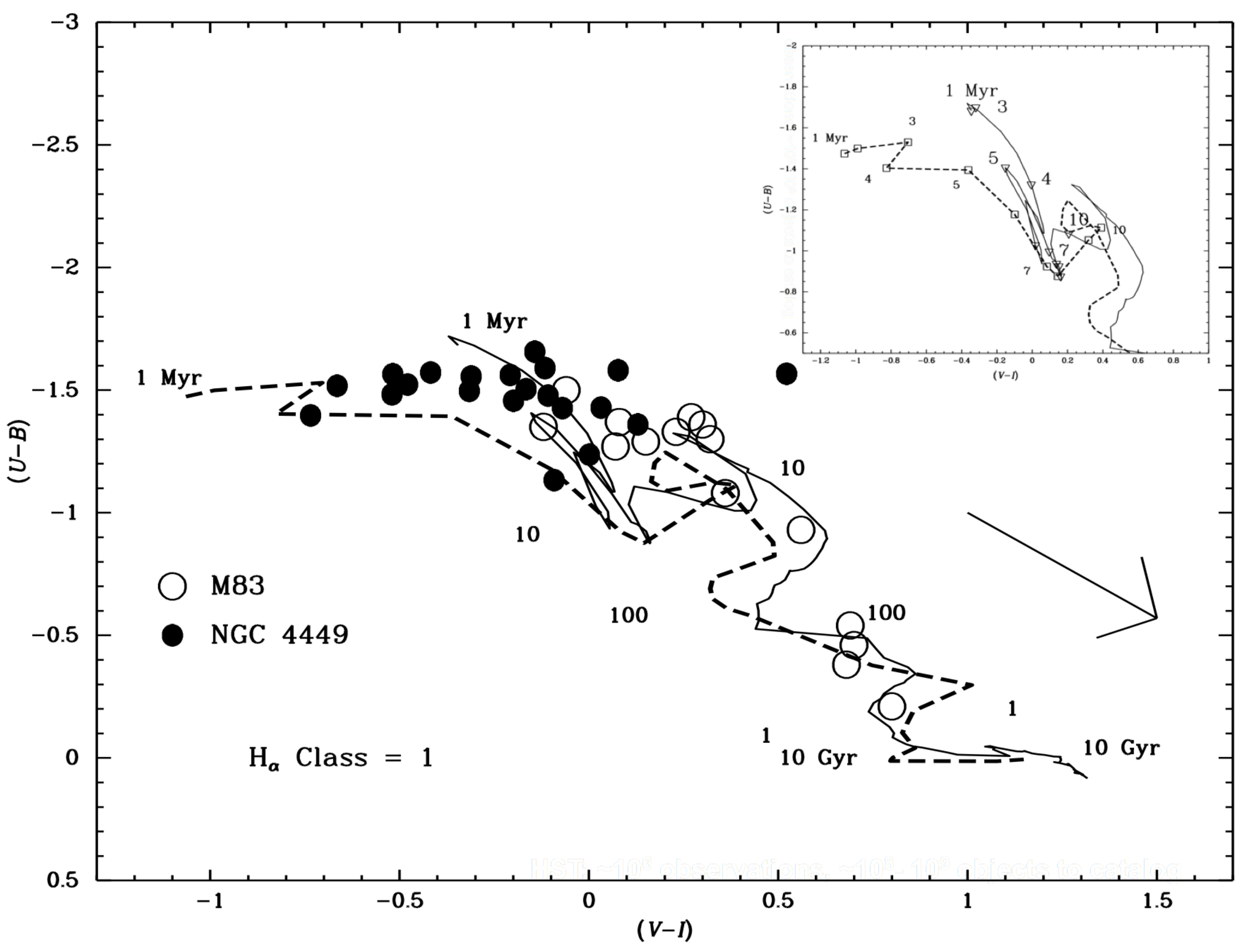}

\end{center}
\caption{V-I vs. U-B color-color diagrams for the H$\alpha$ class 1 clusters in NGC 4449 (filled circles) and M83 (open circles, from Whitmore et al. 2011). The SED tracks and reddening vector are the same as shown in Figure 5. An insert has been included to make it easier to see the details of the models at young ages.
Note that six of the M83 clusters show clear evidence of reddening (i.e., they track down the  reddening vector) but none of the NGC 4449 clusters show clear evidence of reddening (i.e., they can all be explained by SEDs with ages 7 Myr or lower, as expected for objects with H$\alpha$ emission and no reddening).
}
\end{figure}

\begin{figure}
\begin{center}

\includegraphics[width = 5.5 in, angle=-0]{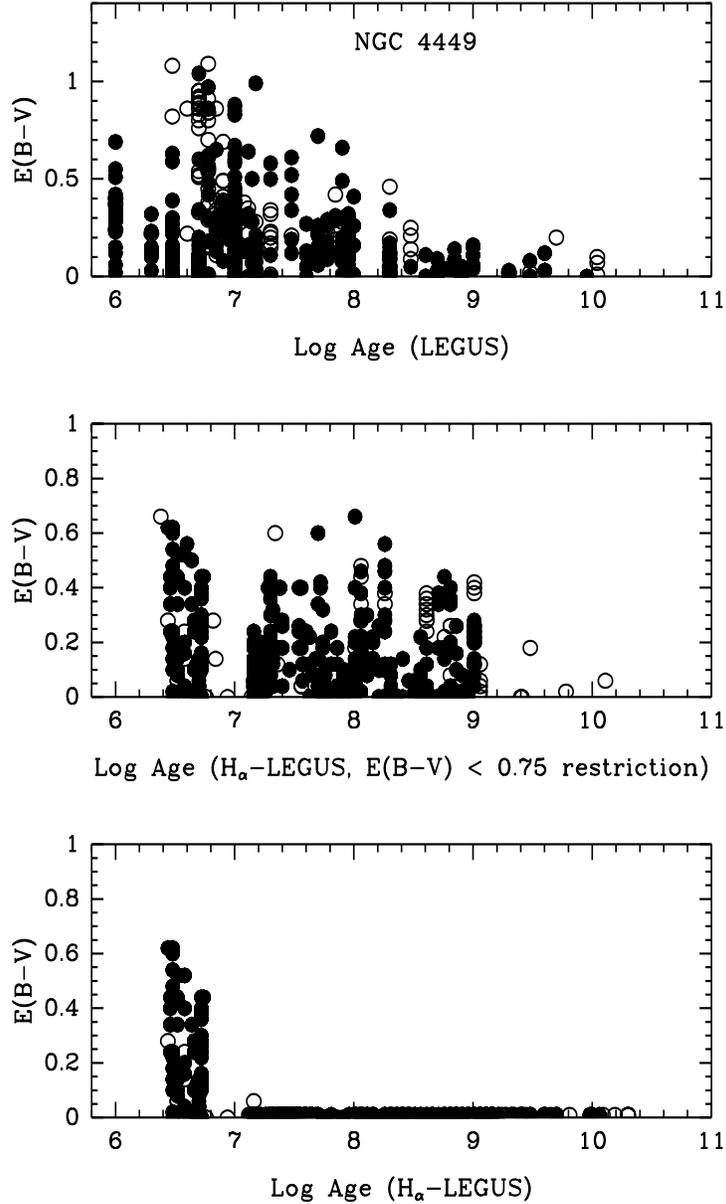}
\end{center}
\caption{E(B-V) reddening values for the LEGUS age dating solutions (upper panel), H$_\alpha$-LEGUS with E(B-V) $<$ 0.75 restriction  (middle panel), and the H$_\alpha$-LEGUS  solution (i.e., where E(B-V) is constrained to be  0 for log Age $>$ 10 Myr) (lower panel). 
Open symbols show clusters with 3 filters (i.e, the outskirts without WFC3 UV or U observations), filled circles show clusters with 5 filters.
Note the large values of E(B-V) derived for the LEGUS solution  (i.e., $\approx$ 1) for many of the clusters with derived values of log Age $\approx$ 6.7 for both  3 and 5 filters. Most of these are actually old globular clusters, as determined by their appearance 
or spectra from Annibali et al. (2017). Also see discussion in \S 5.1}
\end{figure}

\begin{figure}
\begin{center}
\includegraphics[width = 5.5in, angle= 0]{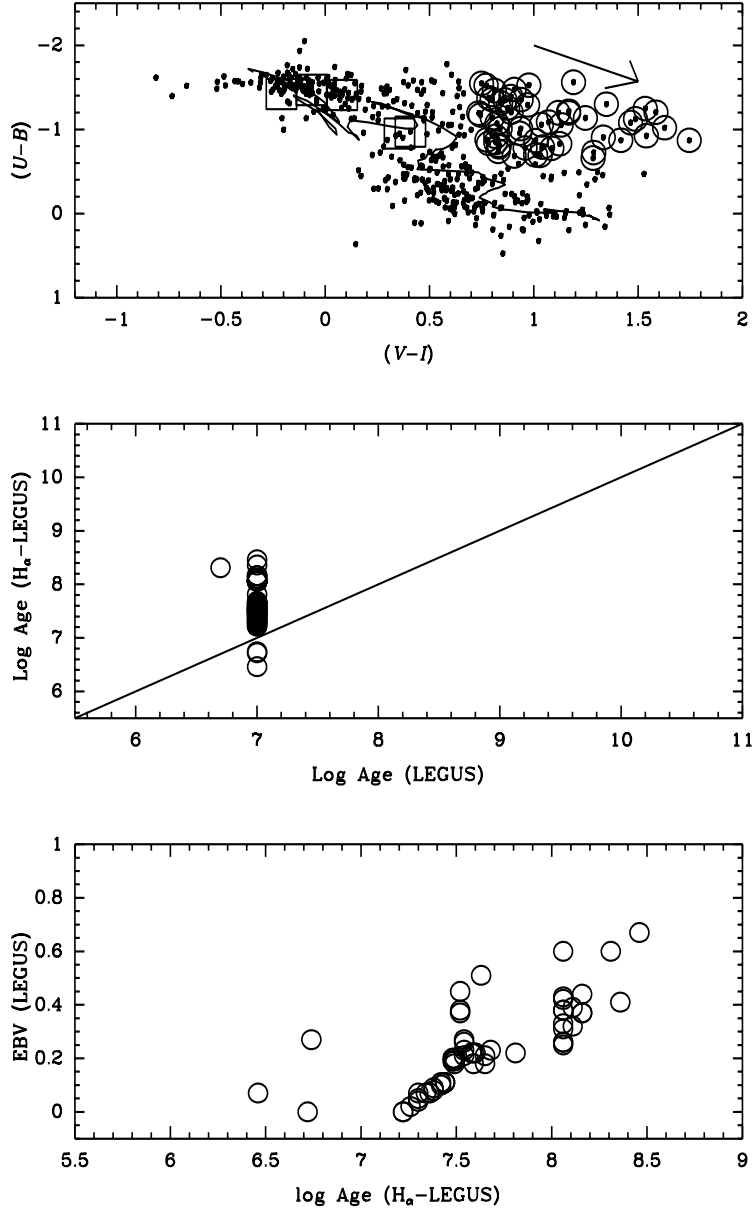}
\end{center}
\caption{The top panel is the same as the upper left panel in Figure~5, but with points with U-B$<-0.6$ and V-I$>0.7$ circled to isolate cases where stochasticity rather than reddening may be important (see also Figure~12 in Johnson et al. 2012).  The five large squares show the objects with blue circles around them (i.e. with no red stars near the center) in Figure~9.  The middle figure shows age estimates for the isolated points from LEGUS and from H$_\alpha$-LEGUS. 
The bottom panel shows that age differences estimated by  
H$_\alpha$-LEGUS are interpreted as large reddening values in  the LEGUS age estimates.
}
\end{figure}

\begin{figure}
\begin{center}
\includegraphics[width = 7.0in, angle= 0]{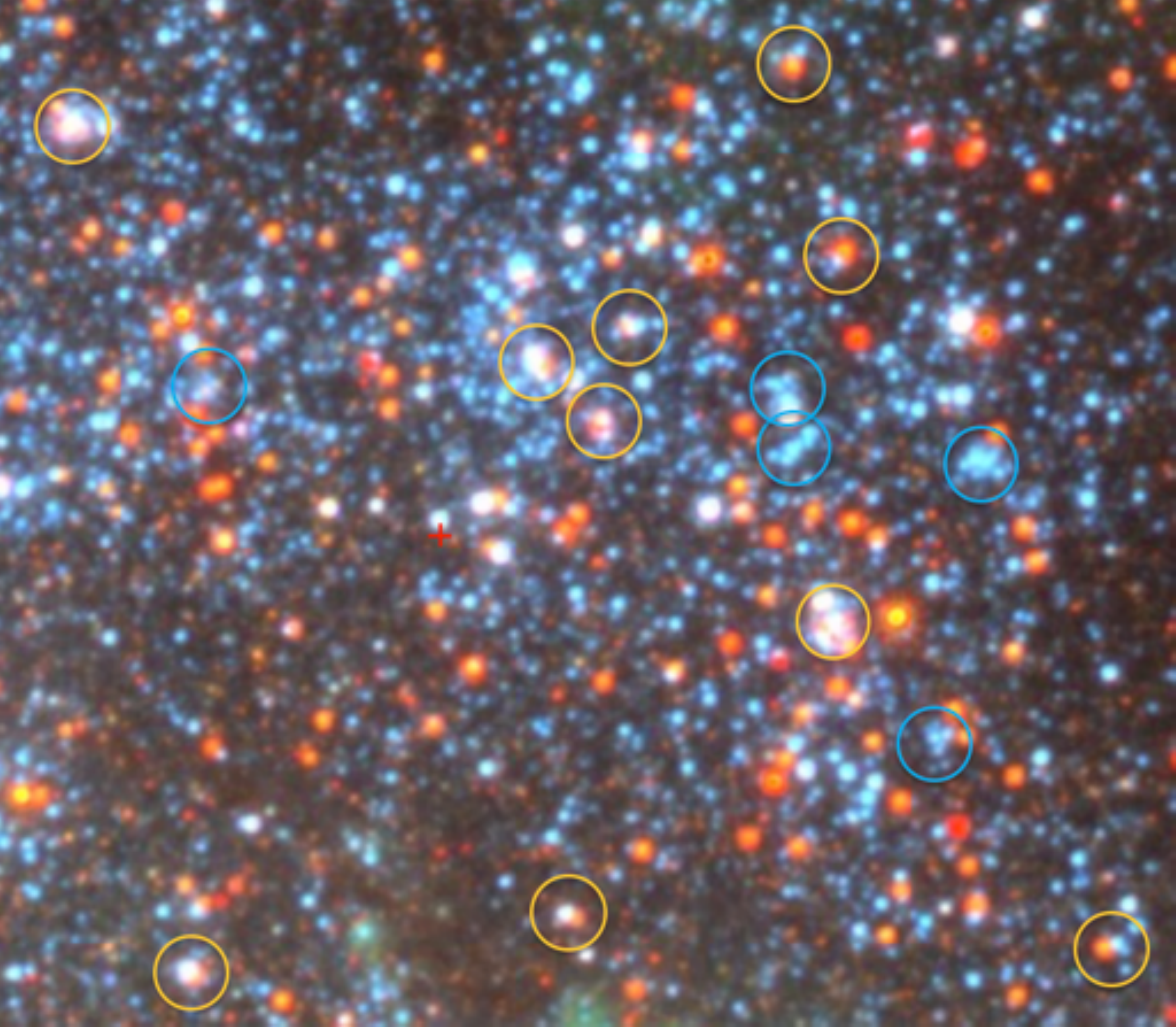}

\end{center}
\caption{A crowded region in NGC 4449 containing 10 of the stochastic candidate objects from Figure 8 (yellow circles), and five category = 3 (compact associations - blue circles) which are not in the
stochastic region of the color-color diagram shown in the top panel of Figure 8 (i.e., they are shown as large squares in Figure 8). Note that {\it all}
of the stochastic candidates have  both red and blue stars near their centers, while none of the compact associations shown by blue circles have bright red stars near their centers. This shows that the position of the stochastic candidates in the color-color diagram is due to the inclusion of a red star, not due to reddening from dust. The region is clearly older than 10 Myr since there is essentially no gaseous emission (green) in the region near the  objects. Note that the circles in Figure 9 are roughly twice as large as the apertures used to make the photometric measurements.
}
\end{figure}

\begin{figure}
\begin{center}
\includegraphics[width = 7.0in, angle= 0]{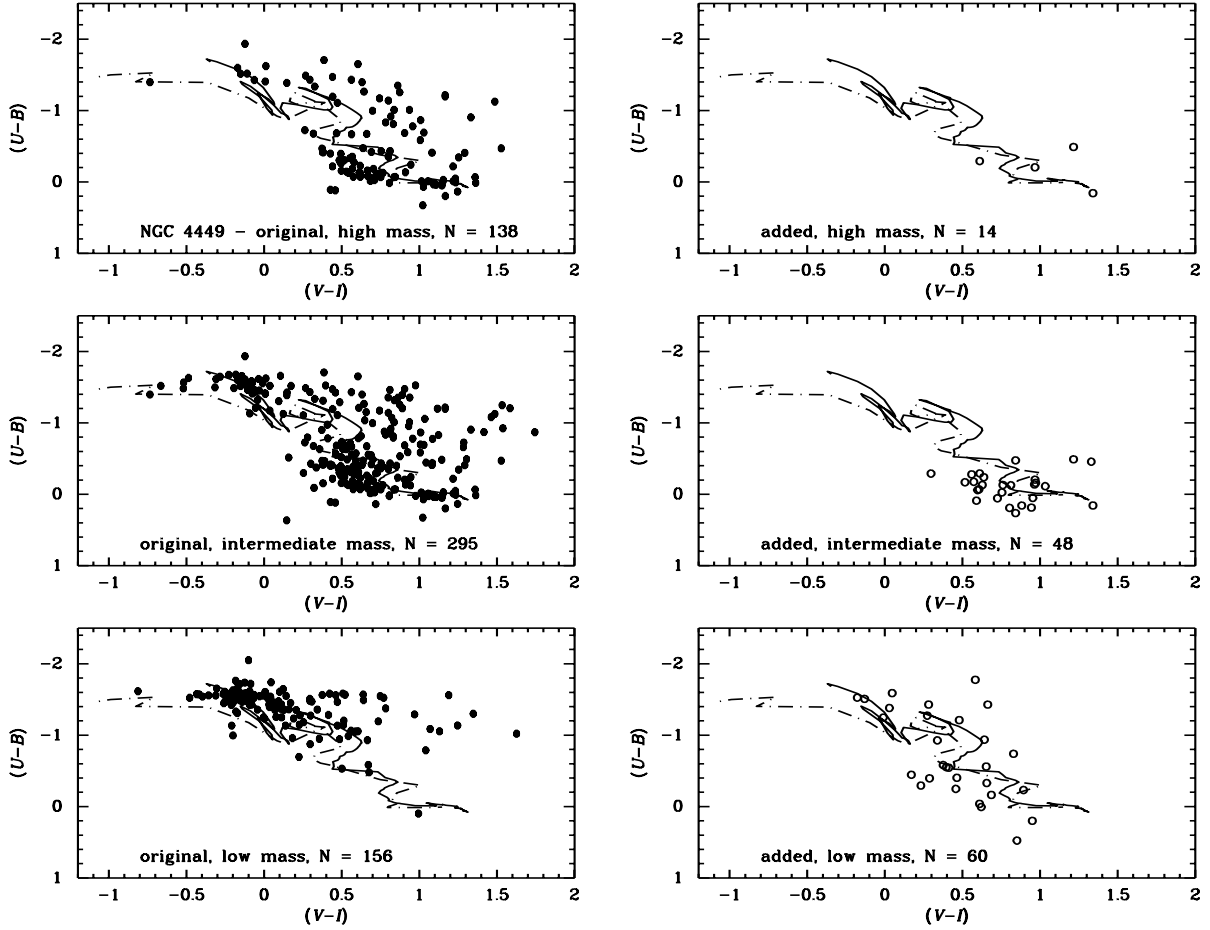}

\end{center}
\caption{V-I vs. U-B color-color diagrams for the originally selected sources (left panels - filled circles) and the clusters that were added (right panels - open circles -  see \S 2.1 for a discussion of the added clusters). The samples are also broken into massive (greater than 10,000 solar mass) at the top, intermediate-mass (greater than 3,000 but less than 10,000 solar masses) in the middle, and low mass (less than 3,000 solar masses) at the bottom. Both the Bruzual-Charlot (solid line) and Yggdrasil (dotted line)  models are included. 
}
\end{figure}

\begin{figure}
\begin{center}
\includegraphics[width = 7.0in, angle= 0]{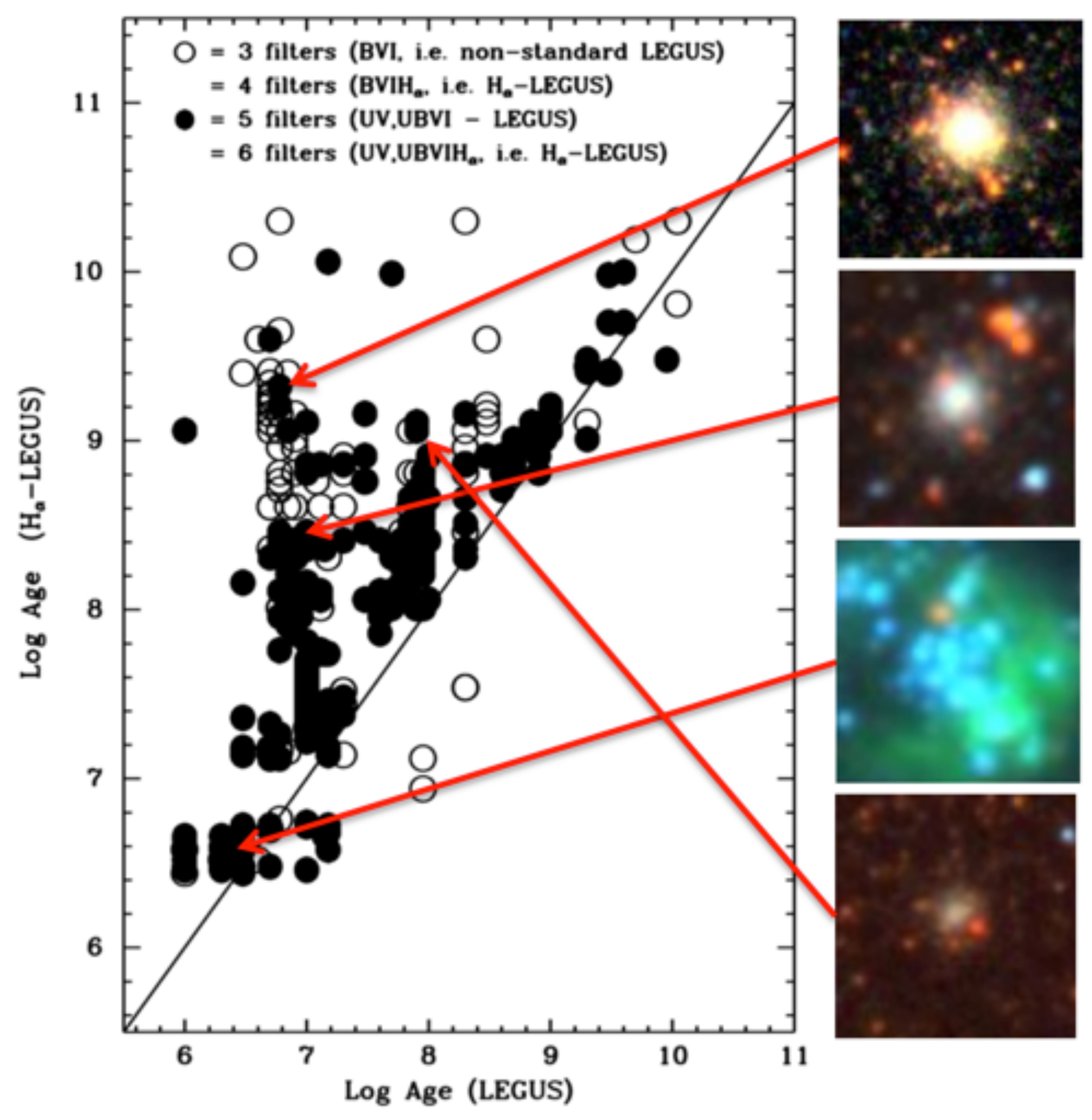}

\end{center}
\caption{Comparison between age estimates using
the LEGUS and the H$_\alpha$-LEGUS  catalogs. 
Small snapshot images show where four typical clusters fall  in the  diagram. Open symbols show clusters with only 3 or 4 filters (i.e, the outskirts without WFC3 UV or U observations). Filled symbols show clusters with  5 or 6 filters. {\bf 
The snapshots have an approximate field of view
of 50 $\times$ 50 pixels.} See \S 4.2 for a discussion.
}
\end{figure}

\begin{figure}
\begin{center}
\includegraphics[width = 6.0in, angle=0]{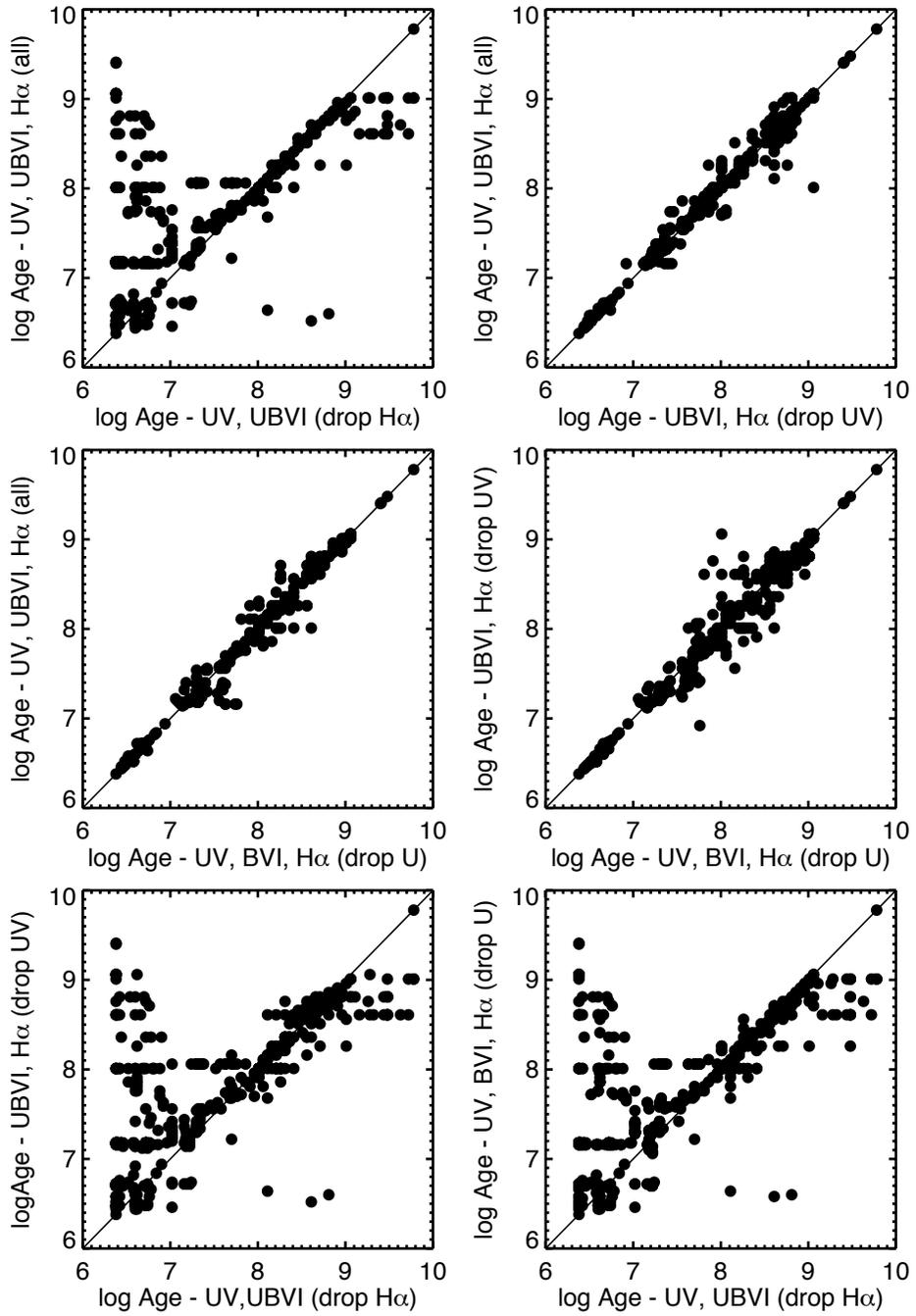}
\end{center}
\caption{Comparisons of log Age estimates using all combinations of  filter choices discussed in \S 4.3.1 . This shows that dropping the H$_\alpha$ filter has the largest effect on the resultant age estimates, i.e., large "chimneys" are present in all the panels involving combinations where H$_\alpha$ is dropped.   }
\end{figure}

\begin{figure}
\epsscale{0.5}
\includegraphics[width = 6.0in, angle=0]{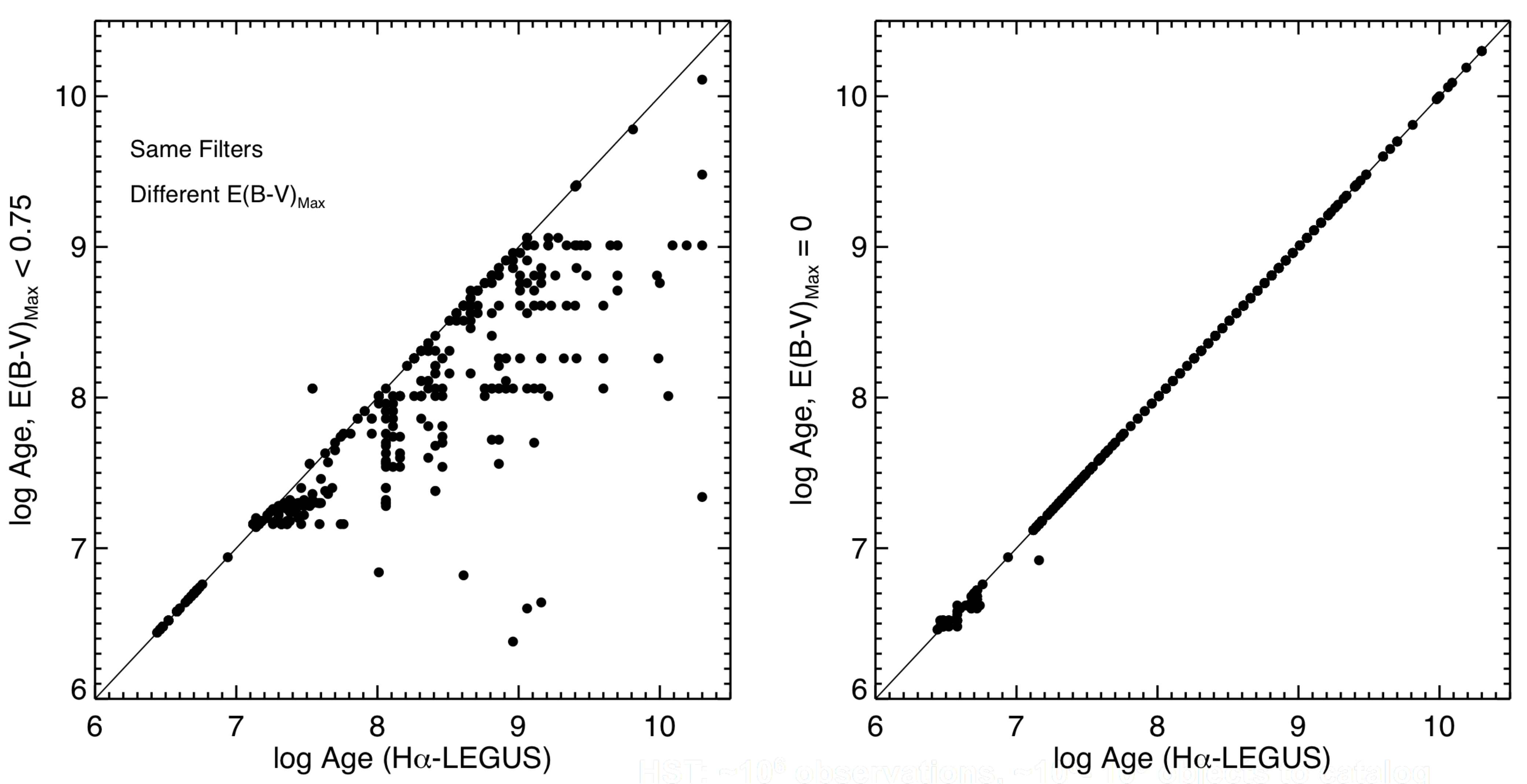}
\caption{Comparisons of log Age estimates using different assumptions about reddening.}
\end{figure}

\begin{figure}
\begin{center}
\includegraphics[width = 6.0in, angle=0]{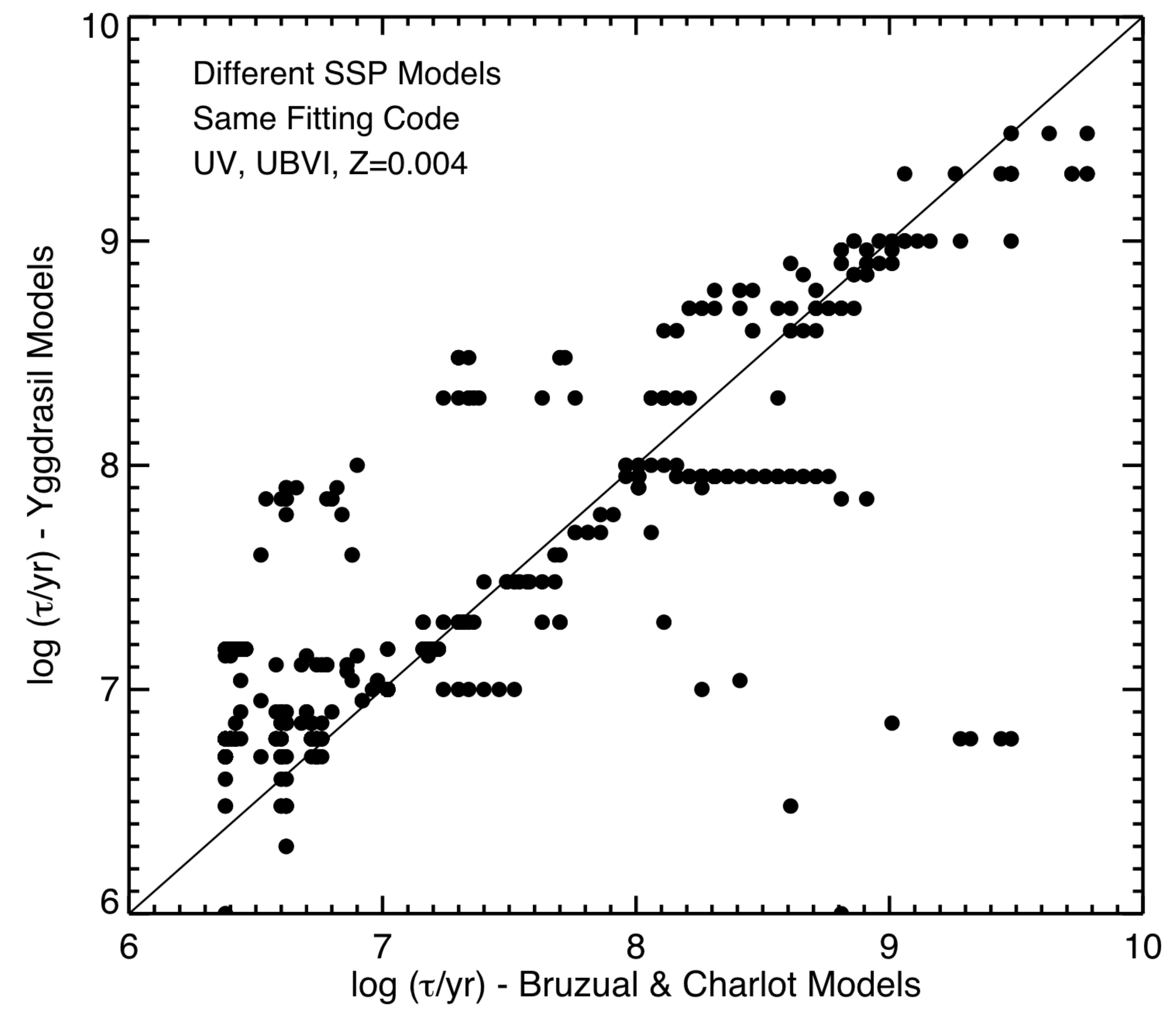}
\end{center}
\caption{Comparisons of log Age estimates using different SSP models. 
}
\end{figure}

\begin{figure}
\begin{center}
\includegraphics[width = 6.0in, angle=0]{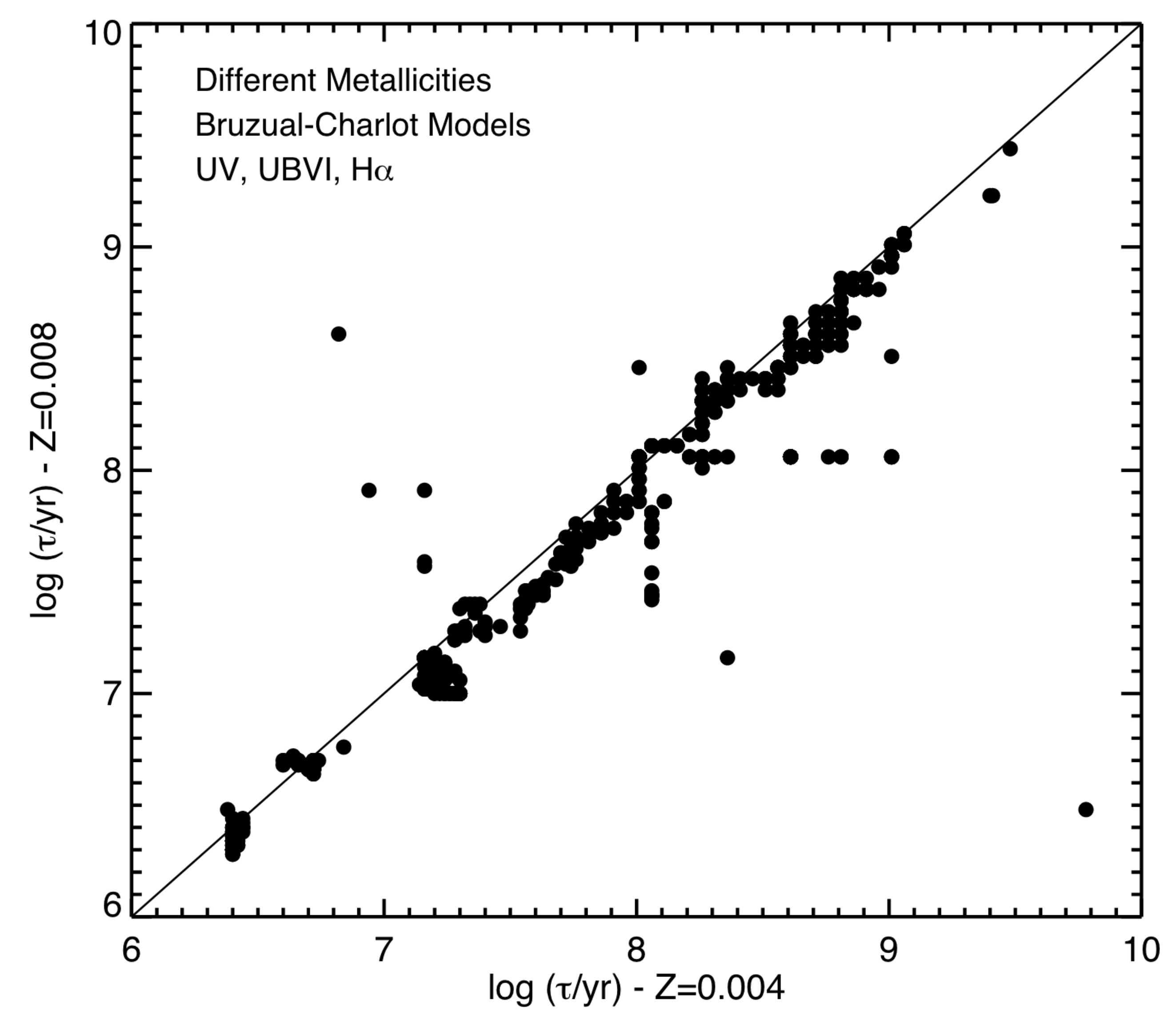}
\end{center}
\caption{Comparisons of log Age estimates using different assumed metallicities. 
}
\end{figure}

\begin{figure}
\begin{center}
\includegraphics[width = 6.0in, angle= 0]{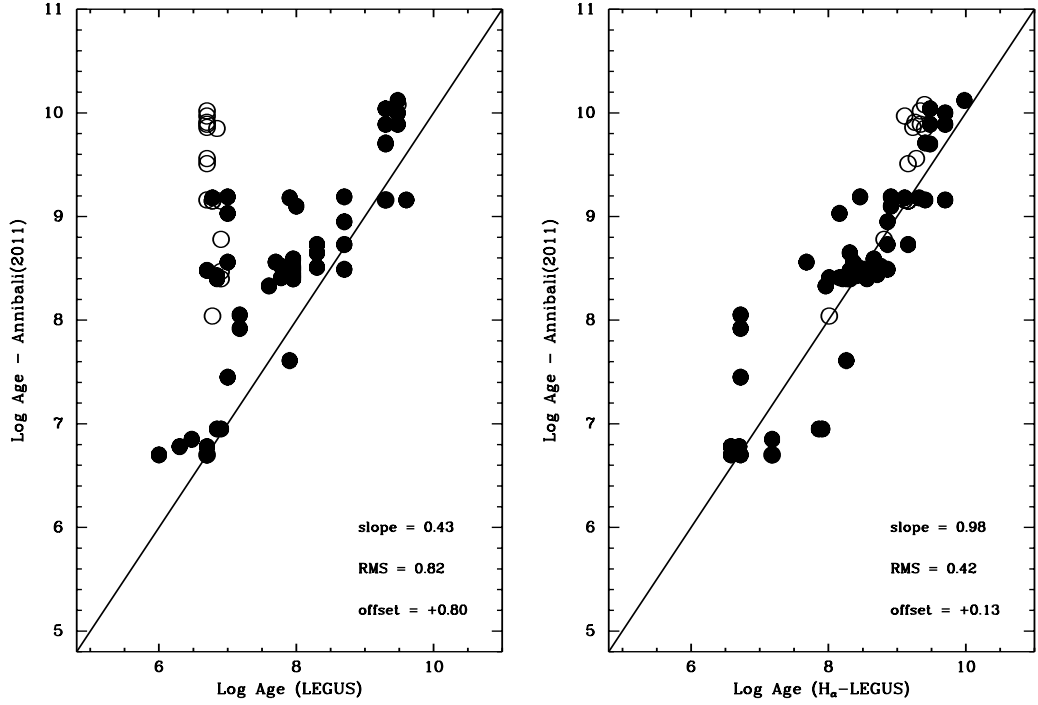}
\end{center}
\caption{Age estimate comparisons between:
(left) - LEGUS and Annibali et al. (2011) solutions, (right) - H$_\alpha$-LEGUS  and Annibali et al. (2011) solutions. 
Open symbols show clusters with only 3 or 4 filters (i.e, the outskirts without WFC3 UV or U observations) while filled circles are clusters with 5 or 6 filters.
Values from linear fits for the slope, RMS scatter, and offsets are provided in each panel. }
\end{figure}

\begin{figure}
\begin{center}
\includegraphics[width = 4.0in, angle=0]{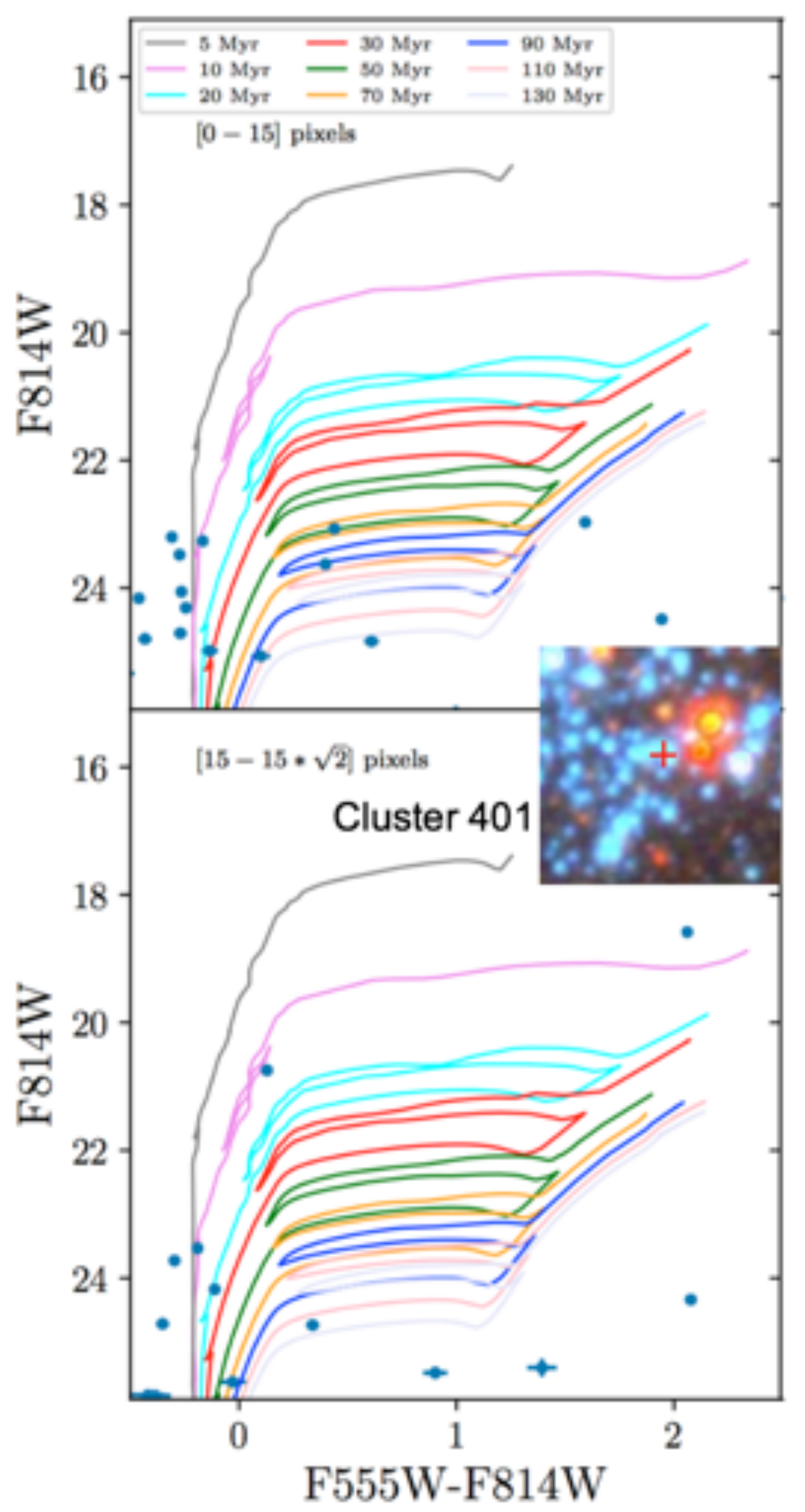}

\end{center}
\caption{Example of color-magnitude diagrams used to estimate age of compact association c3-3144-6144 (alias: cluster 401) using the CMD method.  }
\end{figure}

\begin{figure}
\begin{center}

\includegraphics[width = 6.5in, angle=0]{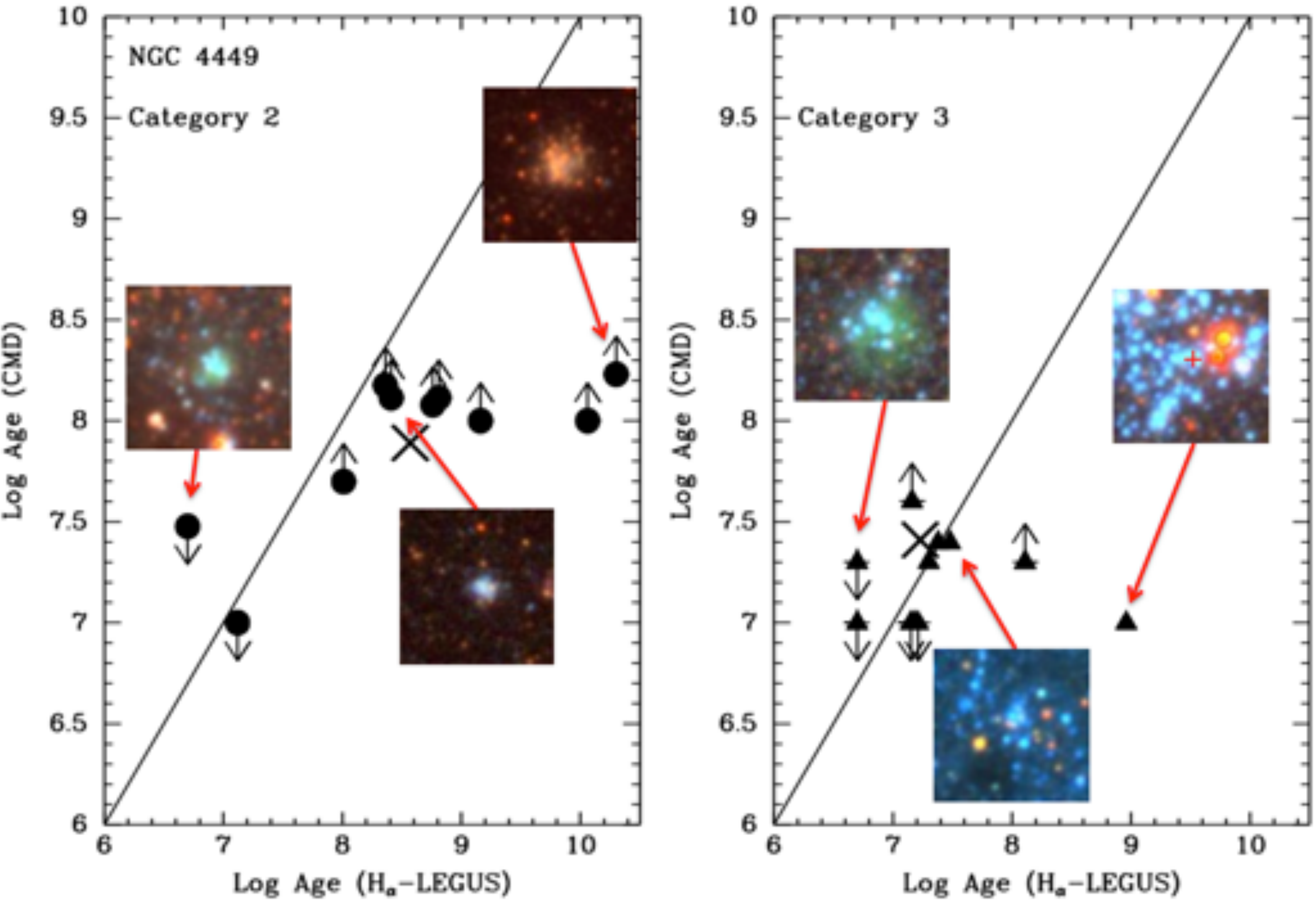}

\end{center}
\caption{Comparison between ages determined by CMD and the H$_\alpha$ - LEGUS method for category 2 and 3 objects. The X marks the location of the means for the distributions. {\bf The arrows show which data points are upper and lower limits, as listed in Table 2.} Snapshots show examples of high, middle, and low points. }
\end{figure}

\begin{figure}
\begin{center}

\includegraphics[width = 5.0in, angle= 0]{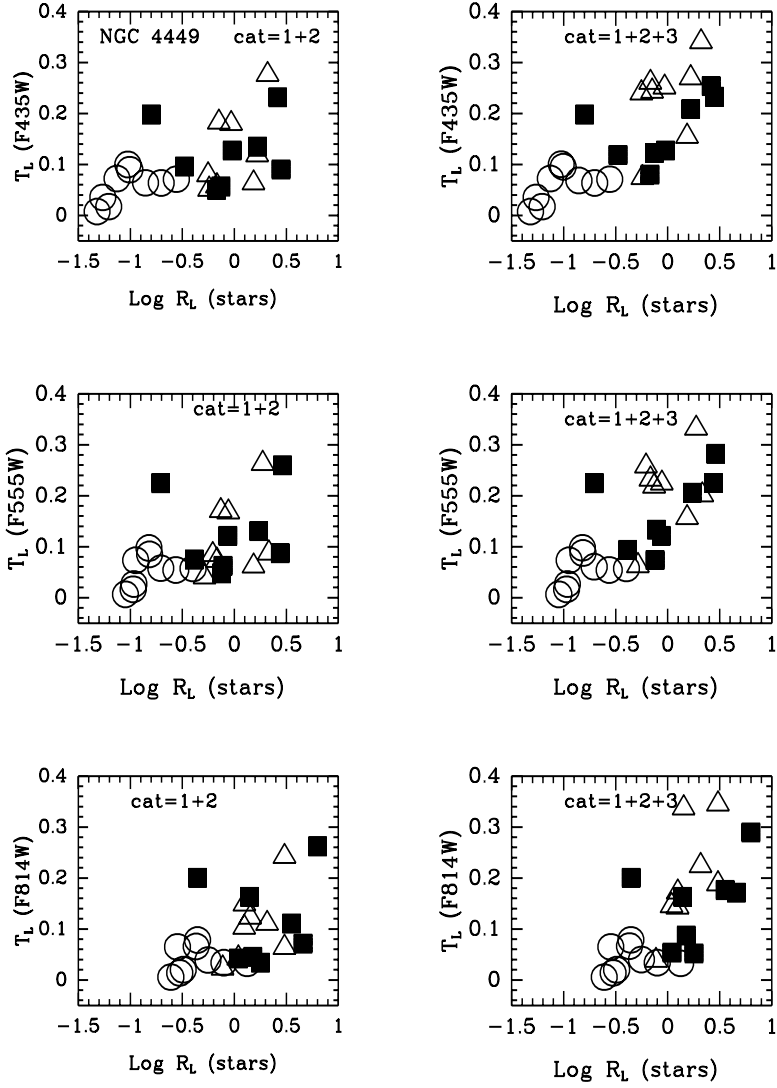}
\end{center}
\caption{Plot of the fraction of light in clusters, T$_L$, versus specific Region Luminosity, R$_L$(stars), for three filters. The left panels show the results for the category 1 + 2 subsample while the right panels show the results for the category 1 + 2 + 3 subsample. Open circles are for regions with H$_\alpha$-LEGUS values of log Age $>$ 8.5; filled squares are for regions with 8.5 $>$ log Age $>$ 7.5; open triangels are for regions with log Age $<7.5$,  based on Table 3. 
}
\end{figure}

\begin{figure}
\begin{center}

\includegraphics[width = 5.0in, angle= 0]{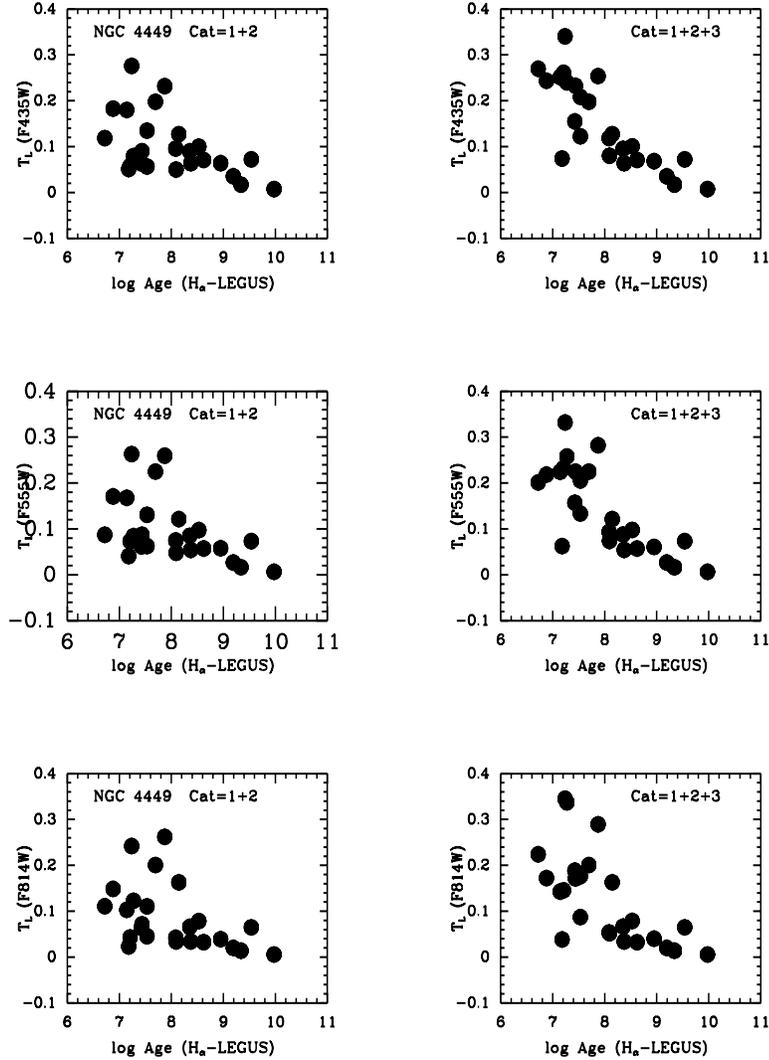} 
 \end{center}
\caption{Plot of the fraction of light in clusters, T$_L$ [using R$_L$ (stars)], versus mean log Age of the clusters in the 25 regions shown in Figure 1. The similarity with Figure~19 demonstrates that specific Region Luminosity and log Age are closely related.
}
\end{figure}

\begin{figure}
\begin{center}

  \includegraphics[width = 8.0in, angle= 0]{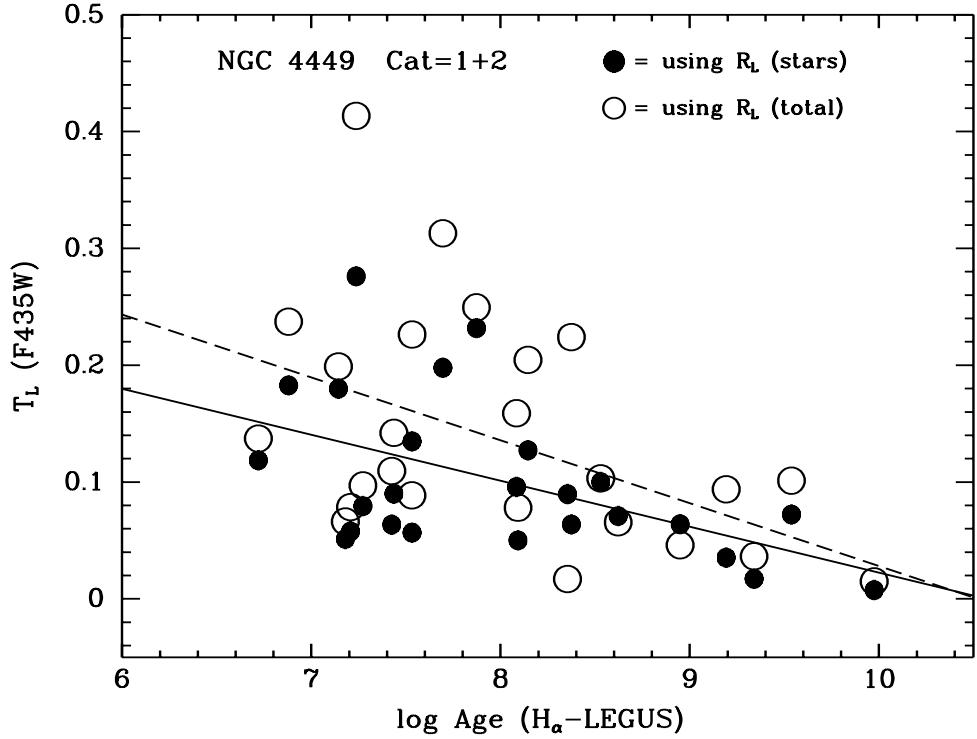} 
\end{center}
\caption{Plot of the fraction of light in clusters, T$_L$, versus mean log Age for the F435W filter. The open circles show the results when T$_L$ is determined using the specific Region Luminosity based on  the pixel values (i.e., R$_L$(total); the solid circles show the results when T$_L$ is determined using specific Region Luminosity based on the luminosity of the stars  from the stellar catalog (i.e., R$_L$(stars).   Linear least-squares fits to the two distributions are included. The best estimate is probably between the two lines, as discussed in the text.
}
\end{figure}

\begin{figure}
\begin{center}

  \includegraphics[width = 6.0in, angle= 0]{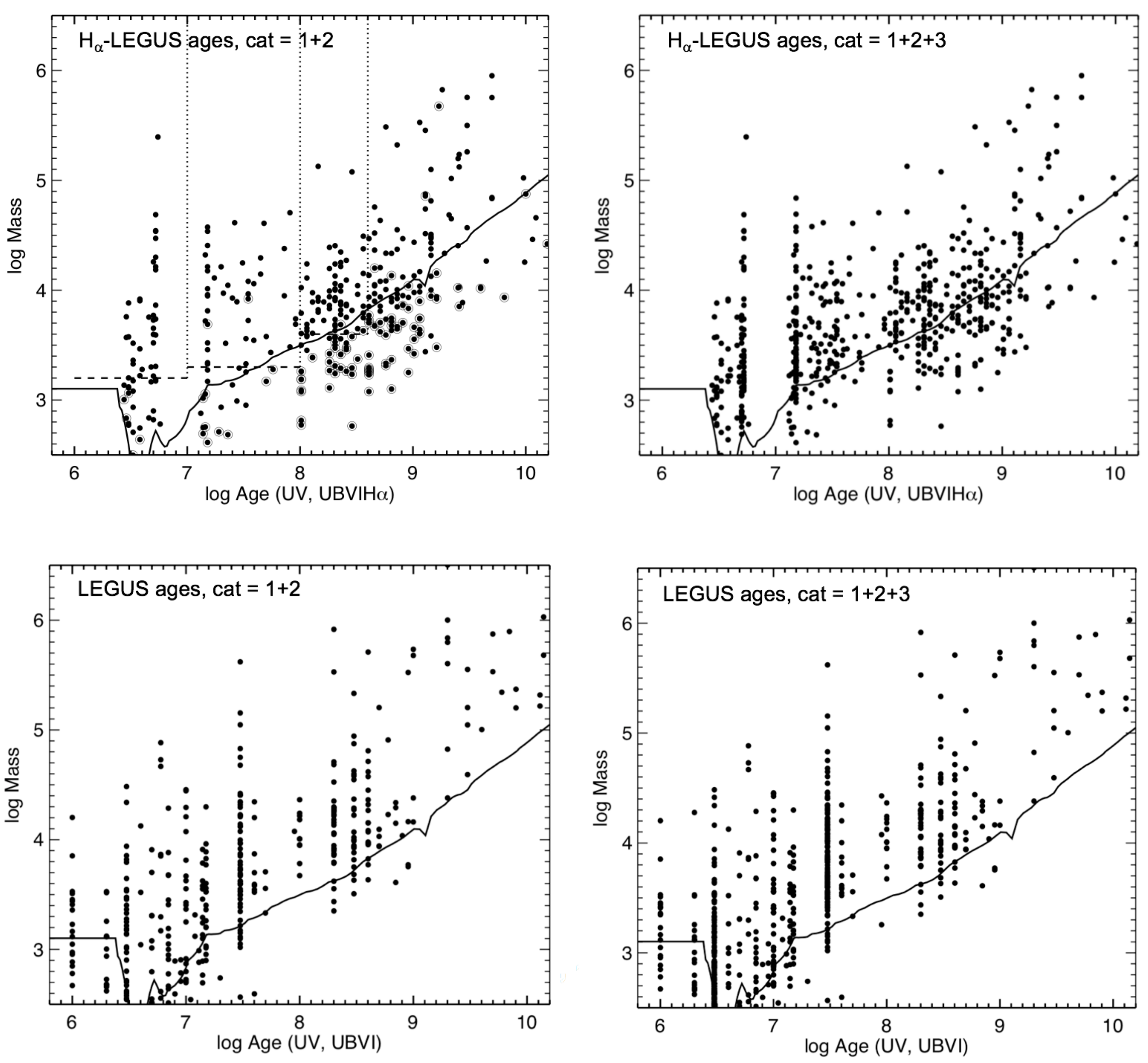}
\end{center}
\caption{Mass - log Age diagrams  for two subsamples (i.e., category 1 + 2 on the left; category 1 + 2 + 3 on the right), and using both the  LEGUS (bottom) and the H$_\alpha$-LEGUS  (top) age-dating methods. The solid line shows the estimated 50 \% completeness limit; the dotted lines show the limits used to make the age and mass functions. The points that are circled in the upper left panel are from added clusters, as discussed in \S 2.1.  Note the apparent enhancement in the number of clusters with log Age values between 8 and 9 in the upper left panel, consistent with the apparent enhancement in the  color-color diagram (Figure 5).  This is discussed in more detail in \S 7.3. 
}
\end{figure}

\begin{figure}
\begin{center}

\includegraphics[width = 6.0in, angle= 0]{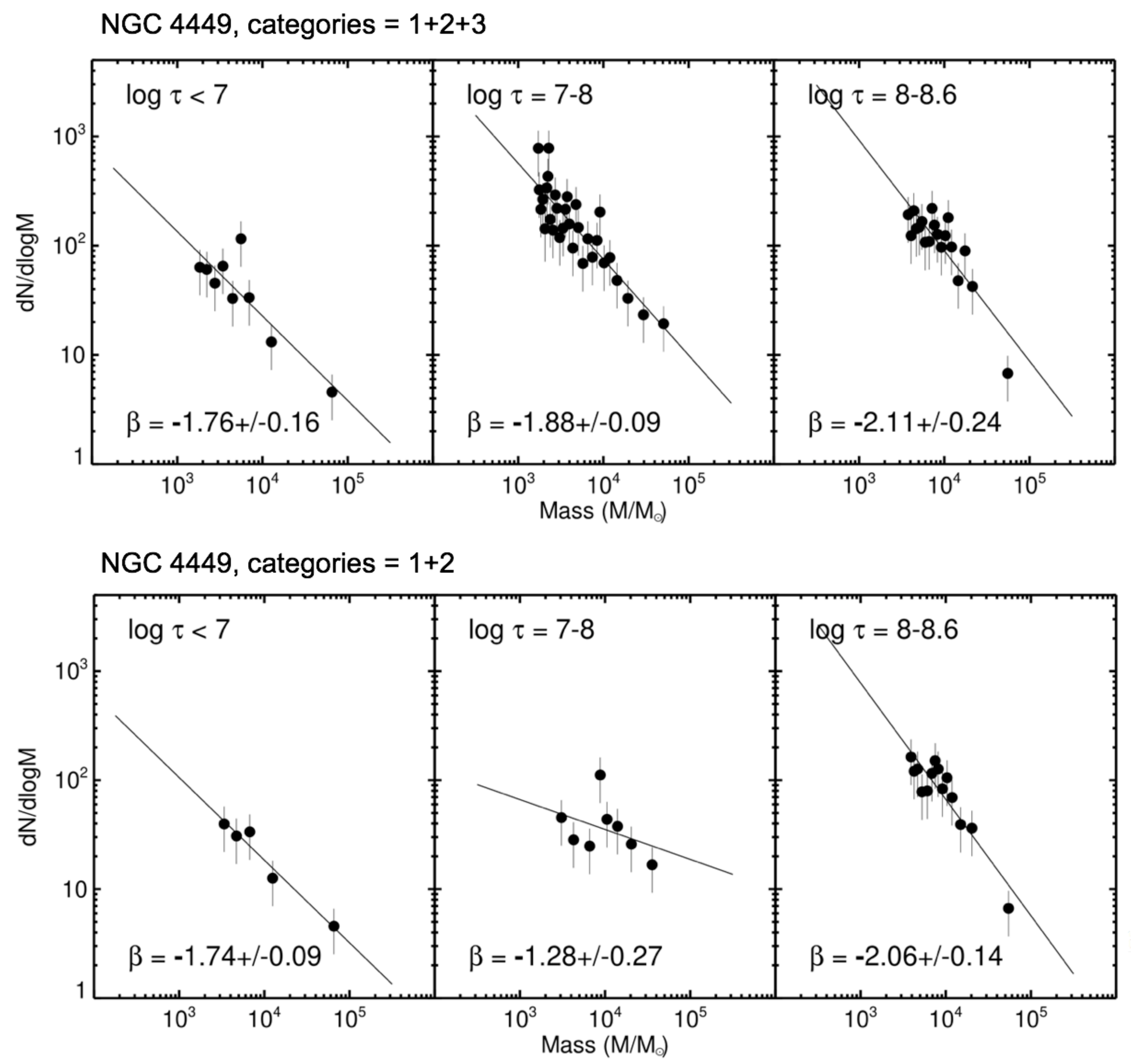}

\end{center}
\caption{Mass functions  for two subsamples (i.e., category 1 + 2 on the top; category 1 + 2 + 3 on the bottom), and three age ranges, as defined in the figure. H$_\alpha$-LEGUS ages are used for all panels. After eliminating  the low and the high values, 
the mean value of $\beta$, the slope of the power law, is $-1.86$, similar to values found in many other studies. }
\end{figure}

\begin{figure}
\begin{center}

\includegraphics[width = 6.0in, angle= 0]{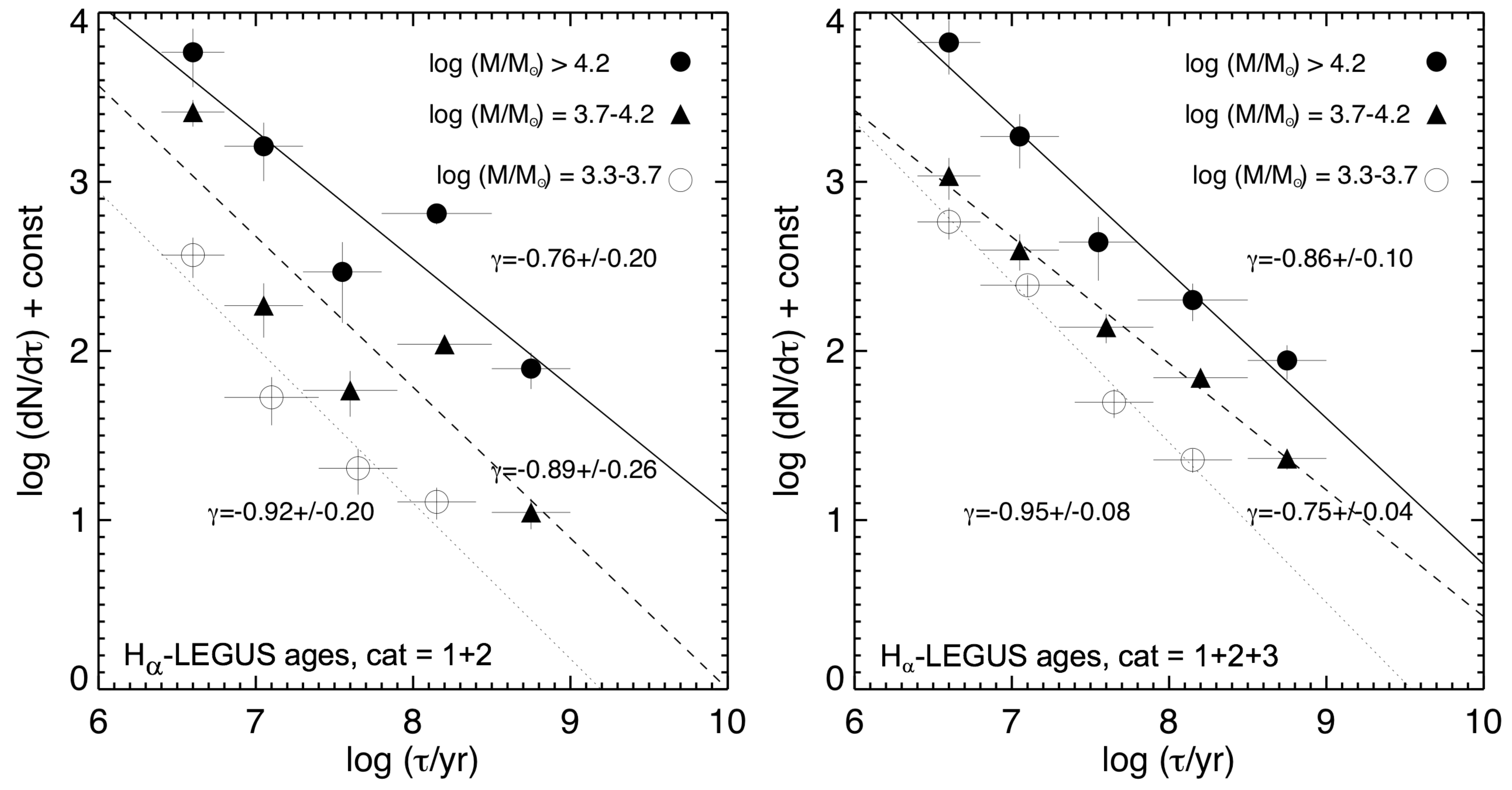}
\end{center}
\caption{
Age distributions  for two subsamples (i.e., category 1 + 2 on the left; category 1 + 2 + 3 on the right), using the   H$_\alpha$-LEGUS age-dating method. The values of $\gamma$, the slope of the power law fit to the age distribution, are relatively stable for all subsamples, ranging from -0.74  to -0.95, with a mean of $-0.85 \pm 0.15$.  There is no apparent dependence  on mass.
}
\end{figure}

\begin{figure}
\begin{center}
 \includegraphics[width = 5.5in, angle= 0]{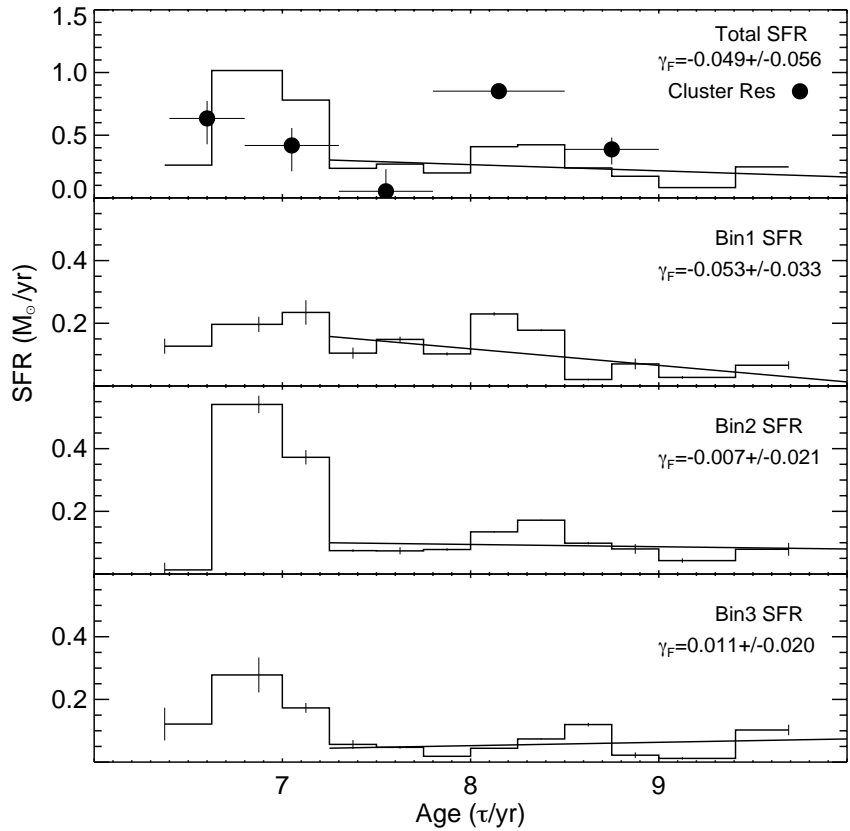}
\end{center}
\caption{A comparison between the star formation histories derived from stars (i.e. the histogram - extraction of three regions used in this study based on Sacchi et al. 2017 data - private communication), and from clusters (the dots - using the H$_\alpha$-LEGUS  ages and the category 1 + 2 subsample for this comparison). The top panel shows the results for the total sample while the next three panels below show the results for the inner to outer thirds of the sample. The cluster age distribution has been divided by the mean slope (i.e., $\gamma = -0.85$) to remove the effects of disruption for this comparison, since the primary goal is to measure the enhancement in the age range 100 - 300 Myr. The two distributions are normalized using the log Age = 7.5 and 8.7 points. 
  }
\end{figure}

\clearpage

\begin{figure}
\begin{center}
\includegraphics[width = 6.5in, angle= 0]{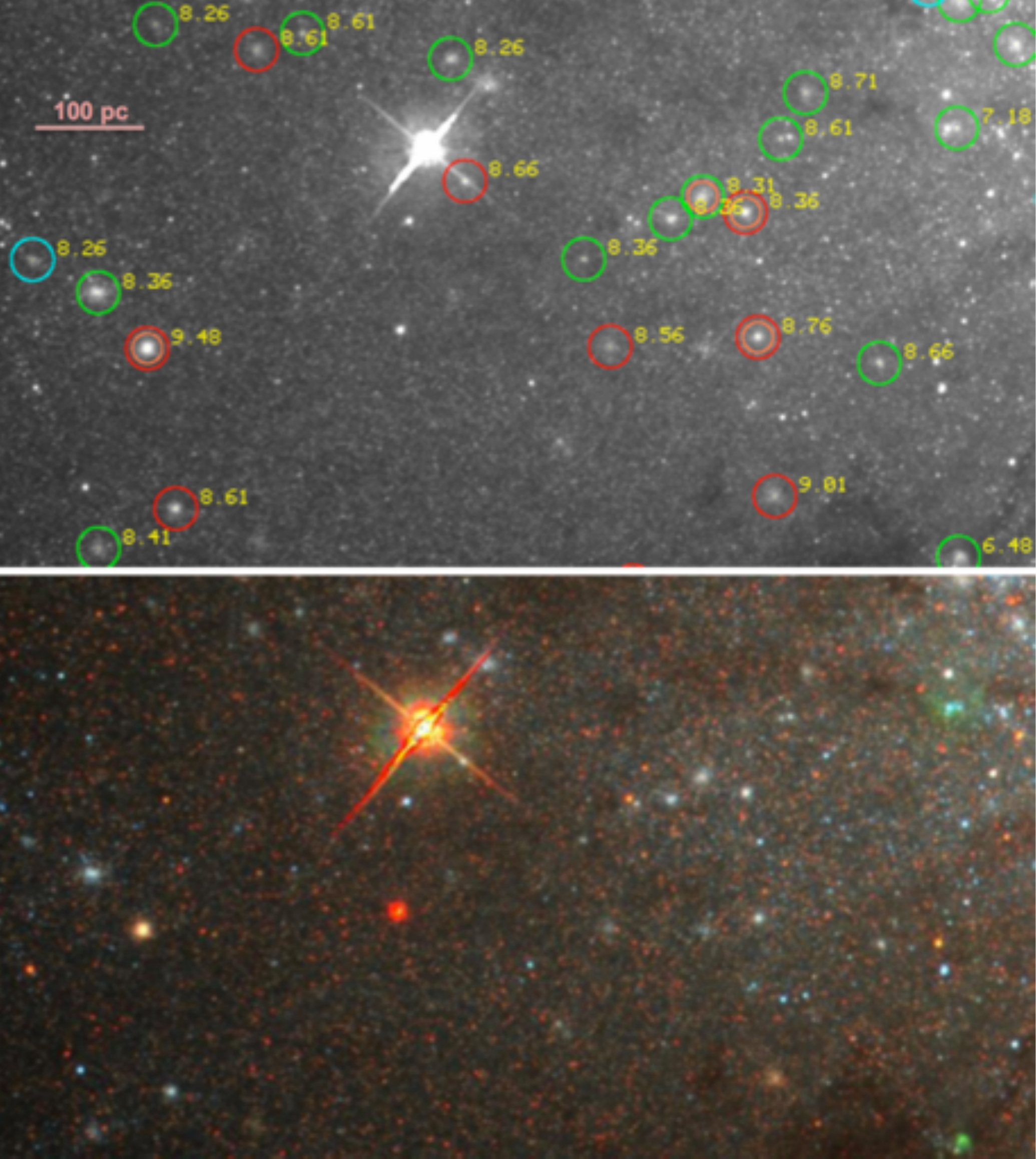}
\end{center}
\caption{Clusters and ages for region 16 (and slightly to the west) from Figure 1, with images and symbols as defined for  Figure 4. Note that 18 of the 22 objects have log Age values  in the range 8.0  to 8.8 (i.e., the burst discussed in  \S 7.3).   Region 23 on the opposite side of the nucleus, has a  similar population of intermediate age clusters. }
\end{figure}

\begin{tiny}
\begin{landscape}
\begin{deluxetable}{lccccccccccc}

\tablecaption{Comparison of log Age Values for Regions\label{tab:alpha}}
\tablewidth{0pt}
\tablehead{
\colhead{Reg. \#}  & \colhead{\# clust.} & \colhead{log Age$^a$} & \colhead{Sigma} & \colhead{log Age$^a$}  
& \colhead{Sigma}  & \colhead{log Age}  
& \colhead{Sigma}  & \colhead{Log R$_L$(st)$^b$,$^c$} & \colhead{T$_L$(st)$^b$} & \colhead{Log R$_L$(tot)$^b$,$^c$} & \colhead{T$_L$(tot)$^b$}
\\
& &  H$\alpha$-LEGUS & &  LEGUS  & & Annibali (2011) & \\
&  &  (yr [median]) & (yr) & (yr) & (yr) &  (yr and \#) & (yr) 
}

\startdata

1  &  4      &  9.19  [9.14] & 0.17     & {\it 7.64}  & 1.23    & 9.16 (1) & ---  & -1.27 & 0.035 & -1.70 & 0.094 \\ 
2  &  2      &  9.34 [9.34]  & 0.08     & {\it 6.77} & 0.10    & 9.70 (2)  &  0.21 & -1.21 &  0.017 & -1.54 & 0.036\\ 
3  &  12      &  6.72 [6.70] & 0.16     & 6.55 & 0.27  & 7.92 (1) &  ---   & 0.22 & 0.119 & 0.16 & 0.137\\ 
4  &  4      &  7.27 [7.16] & 0.67     & 7.00  & 0.00  & --- & ---  & -0.25 & 0.079 & -0.34  & 0.097 \\ 
5  &  2      &  9.98 [9.98]  & 0.47     & 8.41 & 2.31 & ---  & ---  & -1.32  & 0.007 & -1.65 & 0.015 \\

6  &  4      &  9.54 [9.38] & 0.52     & {\it 7.48} & 1.71  & 9.93 (2) & 0.06 & -1.13 & 0.072  & -1.28 & 0.101\\ 
7  &  13      &  7.24 [7.18]  & 0.22     & 6.86 & 0.20 & --- & --- & 0.32 & 0.276  & 0.14  & 0.413 \\ 
8  &  22      &  7.45  [7.38] & 0.26     & 6.94 & 0.27 & --- & --- & 0.19 & 0.064 & -0.04 & 0.109 \\ 
9  &  12      &  7.14 [7.16] & 0.83     & 6.64 & 0.36 & --- & --- & -0.03 & 0.180  & -0.07 & 0.199 \\ 
10  &  5      &  8.35  [8.61] & 1.43     & 7.73 & 1.60 & 8.89 (3) & 1.83 & -1.01  & 0.090  & -0.286 & 0.017\\ 

11  &  20      &  {\it 7.53} [7.18] & 1.09     & {\it 6.83} & 0.65 & 9.18 (1) & --- & -0.13  & 0.057 & -0.32 & 0.089 \\ 
12  &  8     &  7.21 [7.18] & 0.79     & 6.87 & 0.85  & --- & ---  & -0.17  & 0.057  & -0.31 & 0.078\\ 
13  &  20      &  8.08 [8.28]  & 0.86     & 7.54 & 0.87 & 8.41 (2)  & 0.01 & -0.48  & 0.096 & -0.70 & 0.159\\ 
14  &  11      &  8.94 [8.76] & 0.58     & 8.55 & 0.70 & 9.66 (3) & 0.48 & -0.85 & 0.064  & -0.71 & 0.046\\
15  &  73     &  7.53 [7.46] & 0.60    & 7.10 & 0.43  & 8.00 (5)  & 0.75  & 0.22 & 0.135  & -0.04 & 0.226\\ 

16  &  13      &  8.62 [8.66] & 0.23     & 7.92 & 0.66  & 8.51 (4) &  0.03 & -0.56 & 0.071 & -0.52 & 0.065\\ 
17  &  31      &  7.87 [8.06] & 0.75     & {\it 7.22} & 0.66  & 8.81 (8) & 0.43 & 0.42  & 0.232 & 0.39 &  0.249\\ 
18  &  8    &  6.88 [6.72] & 0.23     & 6.63 & 0.19 & --- & --- & -0.15 & 0.183 & -0.26 & 0.237\\ 
19  &  38      &  7.44 [7.18]  & 0.68     & 6.93  & 0.56  & 6.76 (4)  & 0.07 & 0.45  & 0.090  & 0.25 & 0.142 \\ 
20  &  18      &  8.09  [8.04] & 0.73     & 7.46 & 0.73  & 7.89 (3) & 0.81 & -0.17 & 0.050 & -0.36 & 0.078 \\ 

21  &  2      &  8.14 [8.14] & 1.36     & 7.59 & 1.26  & 6.70 (1) & --- & -0.02 & 0.127 & -0.23 & 0.205\\ 
22  &  16      &  8.53 [8.78] & 0.82     & {\it 7.38} & 0.92 & 9.67 (5) & 0.47 & -1.02  &  0.100 & -1.03  & 0.103\\ 
23  &  26      &  8.37 [8.38] & 0.71     & 7.37 & 0.76 & 8.33 (4) & 0.20 & -0.70 & 0.064 & -1.24  & 0.224\\ 
24  &  6      &  7.70 [7.54] & 1.11     & {\it 7.11} & 0.96  & 9.86 (1) & --- & -0.80 & 0.198  & -1.00 & 0.313 \\ 
25  &  2      &  7.18 [7.18] & 0.00     & 6.78 & 0.00 & ---  & --- & -0.24 & 0.051 & -0.35 & 0.066\\

\\

& & mean = 8.01 & = 0.61 & = 7.25 & = 0.73 & = 8.67 & = 0.45 \\

\hline
\enddata

\noindent $^a$ Values in italics show values discrepant by more than 1.5 from Annibali (2011).

$^b$ Values of Log R$_L$ and T$_L$ are for F435W. Values for F555W and F814W are available on request. 

$^c$ Values for the specfic Region Luminosities (i.e., Log R$_L$) are on the same relative scale for both (stars) and (total). 

\end{deluxetable}
\end{landscape}
\end{tiny}


\begin{references}{}
 
\reference{ref1}{}
Adamo, A.K, Kruijssen, J.M.D., Bastian, N., Silva-Villa, E., Ryon, J. 2015, MNRAS, 452, 246

\reference{ref1}{}
 Adamo, A., Ryon, J.~E., Messa, M., et al.\ 2017, \apj, 841, 131 


\reference{ref1}
Annibali, F., Morandi, E., Watkins, L.~L., et al.\ 2018, \mnras, 476, 1942 


\reference{ref1}
Annibali, F., Aloisi, A., Mack, J., et al.\ 2008, \aj, 135, 1900 


\reference{ref1}
Annibali, F., Tosi, M, Aloisi, A. et al.\ 2011, \aj, 142, 129

\reference{ref1}
Annibali F., Tosi, M., Romano, D., et al.  2017, ApJ, 843, 20


\reference{ref1}
Ashworth, G. et al.\ 2017, \mnras, 469, 2464


\reference{ref1}
Bastian, N.  2008, \mnras, 390, 759 

\reference{ref1}
Bastian, N., Adamo, A., Gieles, M., et al.\ 2012, \mnras, 419, 2606 

\reference{ref1}
Billett, O. H., Hunter, D. A., \& Elmegreen, B. G. 2002, AJ, 123, 1454

\reference{ref1}
Bitsakis, T. et al. 2017, \apj, 845, 56

\reference{ref1}
Bitsakis, T. et al. 2018, \apj, 853, 104

\reference{ref1}
Bruzual, G., \& Charlot, S. 2003, \mnras, 344, 1000


\reference{ref1}
Calzetti, D., Lee, J.~C., Sabbi, E., et al.\ 2015, \aj, 149, 51 

\reference{ref1}
Chabrier, G. \ 2003, \pasp, 115, 763







\reference{ref1}
Chandar, R., Fall, S.~M., \& Whitmore, B.~C.\ 2015, \apj, 810, 1 

\reference{ref1}
Chandar, R., Fall, S.~M., Whitmore, B.~C., \& Mulia, A.~J.\ 2017, \apj, 849, 128 


\reference{ref1}
Chandar, R., Whitmore, B.~C., Kim, H., et al.\ 2010, \apj, 719, 966 

\reference{ref1}
Chandar, R., et al.  2019, in preparation




\reference{ref1}
Cignoni, M., Sacchi, E., Aloisi, A., et al.\ 2018, \apj, 856, 62 

\reference{ref1}
Cook, D., Lee, J. C., Adamo, A., et al. 2019, \mnras, 484, 4897






\reference{ref1}
Fall, S. M., Chandar, R., \& Whitmore, B. C. 2005, ApJ, 631, L133 


\reference{ref1}
Fall, S. M., \& Chandar, R.  2012, ApJ, 752, 96

\reference{ref1}
Fitzpatrick, E. L. 1999, PASP, 111, 63

\reference{ref1}
Fouesneau, M., \& Lancon, A. 2010, A\& A,521, A22

\reference{ref1}
Fouesneau, M., Lançon, A., Chandar, R., \&  Whitmore, B. C. 2012, ApJ, 750, 60

\reference{ref1}
Fouesneau, M., Johnson, L. J., Weisz, D. R. Et al., 2014, ApJ, 786, 117




\reference{ref1}
Gelatt, A.~E., Hunter, D.~A., \& Gallagher, J.~S.\ 2001, \pasp, 113, 142 


\reference{ref1}
Glatt, K.,  Grebel, E. K., Koch, A. 2010, A\&A, 517A, 50


\reference{ref1}
Grasha, K. et al.  2017, ApJ, 840, 113

\reference{ref1}
Grasha, K. et al.  2018, \mnras, 481, 1016



\reference{ref1}
Goddard, Q. E., Bastian, N., \& Kennicutt, R. C. \ 2010, \mnras, 405, 857

\reference{ref1} 
Hannon, S. et al.  2019, MNRAS, in press

\reference{ref1}
Harris, J., \& Zaritsky, D.  2004, \aj,  127, 1531

\reference{ref1}
Harris, J., \& Zaritsky, D.  2009, \aj,  138, 1243

\reference{ref1}
Gieles M., Larsen S. S., Scheepmaker, R. A., et al. 2006, A\&A, 446L L9


 \reference{ref1}
 Grasha, K., Calzetti, D., Adamo, A., et al.\ 2017, \apj, 840, 113 

\reference{ref1}
Hollyhead, K., Bastian, N., Adamo, A., et al.\ 2015, \mnras, 449, 1106 

\reference{ref1}
Hunter, D.~A., Wilcots, E.~M., van Woerden, H., Gallagher, J.~S., \& Kohle, S.\ 1998, \apjl, 495, L47 

\reference{ref1}
Hunter, D.~A., van Woerden, H., \& Gallagher, J.~S.\ 1999, \aj, 118, 2184 

\reference{ref1}
Johnson, L.~C., Seth, A. C., Dalcanton, J. J. et al. \ 2012, \apj, 752, 95

\reference{ref1}
Johnson, L.~C., Seth, A. C., Dalcanton, J. J. et al. \ 2016, \apj, 827, 33

 \reference{ref1}
Johnson, L. C., Seth, A., C, Dalcanton, J. J. et. al. 2017, ApJ, 839, 78










\reference{ref 1} 
Karachentsev, I. D., Karachentseva, V. E., \& Huchtmeier, W. K. 2007,
Astron. Lett., 33, 512

\reference{ref 1} 
Kim, H. et al. 2012, ApJ, 753, 26

\reference{ref 1} 
Kim, H. et al. 2019, in preparation 


\reference{ref 1}
Krumholz, M. R., Adamo, A., Fumagalli, M., et al. 2015, ApJ, 812, 147


\reference{ref 1}
Krumholz, M. R., McKee, C. F. \&  Bland-Hawthorn, J. 2019, in press





\reference{ref1}
Kruijssen, J.~M.~D.\ 2012, \mnras, 426, 3008 

\reference{ref1}
Larsen, S.~S.\ 1999, \aaps, 139, 393 

\reference{ref1}
Larsen, S.~S., de Mink, S.~E., \& Eldridge, J. J. \ 2011, \aap, 532, 147


\reference{ref1}
Larsen, S.~S., \& Richtler, T.\ 2000, \aap, 354, 836 

\reference{ref1}
Larsen, S.~S., \& Richtler, T.\ 1999, \aap, 345, 59 


%



\reference{ref1}
Leitherer, C. \& Heckman, T. M. 1995, \apjs, 96, 9

\reference{ref1}
Maiz Apellaniz, J. \& Ubeda, L. \ 2005, \apj, 629, 873


\reference{ref1}
Maiz Apellaniz, J. 2009, \apss, 324, 95

\reference{ref1}
Massey, P., Lang, C. C., Degioia-Eastwood, K., \& Garmany, C. D. 1995, ApJ, 438, 188


\reference{ref1}
McQuinn, K.~B.~W., Skillman, E.~D., Cannon, J.~M., et al.\ 2010, \apj, 721, 297 



\reference{ref1}
Martínez-Delgado, D., Romanowsky, Aaron J., \&  Gabany, R. 2012, \apj, 748, 24

\reference{ref1}
Matthews, A. M, Johnson, K. E., Whitmore, B. C. et al. 2018, in press.


 \reference{ref1}
  Messa, M., Adamo, A., {\"O}stlin, G., et al.\ 2018, \mnras, 473, 996 

\reference{ref1}
Meurer, G.~R., Heckman, T.~M., Leitherer, C., et al.\ 1995, \aj, 110, 2665 




\reference{ref1}
Mok A., Chandar, R., Fall, S. M. 2019, ApJ, 872, 93






\reference{ref1}
Rangelov, B.,  Prestwich, A., \& Chandar, R.  2011, ApJ, 741, 86

\reference{ref1}
Reines, A. E., Johnson, K. E., \& Goss, W. M. 2008, \aj, 135, 2222

\reference{ref1}
Reines, A. E., Nidever, D. L., Whelan, D. G., \& Johnson, K. E. 2010, \apj, 708, 26

\reference{ref1}
Rich, R. M., Collins, M. L. M.,  Black, C. M.,  Longstaff, F. A.,  Koch, A.,  Benson, A., \& Reitzel, D. B. 2012, Nature, 482, 192





\reference{ref1}
Sabbi, E., Calzetti, D., Ubeda, L., et al.\ 2018, \apjs, 235, 23 

\reference{ref1}
Sacchi, E., Cignoni, M, Aloisi, A. et al.\ 2018, \apj, 857, 63

\reference{ref1}
Schlegel D. J., Finkbeiner D. P., \& Davis M.,  1998, \apj, 500, 525





 \reference{ref1}
Silva-Villa, E. \& Larsen, S.S. A \&A, 2011, 529, A24




\reference{ref1}
Sokal, K. R., Johnson, K. E., Indebetouw, Rémy, \& Reines, Amy E. 2015, \aj, 149, 115

\reference{ref1}
Theis, C., \& Kohle, S.\ 2001, \aap, 370, 365 

\reference{ref1}
VandenBerg D. et al. 2013, \apj, 775, 134


\reference{ref1}
Whitmore, B.~C., Allam, S.~S., Budav{\'a}ri, T., et al.\ 2016, \aj, 151, 134 


\reference{ref1}
Whitmore, B.~C., Chandar, R., Bowers, A.~S., et al.\ 2014, \aj, 147, 78 

\reference{ref1}
Whitmore, B.~C., Chandar, R., Kim, H., et al.\ 2011, \apj, 729, 78 

\reference{ref1}
Whitmore, B. C., Chandar, R., \& Fall, S. M. 2007, AJ, 133, 1067

\reference{ref1} Whitmore, B. C., \& Schweizer, F. 1995, AJ, 109, 960

 \reference{ref1}
 Whitmore, B. C., Chandar, R., Schweizer, F. et al. 2010, AJ, 140, 75


\reference{ref1}
Whitmore, B. C.  2004, ASPC, 322, 419

\reference{ref1}
Zackrisson, E., Rydberg, C.-E., Schaerer, D.,Ostlin, G., \& Tuli,
M. 2011, \apj, 740, 13





\reference{ref1}
Zaritsky, D., Harris, J., Thompson, I. B.,  Grebel, E. K., \& Massey, P.
2002, AJ, 123, 855


  

\end{references}
\end{document}